\documentclass[]{imsart}
\usepackage{amsfonts}
\usepackage{amsmath}
\usepackage{titlesec}
\usepackage{amssymb}
\usepackage{amsthm}
\usepackage[english]{babel}
\usepackage[utf8]{inputenc}
\usepackage{algorithm}
\usepackage[noend]{algpseudocode}
\usepackage{epsfig}
\usepackage{amscd}
\usepackage{mfirstuc}
\usepackage{ifpdf}
\usepackage[backref,bookmarksdepth=3,bookmarksnumbered,%
  pdftitle={Hamiltonian Monte Carlo on Symmetric and Homogeneous Spaces via %
    Symplectic Reduction},%
  pdfauthor={Alessandro Barp, A. D. Kennedy, and Mark Girolami},%
  pdfstartview={FitH},pdfpagemode={UseOutlines},%
  pdfpagescrop={110 110 500 720},%
  colorlinks,citecolor=blue,urlcolor=blue]{hyperref}
\startlocaldefs
\def\epspdffile#1{\leavevmode\ifpdf\epsffile{Figures/#1.pdf}%
  \else\epsffile{Figures/#1.eps}\fi}
\newif\ifdraft \draftfalse
\newcommand{\A}{\mathcal A}              % Connection
\renewcommand{\H}{{\mathcal H}}          % Hausdorff measure
\renewcommand{\L}{{\mathcal L}}          % Target measure
\newcommand{\R}{\mathbb R}               % Set of real numbers
\newcommand{\C}{\mathbb C}               % Set of complex numbers
\newcommand{\Z}{\mathbb Z}               % Set of integers
\newcommand{\NN}{\mathbb N}              % Set of natural numbers
\newcommand{\M}{\mathcal M}              % Manifold M
\newcommand{\N}{\mathcal N}              % Manifold N
\newcommand{\MA}{\mathcal A}             % Manifold A
\newcommand{\B}{\mathcal B}              % Fibre bundle
\newcommand{\F}{\mathcal F}              % A typical fibre
\newcommand{\ftf}{\omega}                % Fundamental two form
\newcommand{\fof}{\eta}                  % Liouville form
\newcommand{\definition}[1]{\textbf{#1}} % Definition of term
\newcommand{\defn}{\equiv}               % Definition symbol
\newcommand{\hvf}[1]{\setbox0=\hbox{$#1$}%
  \ifdim\wd0>1em\widehat{#1}\else\hat{#1}\fi} % Hamiltonian vector field
\newcommand{\smooth}[1]{C^\infty(#1)}     % Space of smooth functions
\newcommand{\cotbundle}[1]{T^{*}#1}       % Cotangent bundle
\newcommand{\tanbundle}[1]{T#1}          % Tangent bundle
\newcommand{\field}[1]{\Gamma(#1)}       % Sections of bundle (fields)
\newcommand{\form}[2]{\Lambda^{#1}(#2)}   % Space of k-forms in bundle
\newcommand{\X}[1]{\mathcal X(#1)}       % Space of left invariant vector fields
\newcommand{\liegroup}[1]{\mathop{\rm\makefirstuc{#1}}} % Lie group
\newcommand{\liealgebra}[1]{\mathfrak{\MakeLowercase{#1}}} % Lie algebra
\newcommand{\G}{\mathcal G}              % Lie group G
\newcommand{\g}{\liealgebra G}           % Lie algebra g
\newcommand{\K}{\mathcal K}              % Lie group K
\renewcommand{\k}{\liealgebra K}         % Lie algebra k
\newcommand{\p}{\liealgebra P}           % Lie algebra subspace p
\newcommand{\measure}{\mu}               % Invariant measure
\newcommand{\Lie}[1]{\mathcal L_{#1}}    % Lie derivative
\newcommand{\id}[1]{\mathop{\rm id}_{#1}} % Identity map
\newcommand{\ident}{1}                    % Identity element or matrix
\newcommand{\pull}[1]{{#1}^{*}}           % Pull back
\newcommand{\push}[1]{{#1}_{*}}           % Push forward
\newcommand{\tang}[1]{\partial_{#1}}           % Tangent map
\newcommand{\GL}[1]{\mathop{\rm GL}(#1)}  % General linear group
\newcommand{\T}{T}                        % Generator of Lie algebra
\newcommand{\Diff}[1]{\mathop{\rm Diff}(#1)} % Space of diffeomorphisms
\newcommand{\Ad}{\mathop{\rm Ad}\nolimits} % Adjoint action
\newcommand{\ad}{\mathop{\rm ad}\nolimits} % Adjoint action (ad x)y = [x,y]
\newcommand{\End}[1]{\mathop{\rm End(#1)}} % Endomorphisms
\newcommand{\Aut}[1]{\mathop{\rm Aut(#1)}} % Automorphisms
\newcommand{\metric}{g}                    % Riemannian (pseudo) metric
\newcommand{\isog}[1]{\G(#1)}              % Isometry group
\newcommand{\flow}{F}                      % Local flow
\newcommand{\tr}{\mathop{\rm tr}}          % Trace
\renewcommand{\det}{\mathop{\rm det}}      % Determinant
\newcommand{\vspan}{\mathop{\rm span}}     % Span of linear space
\newcommand{\diag}{\mathop{\rm diag}}      % Diagonal matrix
\newcommand{\sphere}[1]{S_{#1}}            % n-sphere
\newcommand{\hyperboloid}[2]{H_{#1}^{#2}}   % Hyperbolic space
\newcommand{\notnormalin}{\mathop{\not\hskip-0.5ex\triangleleft}}
\newcommand{\In}[1]{\mathop{\rm In}_{#1}}   % Inner automorphism
\newcommand{\dt}{\delta t}                 % Integrator stepsize
\newcommand{\trjlen}{\tau}                 % Integrator trajectory length
\newcommand{\SymmetricGroup}[1]{{\cal S}_{#1}} % Permutation group
\renewcommand{\dfrac}[2]{\frac{\displaystyle#1}{\displaystyle#2}}
\newcommand{\refeq}[1]{(\ref{#1})}         % Equation cross reference
\newcommand{\refsec}[1]{\S\ref{#1}}        % Section cross reference
\newcommand{\refapp}[1]{\S\ref{#1}}        % Appendix cross reference
\newcommand{\U}{V}                         % Potential function
\newcommand{\diffeo}{\cong}                % Diffeomorphism

\renewcommand{\O}{O\!}                     % big-O notation
\newcommand{\from}{\leftarrow}
\newcommand{\assign}{\leftarrow}           % pseudocode assignment
\newcommand{\Vol}{\mathop{\rm Vol}}
\newcommand{\fairbreak}{\penalty-100}
\endlocaldefs
\algloopdefx{Return}[1]{\textbf{return}~#1}
\newcommand{\Yield}[1]{\textbf{yield}~#1}
\newcommand{\algprocname}[1]{\mathop{\textsc{#1}}}
\newcommand{\alglambda}{\textbf{lambda}}
\def\rational#1#2{{\mathchoice{\textstyle{#1\over#2}}%
  {\scriptstyle{#1\over#2}}{\scriptscriptstyle{#1\over#2}}{#1/#2}}}
\def\half{\rational12}		% One half
\def\quarter{\rational14}	% One quarter
\begin{document}
\begin{frontmatter}
\title{Hamiltonian Monte Carlo on Symmetric and Homogeneous Spaces via
  Symplectic Reduction}
\runtitle{HMC on Symmetric and Homogeneous Spaces}
\author{\fnms{Alessandro} \snm{Barp,}\ead[label=e1]{a.barp16@imperial.ac.uk}}
\author{\fnms{A.~D.} \snm{Kennedy,}\ead[label=e2]{adk@ph.ed.ac.uk}}
\and
\author{\fnms{Mark} \snm{Girolami}\ead[label=e3]{m.girolami@imperial.ac.uk}}
\address{\printead{e1}, \printead{e2}, \printead{e3}}
\affiliation{Imperial College London, Alan Turing Institute, Higgs Centre for
  Theoretical Physics, University of Edinburgh}
\runauthor{Barp, Kennedy, and Girolami}

\begin{abstract}
{\narrower\parfillskip=0pt\ The Hamiltonian Monte Carlo method generates
  samples by introducing a mechanical system that explores the target density.
  For distributions on manifolds it is not always simple to perform the
  mechanics as a result of the lack of global coordinates, the constraints of
  the manifold, and the requirement to compute the geodesic flow.  In this
  paper we explain how to construct the Hamiltonian system on naturally
  reductive homogeneous spaces using symplectic reduction, which lifts the HMC
  scheme to a matrix Lie group with global coordinates and constant metric.
  This provides a general framework that is applicable to many manifolds that
  arise in applications, such as hyperspheres, hyperbolic spaces, symmetric
  positive-definite matrices, Grassmannian, and Stiefel manifolds.\par}
\end{abstract}
\end{frontmatter}

\section{Introduction} \label{sec:introduction}

A central problem in statistics involves defining methods that generate samples
from a probability measure over a manifold \cite{Diaconis:2013}.  Typical
examples that arise in applications (such as topological statistics, inference,
machine learning, and computer vision) entail sampling on circles, spheres,
hyperbolic spaces, Grassmannian, Stiefel manifolds, projective spaces, or the
space of covariance matrices~\cite{Girolami:2011,Kunze:2004,Brubaker:2012,%
BarpOates:2018}.

Specific examples of distributions on manifolds arise for example in
directional statistics \cite{Mardia:1999} where spherical geometry plays a key
role, in computational molecular biology to generate protein conformations
\cite{hamelryck2006sampling}, in goodness of fit tests in exponential families
\cite{Diaconis:2013}, to analyse the patterns of crystallographic preferred
orientations \cite{kunze2004bingham}, in Bayesian spectral density estimation,
in random matrix theory \cite{mezzadri2006generate} and of course in molecular
dynamics \cite{hartmann2008ergodic}.

Hamiltonian (Hybrid) Monte Carlo (HMC) \cite{Duane:1987} is a powerful sampling
algorithm, that is usually formulated to sample from distributions over
\(\R^n\), although in physics applications it has long been used for sampling
on Lie group manifolds \cite{kennedy88b,Barp:2018,Betancourt:2017a,%
Kennedy:2012}.  An important characteristic of HMC is that it only requires the
target distribution to be known up to a constant factor.  This is particularly
useful in Bayesian statistics, where one wants to sample from the posterior
distribution whose normalisation constant is often intractable.  Our aim is to
show how it may be used to sample from distributions over a large class of
interesting manifolds.  The HMC algorithm is usually expressed in terms of
local coordinates \cite{Neal:2012,Betancourt:2018,Betancourt:2017b}, but in
general manifolds cannot be covered by a single coordinate patch, so any
coordinate-dependent scheme would require changing coordinates during the
computation of a Hamiltonian trajectory, which would be very inconvenient.  In
some examples there is a single coordinate system that is valid almost
everywhere: for example, on the sphere \(\sphere2\) the familiar spherical
polar coordinates are valid everywhere except at the poles, and the simplest
coordinate system on the torus \(\sphere1\times\sphere1\) is valid everywhere
but on two lines.  Even in these cases ``artificial'' coordinate singularities
often appear on the boundaries of the coordinate patches, which may lead to
numerical instabilities when integrating trajectories that pass through or near
to them.  Moreover, if we embed the manifold in a higher dimensional space, it
is usually necessary to impose constraints for the trajectories to stay on the
manifold, which requires the use of small integration step sizes.  There have
been several articles attempting to address these issues in special
cases~\cite{Lan:2014,Simon:2013,Holbrook:2017}.

In this paper we develop a general methodology to solve this problem in the
case where the manifold \(\M\) is a symmetric space, or more generally a
naturally reductive homogeneous space (defined in \refsec{sec:background}),
using symplectic reduction.  It is thus applicable to all the spaces mentioned
above.  We shall formulate the HMC algorithm on such a space (which may have
nontrivial topology and curvature) by defining a suitable Hamiltonian system in
terms of coordinate-free geometric quantities; and we will then show how this
dynamics may be efficiently implemented by using the embedding of its phase
space in a space of matrices that admits global coordinates.  In our examples
we will display the equations of motion in local coordinates, but this is done
purely for the purpose of illustration.

We shall require that the motion stay on the manifold without the imposition of
constraints through a potential or by the use of Lagrange multipliers
\cite{callaway82a,Brubaker:2012,lelievre2018hybrid}, because these would
require the use of integrators with small steps \cite{Hairer:2006}.  Once such
a suitable Hamiltonian mechanics has been defined, we shall construct a
practical discretization scheme by constructing symmetric symplectic
integrators that are compositions of elementary integration steps that do not
require explicit constraints.  These steps are not entirely trivial to build on
curved manifolds~\cite{Leimkuhler:1996}, so we will explain in detail how they
may be implemented efficiently.

\begin{figure}
  \begin{center}
    \epsfxsize=0.4\textwidth
    \epspdffile{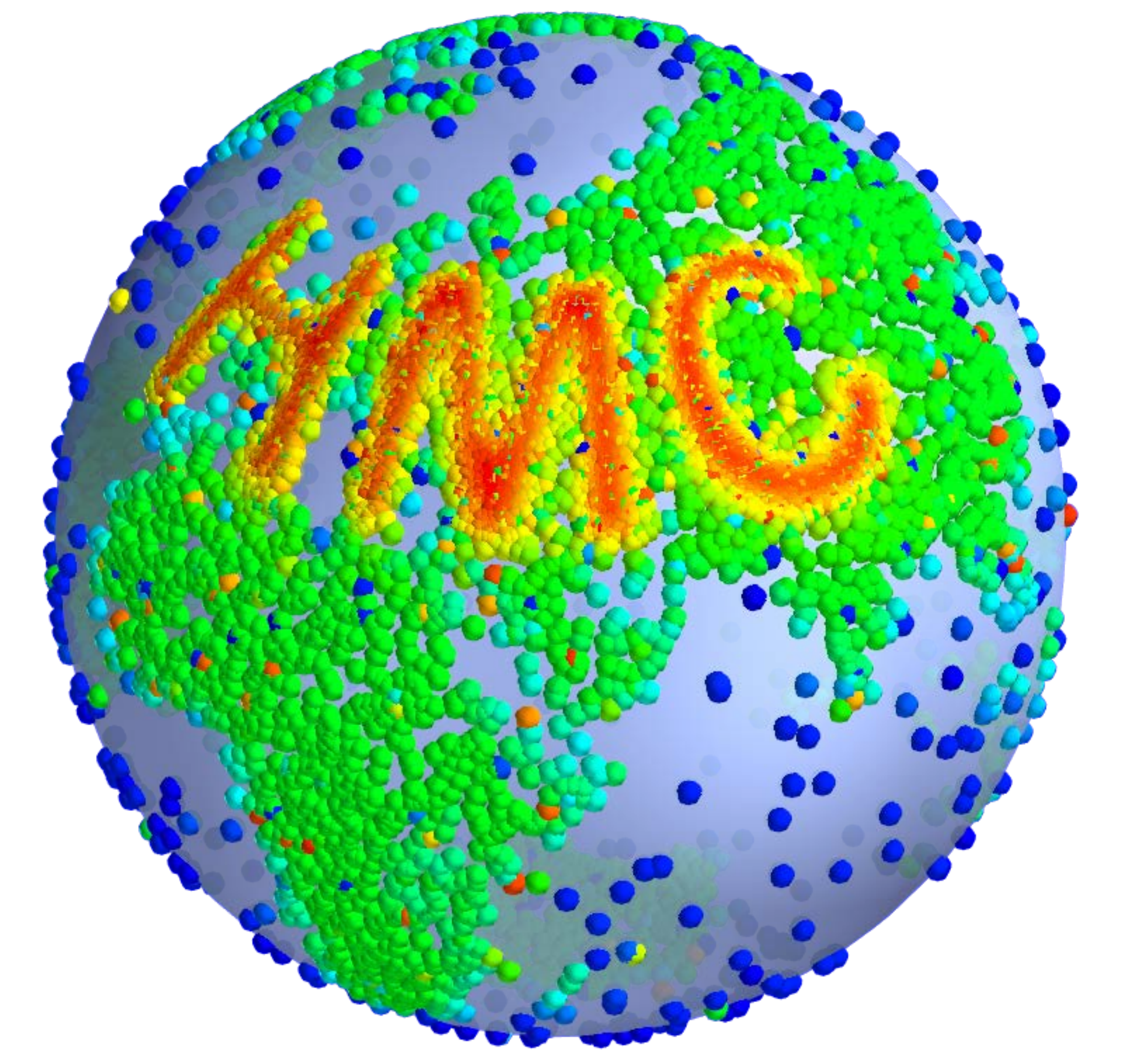}
  \end{center}
  \caption{Sampling the world.  The trajectory length is \(6,371\) km, the step
    size is \(63.71\) km.  We compute \(10,000\) samples from a recognisable
    potential.  The colour shows the magnitude of the potential at that point
    (the three-dimensional figures were generated using~\cite{Mayavi:2011}).}
\end{figure}

Homogeneous spaces contain natural symmetries in the sense that they can be
written (up to isomorphism) as a quotient space \(\M \cong\G/\K\): that is,
they arise by identifying the elements of a Lie group \(\G\) that are related
by transformations belonging to a subgroup \(\K \subset \Diff \G\) (so that
elements of \(\M\) can be viewed as subsets of \(\G\)).  Symmetries play an
important role in mathematics and physics; in the context of Hamiltonian
mechanics, they give rise to conserved quantities (generalised momenta) via
Noether's theorem \cite{Abraham:2008}.  Such symmetries are encoded in terms of
the action of a Lie group on the manifold, and the surfaces corresponding to
conserved momenta, which are given by level sets of a momentum map, is a
foliation of phase space.  The algorithm we propose computes a Hamiltonian
system on \(\G\) that is invariant under the action of \(\K\), and which
preserves the corresponding conserved quantity \(J=0\), where \(J\) is the
momentum map for the action of \(\K\) on \(\G\).  Expressing such a Hamiltonian
system in terms of a matrix representation of the Lie group and its concomitant
Lie algebra not only allows its embedding in a space with a global coordinate
system, but also allows the group multiplication to be implemented using matrix
multiplication rather than requiring the use of transcendental functions.  The
symmetries of the Hamiltonian system on \(\G\) ensure that the mechanics stays
on a submanifold \(J^{-1}(0)\subset T^*\G\), which allow its reduction to a
mechanics on \(\M\diffeo\G/\K\), and enables sampling from the target measure
on~\(\M\).

\subsection{Advantages of the Method} \label{sec:advantages}
To the best of our knowledge, the algorithm we derive is the first one defining
an explicit HMC scheme for a large class of manifolds.  Currently the main
reference is~\cite{Simon:2013}, which informally demonstrates how to deal with
the lack of global coordinates for manifolds isometrically embedded in $\R^n$,
but does not explain how to compute the geodesic flow for these submanifolds
(or the orthogonal projection onto the tangent spaces).  Aside from the
manifolds discussed in~\cite{Simon:2013}, the family of naturally reductive
homogeneous spaces also includes Grassmannian and the space of positive
definite matrices.  Moreover in the important case of Stiefel manifolds, the
metric induced by the quotient is consistent with the symmetries of the
manifold, unlike~\cite{Simon:2013} which pulls-back the Euclidean metric.  This
means that our reference measure is the uniform distribution, which is
particularly suitable for Bayesian applications (in particular the absence of
potential energy the proposed method samples uniformly).

Performing the HMC algorithm on the Lie group \(\G\) rather than \(\M\) will
yield several advantages.  First of all, as mentioned earlier, we will be able
to use the global coordinates provided by the matrix representation.  Moreover,
since the tangent bundle is trivial, \(T\G \cong \G \times \g\), and the metric
will be left-invariant, the velocity update is greatly simplified: we shall see
that we only need to compute the Riemannian gradient at the identity, rather
than at each point of the manifold; in other words our Riemmannian metric on
\(\G\) is determined by a (constant) inner product on~\(\g\).

By choosing an appropriate Riemannian metric, we can write the geodesics in
closed form as a single matrix exponential.  We only need to find the geodesics
starting at the identity, rather than at each point, and no parallel transport
of the velocity along the geodesics is required, in contrast to both
\cite{Holbrook:2017,Simon:2013}.  Moreover if we use an appropriate basis of
\(\g\), the metric becomes Euclidean and in particular no inverse metrics
appear.  Hence, our algorithm does not reduce to \cite{Holbrook:2017} when
\(\M\) is the space of positive definite matrices.

We may then perform HMC on \(\M\) by exponentiating the Lie algebra without
having to introduce constraints, which allows us to use numerical integrators
with large steps.  The constraint that the algorithm stays on \(J^{-1}(0)\) (and
thus \(\M\)) are automatically satisfied since, unlike in \cite{Simon:2013},
the geodesic on \(J^{-1}(0)\) reduce to geodesics on \(\M\), and no additional
orthogonal projection to the tangent space is required: the geometry of the
bundle \(\G \to \G/\K\) and the symmetries of the Hamiltonian allow us to
naturally implement the constraints of~\(\M\).

From the expression for the symplectic structure in terms of non-vanishing
1-forms, we can easily construct Shadow Hamiltonians and higher order
integrators, which for some cases may be computationally
cheaper~\cite{Kennedy:2012,izaguirre2004shadow, skeel2001practical}.  Such
integrators may involve a carefully choreographed sequence of \(\hvf V\) and
\(\hvf T\) steps, and the use of ``second derivatives" or ``force gradient"
steps such as \({\hvf{\{V,\{V,T\}\}}}\) may also be advantageous \big(here
\(\{{V,T}\} = -\omega(\hvf V,\hvf T)\) is the \definition{Poisson bracket}\big).

Finally, we will show that our method can also be used to compute Hamiltonian
trajectories on pseudo-Riemannian manifolds.

\subsection{Description of the Method} \label{sec:description}
The following commutative diagram summarises the geometric structures involved
in this method of sampling from probability measure on a naturally reductive
homogeneous manifold~\(\M\).
\begin{equation*}
  \begin{CD}
    \cotbundle{\GL n} @<\rho\otimes \tang{}\rho << \cotbundle\G
      @>J^{-1}(0)/\K>> \cotbundle(\G/\K) @>\psi \otimes \tang{}\psi>> T^{*}\M \\
    @V\pi VV @V\pi VV @V\pi VV @V\pi VV \\ 
    \GL n @<<\rho< \G @>>\pi> \G/\K @>>\psi> \M \\
   \end{CD}
\end{equation*}
Since \(\M\) is a homogeneous space it is diffeomorphic to the quotient
\(\G/\K\) (bottom right of the diagram) of the Lie group \(\G\) that acts on
\(\M\) by the stabilizer subgroup \(\K\) of this action.  We set \(\mu\) to be
the natural volume (uniform) measure on \(\M\) generated by a \(\G\)-invariant
Riemannian metric, and assume the target density has the form \(e^{- \U}\mu\)
up to a constant factor (in the Bayesian paradigm, \(\U\) is the negative
log-likelihood).

The HMC algorithm generates samples from the density \(e^{-V}\) on \(\M\) by
sampling from an extended distribution \(e^{-H}\) on phase space \(T^*\M\)
using the Hamiltonian flow of a Hamiltonian \(H\).  The method we propose
computes the Hamiltonian flow of \(H\) using a Hamiltonian flow on \(\G\).

In order to construct the Hamiltonian system on \(\G\) we start by noting that
there is a canonical phase space, namely the cotangent bundle \(\cotbundle\G\)
endowed with a symplectic structure by the fundamental two-form \(\omega =
-d\eta\) where \(\eta\) is the Liouville form defined in terms of the
left-invariant Maurer--Cartan form \(\theta\) on~\(\G\).  The potential energy
\(\U\) on \(\M\) pulls back to a potential energy on \(\G\), while a kinetic
energy on \(\G\) can be induced by the \(\G\)-invariant metric on \(\M\),
together with the requirement that \(\G/\K\) is naturally reductive.  By
construction, the resulting Hamiltonian on \(\G\) is (right) \(\K\)-invariant.

Once we have a Hamiltonian system on \(\cotbundle\G\) we can compute the
Hamiltonian system on \(\cotbundle \M\) by symplectic reduction.  More
precisely, to construct numerical trajectories on \(\cotbundle(\G/\K)\) we
restrict the initial momenta corresponding to \(\K\) to be zero (so \(J=0\)).  
Then, the symmetries of the Hamiltonian ensure the dynamics in \(\cotbundle\G\)
is restricted to the level set \(J^{-1}(0)\) of the momentum map~\(J\)
corresponding to the Noether momenta for \(\K\)-transformations.  The
symmetries of the Hamiltonian and the restriction to \(J^{-1}(0)\), avoid the
appearance of analogues of centrifugal forces, and it follows from the
symplectic reduction theorem \cite{Marsden:1974,Abraham:2008} that this
dynamics on \(\cotbundle{\G}\) reduces to a well-defined dynamics on
\(\cotbundle{(\G/\K)}\diffeo\cotbundle\M\).

In order to efficiently implement the numerical Hamiltonian dynamics on
\(\cotbundle\G\) we use the embedding provided by a faithful representation
\(\rho\) of \(\G\) in \(\GL n\), the space of non-singular \(n\times n\)
matrices, and the concomitant embedding of \(\cotbundle\G\) in \(\cotbundle\GL
n\) (the leftmost columns in the diagram).  This has two significant benefits:
the space of \(n\times n\) matrices is diffeomorphic to \(\R^{n^2}\) which
provides a global coordinate system, and group multiplication corresponds to
the multiplication of representation matrices.  Thus while we have to pay for
the larger dimension of the space of matrices the computations required within
it are relatively cheap; and we may concentrate our efforts on efficiently
implementing the only transcendental operation required, the exponentiation of
matrices in the representation of the Lie algebra \(\g\) to those in the
representation Lie group~\(\G\).

Finally, it is worth mentioning that the procedure of symplectic reduction
provides a Hamiltonian system on \(\M\), even when the kinetic energy on \(\G\)
is not a Riemannian metric, and we can use any (symmetric) symplectic
integrator to perform approximate numerical integration of the trajectories in
\(T^*\M\) using trajectories in \(J^{-1}(0)\).  For example in
\refapp{sec:h2111} we will compute Hamiltonian trajectories on the Lorentzian
single sheeted hyperbolic space (which is an anti-de Sitter space).  However
this does not lead to an HMC algorithm for \(\M\).  Indeed HMC also requires a
momentum refreshment Markov step (Gibbs sampler), and this is only possible if
the distribution \(e^{-T}\) is normalizable, which is not necessarily the case
when the metric is pseudo-Riemannian.  However, as long as the kinetic energy
is Riemannian on an appropriate subspace (called \(\p\) later) of \(\g\), which
will always be the case if the kinetic energy on \(\G\) is induced by a
Riemannian metric on \(\M\), this will indeed lead to an HMC algorithm.

\subsection{Structure of paper} \label{sec:structure}
In~\refsec{sec:background} we give an overview of the differential geometric
language that is used in the paper.  Section \refsec{sec:red-hom-space} gives
definitions of symmetric and homogeneous spaces.  In the supplementary
materials we discuss these spaces in more detail and introduce all the
necessary geometry to fix our notation and make this paper self-contained.
In~\refsec{sec:metric} we explain the relation between inner products on \(\g\)
and \(\G\)-invariant metrics on \(\M\).  In~\refsec{sec:dyn-LieHom} we derive
the equations of motion on Lie groups and explain how to set up the dynamics to
sample on the reduced homogeneous spaces.  In~\refsec{sec:pseudo} we discuss
the resulting algorithm in detail.  In~\refsec{sec:sphere} we provide an
illustration of our algorithm for the case of the two-sphere, and its related
hyperbolic spaces are treated in~\S\ref{sec:par-quotient} of the supplementary
materials.  We explain how matrix exponentials can be efficiently computed
in~\refsec{sec:mat-exp} and \refapp{sec:sup-exp}.  Finally,
in~\refsec{sec:conclusions} we summarise what has been achieved.  The proofs of
the various results are provided in~\refapp{proofs}.

\section{Background} \label{sec:background}

\subsection{Reductive Homogeneous Spaces} \label{sec:red-hom-space}

Homogeneous spaces are quotient spaces with natural symmetries arising from the
action of a Lie group.  Typically, homogeneous spaces are induced by a manifold
\(\M\) upon which a group \(\G\) acts transitively by an action which we denote
as \(g\cdot q\) for \(g\in \G\) and \(q\in\M\) (transitivity means \(\G\cdot q
= \M\)).  Let \(\K\) be the stabilizer subgroup that fixes a specified point
\(p\in \M\), \(g\cdot p=p\) for all \(g\in\K\).  The left action of \(\G\) on
itself \(L_gh=gh\) defined by group multiplication induces a left action of
\(\G\) on~\(\G/\K\).
%%%%%%%%%%%%%%%%%%%%%%%%%%%%%%%%%%%%%%%%%%%%%%%%%%%%%%%%%
%%                                                     %%
%% Proof (if desired)                                  %%
%%                                                     %%
%%%%%%%%%%%%%%%%%%%%%%%%%%%%%%%%%%%%%%%%%%%%%%%%%%%%%%%%%
%%
%% If \(k\in\K_{g\cdot q}\) then \(k\) leaves the point \(g\cdot q\) fixed,
%% \(k\cdot(g\cdot q) = (kg)\cdot q = g\cdot q\), whence \(g^{-1}kg\in\K\)
%% so \(\K_{g\cdot q} = g\K g^{-1}\).
%%
It follows that the map \(\psi:\G/\K\to\M\) defined by \(\psi:g\K\mapsto g\cdot
p\) is an (equivariant) diffeomorphism, \(\M\diffeo\G/\K\) (see
\refapp{homogeneous-spaces}), and we can think of the point \(g\cdot p \in \M\)
as the subset \(g\K \subset \G\).  When defining HMC on \(\M\) through
symplectic reduction, we will use this isomorphism to view \(\M\) as the
homogeneous space~\(\G/\K\).  The canonical projection \(\pi:\G\to\G/\K\),
\(\pi:g\mapsto g\K\), will then be used to reduce the mechanics on \(\G\) to a
mechanics onto~\(\M\).

In order to have a simple relation between metrics on \(\M\) and inner products
on \(\G\), we shall require that the homogeneous space \(\G/\K\) is
\definition{reductive} \cite{Deng:2012}, which means that there is a
decomposition of the Lie algebra \(\g=\k\oplus\p\) with
\(\Ad_\G(\K)\p\subseteq\p\) (see \refapp{lie-groups}).  This is always true
when \(\G\) is the isometry group of a Riemannian manifold
\(\M\)~\cite{Zdenek:2008}.  With this assumption the tangent space at
\(p\in\M\) is isomorphic to~\(\p\), while \(\k\), the lie algebra of \(\K\),
defines directions of symmetry: moving along \(\k\) in \(\G\) does not result
in any projected motion on~\(\M\).

\subsubsection{Symmetric Spaces} \label{sec:symspaces}

Symmetric spaces~\cite{Helgason:2001} form an important class of naturally
reductive homogeneous spaces and arise frequently in statistics and physics.
These include the sphere \(\sphere2\), Grassmannian manifolds, and the space of
positive definite matrices.  Riemannian symmetric spaces are connected
Riemannian manifolds (see \refapp{riemannian}) with the property that for any
\(x\in\M\) there is an isometry \(\sigma_x\) of \(\M\) such that
\(\sigma_xx=x\) and \(\tang x\sigma_x=-\id{\tanbundle_x\M}\), where
\(\tang{x}\) denotes the tangent map at \(x\).  It can then be shown that the
connected component \(\isog\M_0\) of the isometry group of \(\M\) acts
transitively on \(\M\), thus if \(\K_p\) is the isotropy group at \(p\in\M\),
then as before we can identify \(\M\diffeo\isog\M_0/\K_p\).  Homogeneous spaces
arising this way are automatically reductive, in fact there exists a subspace
\(\p\) complementary to \(\k\), \(\g_0(\M) = \k\oplus\p\), with
\([\k,\k]\subseteq\k\), \([\k,\p]\subseteq\p\), and \([\p,\p]\subseteq\k\).

%%%%%%%%%%%%%%%%%%%%%%%%%%%%%%%%%%%%%%%%%%%%%%%%%%%%%%%%%
%%                                                     %%
%% Why is the following paragraph relevant? It is not  %%
%% needed for HMC.  Furthermore, it surely isn't true  %%
%% for the symmetric space H1.                         %%
%%                                                     %%
%%%%%%%%%%%%%%%%%%%%%%%%%%%%%%%%%%%%%%%%%%%%%%%%%%%%%%%%%
%%
%% Moreover if \(X\in\p\), it can be shown \(e^X\cdot p = \exp_p(d_1\pi(X))\),
%% where \(\exp_p:T_p\M\to\M\) is a surjection (bijection?) when \(\M\) is a
%% symmetric space, called the Riemannian exponential map (see
%% \cite{Holmelin2005symmetric} results 2.4 and 3.3).  It follows that
%% \(e^\p\cdot p =\M\), and we can ``parameterize" \(\M\) using the vector
%% space~\(\p\).

\subsection[G-Invariant Metrics on G/K]%
           {\(\G\)-Invariant Metrics on \(\G/\K\)} \label{sec:metric}
In order to implement HMC we need to choose a Riemannian metric
\cite{betancourt2013general}.  On Lie groups there is a bijection between inner
products on \(\g\) and left-invariant metrics on \(\G\), as well as between
\(\Ad \G\)-invariant inner products on \(\g\) and bi-invariant metrics on
\(\G\) (see \refapp{lie-groups}).  A similar result holds for reductive
homogeneous spaces: there is a bijection between \(\G\)-invariant metric on
\(\G/\K\) and \(\Ad_\G\K\)-invariant inner products on \(\p\).  This bijection
allows us to pull back the Riemannian metric on \(\M\) to an inner product on
\(\p\), which we will use to define the kinetic energy.  By construction the
canonical projection \(\pi:\G\to \G/\K\) is an isometry when restricted to
\(\p\), so that geodesics will be projected to geodesics (see
\refapp{proof-of-isometry} for the proof).  To make our HMC scheme as simple as
possible, we will use an \(\Ad_{\G}\K\)-invariant non-degenerate quadratic form
on \(\g\) for which \(\p\) and \(\k\) are orthogonal.  For an important class
of homogeneous spaces, called naturally reductive, this enables us to compute
the (horizontal) geodesics on \(\G\) (and \(\M\)) by exponentiating matrices in
(a representation of) \(\p\).  Naturally reductive homogeneous spaces include
homogeneous spaces with an \(\Ad \G\)-invariant metric and symmetric spaces.

\begin{figure}
  \begin{center}
    \epsfxsize=0.25\textwidth
    \epspdffile{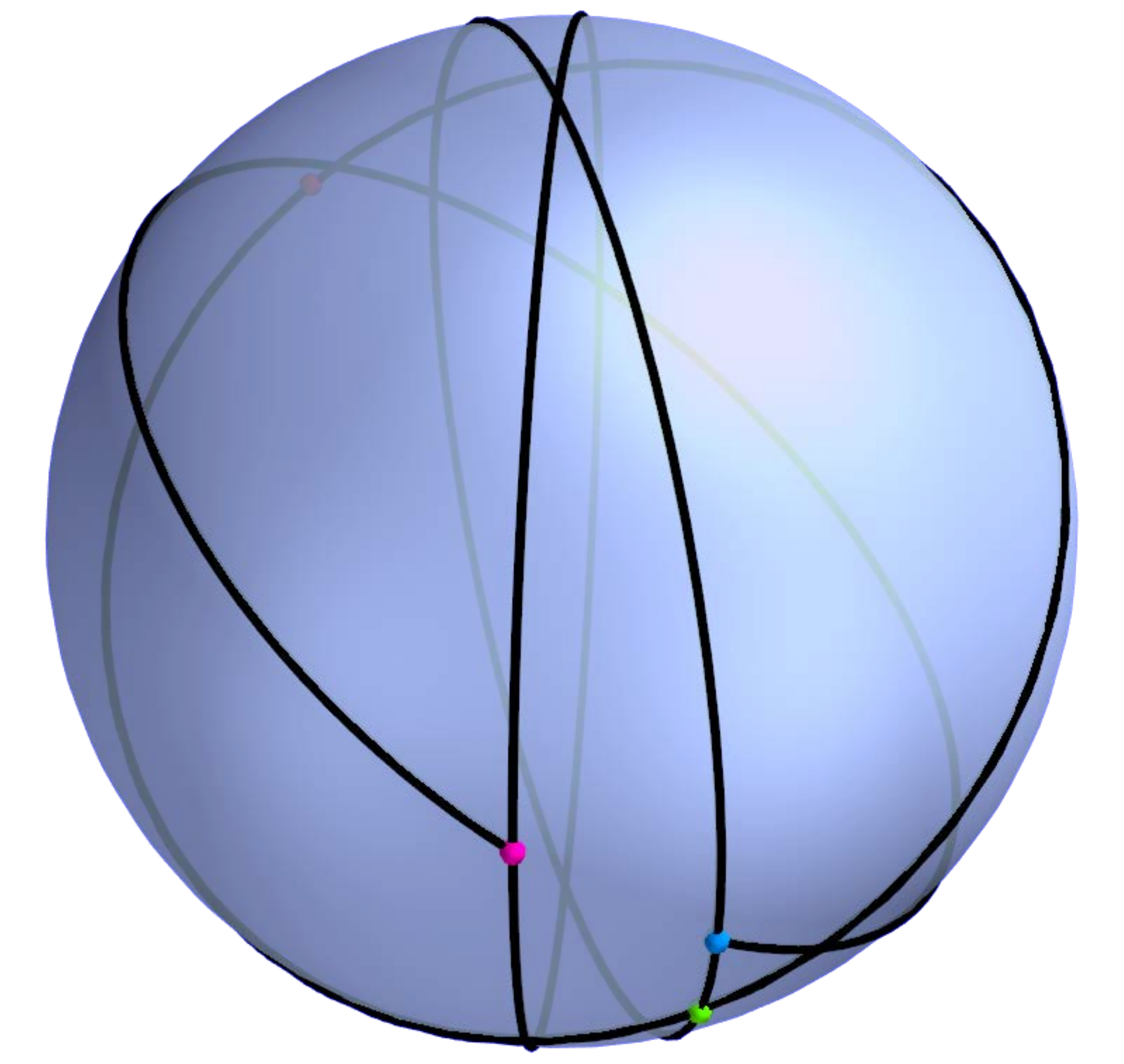}
    \qquad\qquad
    \epsfxsize=0.25\textwidth
    \epspdffile{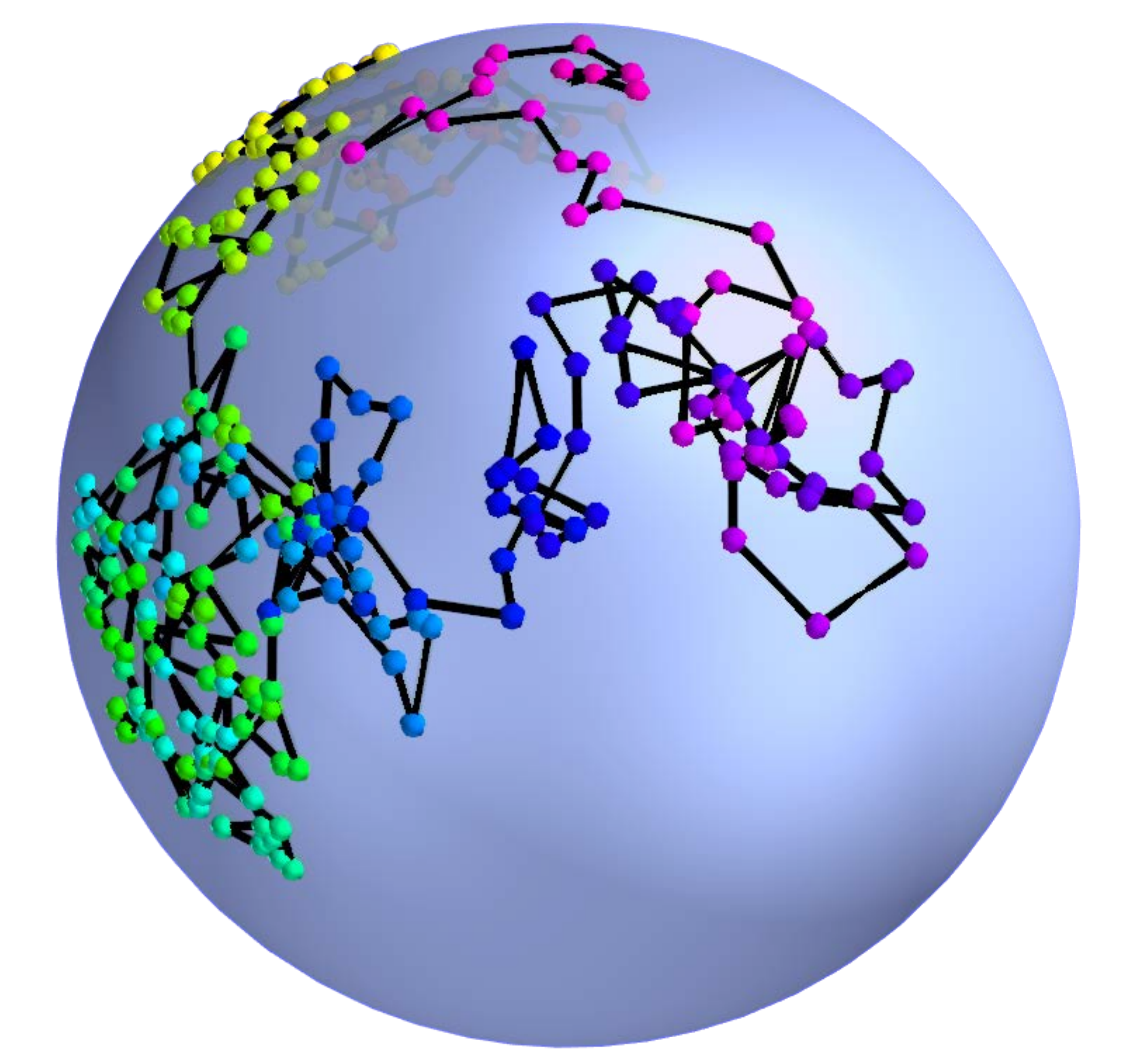}
  \end{center}
  \caption{Geodesics (\(\U=0\)) on the unit sphere obtained from HMC, with
    kinetic energy defined by the canonical (\(\liegroup{SO}(3)\)-invariant)
    metric.  This is implemented by reducing a free mechanics on
    \(\liegroup{SO}(3)\) with the associated
    \(\Ad_{\liegroup{SO}(3)}\liegroup{SO}(2)\) inner product on \(\p\).  The
    figure on the left has \(4\) trajectories of length \(\trjlen=10\), and
    that on the right has \(300\) trajectories of length \(\trjlen=0.1\).  The
    colours are chronological.}
\end{figure}

In section \refapp{proof-of-relation-Killing} we show how Killing fields may be
used to characterise the relation between the metrics on \(\G/\K\) and \(\p\).

\section{Dynamics on Homogeneous Space} \label{sec:dyn-LieHom}

\subsection{Dynamics on Lie Groups} \label{sec:dynamics} 

The phase space for Hamiltonian dynamics on a Lie group \(\G\) is the cotangent
bundle \(\cotbundle\G\), and as on any cotangent bundle there is a natural
one-form \(\fof\) called the \definition{Liouville\/} form.  A useful property
of Lie groups is that we can construct \(\fof\) using the non-vanishing 1-forms
\(\theta^i\) of the Maurer--Cartan frame, which is defined by choosing a basis
of the Lie co-algebra \(\g^*\) (see \refapp{lie-groups}), as we now show.

Given the Maurer-Cartan form \(\theta: g \mapsto \theta_g \defn \tang
gL_{g^{-1}}\) and a frame \(\theta^i\) (i.e., a basis of \(\g^*\)), define the
momentum fibre coordinates \(p_i:T^{\ast}\G \to \R\), by first defining
\(p\defn\pull{( \theta^{-1})}: \cotbundle\G \to \g^{*}\), and then setting \(p=
\theta^i_1p_i\).  Here the inverse is taken pointwise \(\theta^{-1}_g \defn
(\theta_g)^{-1}:\g\to T_g\G\), and the inverse function theorem implies
\(\theta_g^{-1} = \tang 1L_g\).  Thus, setting \(p_g\defn p|_{T_g\G}\), for any
\(\alpha_g\in T_g^{*} \G\),
\begin{equation} 
  p_g(\alpha_g) = (\theta^{-1}_g)^{\ast}(\alpha_g)
  = \alpha_g\circ\theta^{-1}_g
  = L_g^{\ast}\alpha_g\in\g^{\ast}.
  \label{eq:momentum-coord}
\end{equation}  

We define the Liouville 1-form on \(\cotbundle\G\) as \(\fof\defn
p_i\pi^{*}\theta^i\).  On \(\G=\R^n\) we have \(\theta^i =dx^i\), so that
\(\fof\) may be viewed as a generalisation of the usual Liouville form
\(p_idx^i\).  By analogy with \(\R^n\) we might guess that the corresponding
symplectic structure should be \(dp_i \wedge \theta^i\), but this is not
closed.  The corresponding symplectic structure \(\ftf\) is in fact given by
\begin{equation*}
  \ftf \defn -d\fof = -d(p_i \pi^{*}\theta^i)
  = \pi^{*}\theta^i\wedge dp_i
    + \half p_ic^i_{jk}\pi^{*}\theta^j\wedge\pi^{*}\theta^k
\end{equation*}
where \(c^{i}_{jk}\) are the structure constants of the Lie algebra and we used
\refeq{eq:Maurer-Cartan-relations}.  The term in the fundamental two-form
explicitly depending on the structure constants will not appear in the
Hamiltonian vector fields derived below, but is required to ensure that
\(\ftf\) is closed, \(d\ftf=0\).  Moreover, it is important for the
construction of Poisson brackets and higher derivative (force-gradient)
integrators~\cite{Kennedy:2012}.

Note the Liouville form does not depend on the choice of frame, and in fact may
be written in a frame-independent manner as \(\fof=p(\pi^{*} \theta)\) (see
also \refapp{proof-symplectomorphism}).  The definition of \(p\) suggests it
transforms objects in the opposite way of \(\theta\).  This intuition is
formalised by the following important symmetry (derived in
\refapp{proof-symmetriesfof}) which proves \(\fof\) is the canonical Liouville
form: for any 1-form \(\beta: \G \to T^*\G\)
\begin{equation}
  \pull{\beta}\fof = \beta.
  \label{eq:symmetriesfof} 
\end{equation}     
  
By construction the Liouville form \(\fof\) is left-invariant, that is, for
each group element \(g\), the induced map \(\pull{L_g}\) on phase space is a
symplectomorphism.  The analogous result holds for the right action, and more
generally for any map induced by a diffeomorphism of \(\G\), as we show in
\refapp{proof-symplectomorphism}.  Let us restrict the domains of the functions
\(p_i\) to \(\g^{*}\), and use these as fibre coordinates (more precisely, we
identify~\(\cotbundle\G\cong\G\times\pull\g\)).

We introduce a Hamiltonian function \(H=T+\U:\cotbundle\G\to\R\) on this phase
space, as the sum of a potential energy \(\U:\G\to\R\) and a kinetic energy
\(T: \g^* \to \R :p \mapsto \half\langle p,p\rangle^{-1}\), with
\(\langle\cdot, \cdot\rangle^{-1}\) an inner product on \(\g^{*}\), which for
now we assume to be induced by an \(\Ad \G\)-invariant inner product
\(\langle\cdot, \cdot\rangle\) on \(\g\).  We will see in
\refsec{sec:reduced-dynamics} that \(H\) is the pull-back of the Hamiltonian on
\(T^*\M\).  The inner product on \(\g\) enables us to identify \(\g
\diffeo\g^{*}\) and to define dual momentum (i.e., velocity) coordinates
\(p^i\defn \partial_{p_i}T = g^{ij}p_j\) (which are the coordinates of the
associated vector), where \(g^{ij}\defn \langle\theta^i,\theta^j\rangle^{-1}\).

The mechanical trajectories are the integral curves of the Hamiltonian vector
field \(\hvf H\) (\refapp{hamiltonian-dynamics}).
%%%%%%%%%%%%%%%%%%%%%%%%%%%%%%%%%%%%%%%%%%%%%%%%%%%%%%%%%
%%                                                     %%
%% Derivation of Hamiltonian vector fields (if needed) %%
%%                                                     %%
%%%%%%%%%%%%%%%%%%%%%%%%%%%%%%%%%%%%%%%%%%%%%%%%%%%%%%%%%
%%
%% The Hamiltonian vector field \(\hvf T\) satisfies \(dT(X) = \ftf(\hvf T,X)\)
%% so
%% \begin{align*}
%%   \partial_{p_i}T\,X_{p_i}
%%   &= \ftf\left(\hvf T_{q^i}e_i + \hvf T_{p_i}\partial_{p_i},
%%                X_{q^i}e_i +  X_{p_i}\partial_{p_i}\right)
%%   = \hvf T_{q_i} X_{p_i} - \hvf T_{p_i} X_{q_i}
%%       +\half p_i c^i_{jk} \hvf T_{q_j} X_{q_k} \cr 
%%   &\implies \hvf T_{q^i} = 0
%%   \implies \hvf T = \partial_{p_i}T\,e_i = p^ie_i,\cr
%%   e_i(\U)\,X_{q^i}
%%   &= \ftf\left(\hvf \U_{q^i} e_i + \hvf \U_{p_i} \partial_{p_i},
%%                X_{q^i} e_i +  X_{p_i} \partial_{p_i}\right)
%%   = \hvf \U_{q_i} X_{p_i} - \hvf \U_{p_i} X_{q_i}
%%       +\half p_i c^i_{jk} \hvf \U_{q_j} X_{q_k} \cr
%%   &\implies \hvf \U_{q^i} = 0 \implies \hvf \U = -e_i(\U)\,\partial_{p_i}, 
%% \end{align*}
%% where
%% \begin{equation*}
%%   X = X_{q^i} e_i + X_{p_i}\partial_{p_i}, \qquad
%%   \hvf T = \hvf T_{q^i} e_i + \hvf T_{p_i}\partial_{p_i},
%%   \qquad\mbox{and}\qquad
%%   \hvf \U = \hvf \U_{q^i} e_i + \hvf \U_{p_i}\partial_{p_i}.  
%% \end{equation*}
%%
The Hamiltonian vector fields corresponding to \(T\) and \(\U\) are given by
\(\hvf T = p^ie_i\) and \(\hvf\U=-e_i(\U)\,\partial_{p_i}\) where \(e_i\)
denote the left-invariant dual vector fields,~\(\theta^i(e_j) =
\delta^i_j\)~\cite{Kennedy:2012}.

As we wish to compute Hamiltonian trajectories globally we express the dynamics
in terms of motion on the submanifold of \(\R^{n^2}\) corresponding to the
image of a faithful (injective) representation \(\rho:\G\to\GL{n,\R}\) of the
Lie group (\refapp{lie-groups}); if \(\G\) is a matrix group then the defining
representation is often convenient for this purpose, so that the reader may
take \(\rho\) to be the inclusion.  Since the cotangent bundle over a Lie group
is trivial, \(\cotbundle\G\diffeo \G\times\g^{*} \diffeo \G\times\g\), we can
identify \(\G \times \g\) with \(\rho(\G)\times \tang
1\rho(\g)\diffeo\tanbundle\rho(\G)\).

The potential energy \(\U\) defines a function \(\U \circ \rho^{-1}\) on the
image of the representation \(\rho(\G)\).  If \(\rho(\G)\) is not an open subset
of \(\GL n\), it will be useful to use an arbitrary differentiable extension
\(\U_{\rho}:\GL n \to \R\) of \(\U \circ \rho^{-1}\) (so \(\U_{\rho}=\U \circ
\rho^{-1}\) on \(\rho(\G)\)) in order to differentiate the potential energy
with respect to global coordinates.  Similarly, using \(\tang 1 \rho\) we can
define an inner product \(\langle \cdot, \cdot \rangle_{\rho}\) on \(\tang 1
\rho (\g)\), satisfying \((\tang 1\rho)^{*}\langle\cdot,\cdot\rangle_{\rho} =
\langle\cdot,\cdot\rangle\), from which we construct a kinetic energy
\(T_{\rho}:\tang 1\rho(\g) \subseteq\R^{n^2}\to\R\).  By construction we have
\(T= T_{\rho}\circ \tang 1\rho\) and \(\U = \U_{\rho} \circ \rho\).

The map \(\tang1\rho\) identifies the vector \(p^i e_i(1)\) with its
representation \(P \defn p^i T_i\), where \(\T_i\defn (e_i\rho)(\ident)\) are
called the generators and form a basis of \(\tang 1\rho(\g)\).  We set \(g_{ij}
\defn\langle T_i,T_j\rangle_{\rho}\).
 
The Hamiltonian trajectories on \(\cotbundle \G\) are the integral curves of
the vector field \(\hvf H =\hvf T+\hvf\U\), which we approximate by
constructing \definition{symplectic integrators\/} built of interleaved
segments of the integral curves of \(\hvf T\) and~\(\hvf\U\).  The rate of
change of functions \(P\) and \(x\) along a curve \(c:\R\to\cotbundle\G\) is
\begin{equation} \label{eq:gradV}
  \dot P(t) = \hvf\U P
    = -\tr\bigl(\partial_x\U_{\rho}\cdot\rho(t)\cdot T_i\bigr)T^i
  \qquad\mbox{and}\qquad
  \dot\rho(t) = \hvf T \rho = \rho(t)\cdot P(t)
\end{equation}
respectively, where \(x_{ab}:\GL{n,\R}\to\R\) are the canonical matrix
coordinates, \(T^i \defn g^{ij}T_j\), and \((\partial_x\U_{\rho})_{ab} \defn
\partial\U_{\rho}/\partial x_{ba}\) (see \refapp{proof-of-motion} for proof).
Integrating these equations exactly (with constant \(\rho\) and \(P\)
respectively) yields
\begin{equation*}
  P(t) = P(0) - t\tr\bigl(\partial_x\U_{\rho}\cdot\rho(0)\cdot T_i\bigr)T^i
  \qquad\mbox{and}\qquad
  \rho(t) = \rho(0)\cdot\exp\bigl(tP(0)\bigr);
\end{equation*}
note that this involves right-multiplication by \(e^{tP(0)}\) of the matrix
\(\rho(0)\) representing the initial point in~\(\G\), this is a consequence of
using left-invariant vector fields (which generate right translations).  These
separate \(\hvf\U\) and \(\hvf T\) updates may be used to construct symplectic
integrators, such as the leapfrog integrator \(e^{\half\dt\,\hvf\U }
e^{\dt\,\hvf T } e^{\half\dt\,\hvf\U }\)
\begin{align}
  P\bigl(t + \half\dt\bigr) &=P(t)
    - \half\dt\,\tr\bigl(\partial_x\U_{\rho}\cdot\rho(t)\cdot T_i\bigr)T^i
      \nonumber \\[1ex]
  \rho(t + \dt) &=\rho(t)
    \cdot \exp\Bigl(\dt\,P\bigl(t+\half\dt\bigr)\Bigr)
      \label{eq:leapfrog} \\[1ex]
  P(t + \dt) &=P(t + \half\dt)
    - \half\dt\,\tr\bigl(\partial_x \U_{\rho}\cdot\rho(t+\dt)\cdot T_i\bigr)T^i.
      \nonumber
\end{align}

As an example let us derive the force term for \(\liegroup{SO}(n,\R)\), with
\(\rho\) defined as the inclusion in \(\GL n\) and normalise the generators
\(T_i\) such that \(g_{ij}=-\tr(T_iT_j)=\tr(T_i^{\top}T_j)=2\delta_{ij}\) 
\begin{align*}
  \dot P(t) &= -\tr\bigl(\partial_x\U_{\rho}\cdot\rho(t)\cdot T_i\bigr)T^i 
  = -\tr\bigl((\rho^{\top} \partial_x\U_{\rho}^{\top})^{\top} T_i\bigr)T^i\\
  & =
  - \half\tr\bigl((\rho^{\top} \partial_x\U_{\rho}^{\top})^{\top} T_i\bigr)T_i
  =- \frac{\tr\bigl((\rho^{\top} \partial_x\U_{\rho}^{\top})^{\top} T_i\bigr)}{g_{ii}}\,T_i \\
  &= -\MA(\rho^{\top}\cdot \partial_x\U_{\rho}^{\top})=\MA( \partial_x\U_{\rho}\cdot \rho),
\end{align*}
where \(\MA\) projects onto \(\tang 1 \rho\bigl(\liealgebra{so}(n,\R)\bigr)\),
the subspace of antisymmetric matrices spanned by the generators~\(T_i\).  If
for example \(\U_{\rho}= \tr(B\cdot x)\) for a constant matrix \(B\),
then~\(\partial_x\U_{\rho}=B\).

\subsection{Dynamics on Naturally Reductive Homogeneous Manifolds} 
\label{sec:reduced-dynamics}

The construction of a Hamiltonian system on a Lie group \(\G\) was
straightforward because we were able to take the trivial cotangent bundle with
the Liouville form and express it in terms of the global left-invariant
Maurer--Cartan forms on~\(\G\).  It is more involved to construct a Hamiltonian
system that respects the desired symmetries on homogeneous spaces: an obvious
difficulty is that there are in general no global invariant vector fields.
Indeed, on \(\sphere2\) there are no non-vanishing vector fields (the ``hairy
ball'' theorem).  Fortunately, there is a simple construction of Hamiltonian
systems generated by the quotient structure.  If we construct a
\(\K\)-invariant Hamiltonian on \(\G\) and set the momenta corresponding to
motion in the \(\K\)-direction to zero, then the motion on \(\G\) projected
onto the coset space \(\G/\K\) has all the required properties.

To describe the symmetries associated with the right action of \(\K\) on
\(\G\), we introduce the map \(J: T^{*}\G\to\k^{*}\), with \(J(\alpha_g) =
\left.\pull{L_g}\alpha_g\right|_\k\).  Thus \(J\) describes the change in
momenta in the direction of the infinitesimal generators of the action of
\(\K\) on \(\G\).  The function \(J\) is called the \definition{momentum map},
and plays the role of the conserved quantity in Noether's theorem: if a
Hamiltonian is invariant under the action of the group \(\K\), then \(J\) is
conserved along the trajectories of the Hamiltonian Hamiltonian vector
field.  The definition above follows from the standard definition of the
momentum map \(J(\alpha_g)(\xi) \defn \alpha_g\bigl(\xi_\G(g)\bigr)\).  Here
\(\xi_\G\in\field{\tanbundle\G}\) is the left-invariant vector field generated
by \(\xi\in T_1\K\), i.e., \(\xi_{\G}(g)= \tang 1L_g\xi\) (more generally it is
the Killing field associated to \(\xi\in\k\) by the action of \(\K\) on
\(\G\)).  Hence \(\alpha_g\bigl(\xi_\G(g)\bigr) = \alpha_g(\tang 1L_g\xi) =
\pull{L_g}\alpha_g(\xi)\) and it follows indeed that \(J(\alpha_g) =
\left.\pull{L_g}\alpha_g\right|_\k\).

Since \(\K\) acts symplectically, freely and properly on \(\cotbundle\G\), it
follows from the Marsden--Weinstein reduction (see \cite{Marsden:1974} and
\cite{Abraham:2008} Theorems 4.3.1 and 4.3.5) that there exists a unique
reduced symplectic structure \(\omega_0\) on the reduced space
\(J^{-1}(0)/\K\).  In \refapp{proof-bundle-isomorphism} we prove that the
symplectic manifold \(J^{-1}(0)/\K\) is isomorphic to \(T^{*}(\G/\K)\).
Moreover, upon identifying \(J^{-1}(0)/\K\diffeo T^{*}(\G/\K)\) it follows that
\((\tang{}\pi)^{*}\omega_0=\omega|_{J^{-1}(0)}\), i.e., \(\omega_0\bigl(\tang
q\,\penalty-500\tang {}\pi(v), \tang q\,\tang {}\pi(u)\bigr) = \omega(v,u)\)
for all tangent vectors \(u,v\in T_qJ^{-1}(0)\).  Finally, the flow generated
by a \(\K\)-invariant Hamiltonian \(H\) is projected to a Hamiltonian flow on
reduced space generated by the reduced Hamiltonian \(H_0\), which satisfies
\(H_0 \circ \tang {}\pi = H\) for any momenta in \(J^{-1}(0)\).

Let us now apply this result to our situation.  Consider the kinetic energy
\(\tilde T\) on \(\M\diffeo\G/\K\) defined in terms of a \(\G\)-invariant
Riemannian metric, and a target density \(\exp(-\tilde \U)\) on \(\M\).
From~\refsec{sec:metric} we can choose an \(\Ad_\G\K\)-invariant inner product
on \(\g=\p\oplus \k\) for which \(\p\) and \(\k\) are orthogonal and the inner
product on \(\p\) is the pull-back of the \(\G\)-invariant Riemannian metric on
\(\G/\K\).  This induces an inner product on \(\g^{*}\) which defines a kinetic
energy~\(T\) on \(\g^*\).  Furthermore, the potential \(\tilde\U\) may be
extended to a function \(\U:\G\to\R\) that is coset-independent, \(\U(g) \defn
\pull{\pi} \tilde\U (g) = \tilde \U(g\K)\).  The Hamiltonian system with
\(H\defn T+\U\) on \(\cotbundle\G\) is then \(\K\)-invariant (see
\refapp{proof-K-invariance} for proof) and satisfies all the conditions of a
simple mechanical system with symmetry group \(\K\), as defined in
\cite{Abraham:2008} on page 341.  In particular the reduced Hamiltonian is
precisely the Hamiltonian \(\tilde T +\tilde V\) on \(T^*\M\) (see also Theorem
4.5.6~\cite{Abraham:2008}).

Although in theory we could sample directly from the reduced space
\(\cotbundle\M\), to avoid having to introduce local coordinates and the use of
complicated transcendental functions to implement group multiplication, we
instead ``sample'' from \(J^{-1}(0)\subset\cotbundle\G\) and then take the
quotient by the isotropy group \(\K\) to obtain samples from \(\cotbundle\M\).
In practice the requirement for momenta to remain in \(J^{-1}(0)\) is imposed
by simply setting the initial momentum to be in \(\p^*\), so that it has no
components in the direction of the isotropy to zero (see
\refapp{proof-inverse-at-zero} for proof).

Finally, we note that while the equations of motion \refeq{eq:leapfrog} were
derived using an \(\Ad \G\)-invariant inner product, these are unchanged for an
\(\Ad_{\G}\K\)-invariant inner product on a naturally reductive space, since
the geodesics in the horizontal direction \(\p\) are still given by Lie-group
exponentiation \cite{figueroa2005homogeneity}.

\section{Algorithm} \label{sec:pseudo}

Consider a naturally reductive homogeneous space \(\M\diffeo\G/\K\from\G:\pi\)
with a \(\G\)-invariant metric and a potential \(\tilde V:\M\to\R\); let
\(\langle\cdot, \cdot\rangle\) be the associated \(\Ad_\G\K\)-invariant inner
product on \(\p\), and \(V=\tilde V \circ \pi\) the extended potential.  Fix a
basis of \(\p\) and let \(T\) be the matrix associated to \(\half
\langle\cdot,\cdot\rangle\).  It is convenient to choose an orthonormal basis
so that \(T=\half I\).  We define the Hamiltonian \(H = V+T\) on \(\G\times
\g\), and let \(\rho:\G\to\GL{n,\R}\) be an injective representation (which can
just be the inclusion when \(\G\) is a matrix group).

In order to implement HMC we need to tune two parameters, the time step
\(\delta t\) and the trajectory length \(\trjlen\), see for example
\cite{betancourt2013generalizing,bou2017randomized,Neal:2012}.  The trajectory
length should be taken sufficiently long to minimise the mixing time, and the
step size should be adjusted to give a reasonable Metropolis acceptance rate
(the cost of the method is not very sensitive to the precise acceptance rate,
typically anywhere between 50\% and 80\% is adequate).  Given the current phase
\((\rho_0,P_0')\) the HMC algorithm proceed as follows:
\begin{itemize}
\item Sample \(P_0\) with respect to the measure \(e^{-T}\lambda\), where
  \(\lambda\) is the Lebesgue measure on~\(\p\).
\item For \(i \in [1,L]\), map \((\rho_{i-1},P_{i-1})\) to \((\rho_{i},P_{i})\)
  by performing a Leapfrog update as described in \refeq{eq:leapfrog}, with
  initial phase \((\rho_{i-1}, P_{i-1})\) and time step \(\delta t\).  The
  second step requires exponentiating an element of the Lie algebra
  \refsec{sec:mat-exp}.  This yields a proposal phase (\(\rho^{*},P^{*})\)=
  (\(\rho_L,P_L)\).
  
  It is advisable to correct the matrices representing Lie group elements after
  each update to ensure that they lie in the image of the group so as to reduce
  the effects of numerical floating point rounding errors.  Such projections
  are identity operations in exact arithmetic.
\item We accept the proposed state with probability 
  \begin{align*}
    \min\Bigl(1,\exp\bigl(-H(\rho^*,P^*) + H(\rho_0,P_0)\bigr)\Bigr),
  \end{align*}
  otherwise the proposal is rejected and the new phase is~\((\rho_0, P_0)\).
\item The procedure defines a sequence of samples \((\rho_i)\) on \(\G\) which
  can be projected to samples \(\bigl(\pi(\rho_i)\bigr)\) on~\(\M\).
\end{itemize}

\subsection[Ad(G)(K) Inner Products]%
           {\(\Ad_{\G}\K\) Inner Products} \label{sec:innerprod}

For many homogeneous spaces in which $\G$ is a matrix group the standard inner
product \((A,B) \mapsto \tr(A^TB)\) on $\g$ is often \(\Ad_{\G}\K\)-invariant.
Let us verify this for the inner products we used in our computations.  For the
sphere \(\sphere2\diffeo\liegroup{SO}(3)/\liegroup{SO}(2)\) as parametrised in
\refsec{sec:sphere}, with the defining representation \(\rho:g \mapsto g\).
The projection \(\pi(g) = g\K = g\cdot p_0\) simply maps the matrix \(g \in
\liegroup{SO}(3)\) to its first column.  The standard matrix inner product on
\(\p\) is \(\Ad_\G\K\)-invariant since the Cartan-Killing form is invariant
(\ref{metric}) and~\(\tr(A^TB)=-\tr(AB)\).

For the two sheeted hyperbolic space (\refapp{sec:h212}), since \(\Ad_kA = k A
k^{-1}\) ( this is true for any matrix group), and \(k^{-1} = k^T\) for \(k \in
\K = \liegroup{SO}(2)\), from which the \(\Ad_\G \K\)-invariance follows:
\begin{align*}
  \tr\bigl((\Ad_kA)^T\Ad_kB\bigr)
  &= \tr\left(k^{-T}A^Tk^T kBk^{-1}\right) \\
  &= \tr\left(kA^Tk^{-1} kBk^{-1}\right)
  = \tr(A^TB).
\end{align*}
On the single sheeted hyperbolic space (\refapp{sec:h2111}), we use the inner
product \(A,B \mapsto \tr(AB)\), which is \(\Ad_{\G}\K\)-invariant but
pseudo-Riemannian, rather than Riemannian.

\subsection{Pseudocode} \label{sec:pseudocode}

We begin by constructing the symbolic characteristic polynomial for an
\(n\times n\) matrix \(A\) as given by equation~\eqref{eq:dettr}.  This is
illustrated in Algorithm~\ref{alg:charpoly}, which returns a symbolic
polynomial in the variable \(\lambda\) with coefficients that are expressed in
terms of symbolic powers of traces \(\tr(A^j)\) with \(j\in\NN\) and \(A\) a
symbol.  In line~\ref{line:partitions} we iterate through the partitions \(P\)
yielded by a procedure \(\mathop{\rm partitions}(n)\) that generates all the
partitions of the integer \(n=\dim A\), and in line~\ref{line:jr} we iterate
through the pairs \((j,r)\) of positive integers in the partition~\(P\) where
\(n = \sum_{(j,r)\in P} j^r\).  Line~\ref{line:factor} computes the factor
\(f\) corresponding to \((j,r)\).  The term \(t\) in the summation in
equation~\eqref{eq:dettr} has its sign flipped in line~\ref{line:flipsign} if
necessary.  In line~\ref{line:expand} the resulting expression is expanded as a
polynomial in~\(\lambda\), so that its coefficients may be extracted later:
\(\mathop{\rm expand}\) is assumed to make use of the linearity of~\(\tr\).

This procedure is only evaluated once when initializing the program, and in the
absence of a suitable symbolic manipulation language the characteristic
polynomial may of course be evaluated ``by hand''.

\begin{algorithm}[ht]
  \begin{algorithmic}[1]
    \Procedure{CharPoly}{$n,\lambda, A$}
     \State \(c\assign0\)
     \For{\(P\in\mathop{\rm partitions}(n)\)} \label{line:partitions}
       \State \(t\assign1\)
       \State \(s\assign0\)
       \For{\((j,r)\in P\)} \label{line:jr}
         \State \(s\assign s + r(j+1)\)
         \State \(f\assign\Bigl(\tr\bigl((A-\lambda)^j\bigr)\Bigr)^r/(r!j^r)\)
           \label{line:factor}
         \State \(t\assign t\times f\)
       \EndFor
       \If{\(s\bmod 2 = 1\)} \(t\assign-t\) \label{line:flipsign}
       \EndIf
       \State \(c\assign c+t\)
     \EndFor
     \Return \(\mathop{\rm expand}(c)\) \label{line:expand}
   \EndProcedure
  \end{algorithmic}
  \caption{Symbolic computation of the characteristic polynomial of an
    \(n\times n\) matrix~\(A\).}
  \label{alg:charpoly}
\end{algorithm}

\(\algprocname{CharPoly}\) is used in the procedure \(\algprocname{ExpPoly}\)
given in Algorithm~\ref{alg:exp_poly} that symbolically computes an anonymous
function (\(\lambda\)-expression) that evaluates the Taylor expansion of
\(e^A\) of degree~\(N-1\).  It first computes the symbolic Taylor series
expansion of the exponential function \(e^\lambda\) truncated to degree
\(N-1\), and then in line~\ref{line:rem} it divides this polynomial by the
characteristic polynomial \(cp\) with respect to the symbolic variable
\(\lambda\) to give a polynomial \(e\) in \(\lambda\) (and traces of powers of
\(A\)) of degree~\(\min(N,n)-1\).  It then extracts the coefficient of
\(\lambda^k\) in line~\ref{line:coeff} into \(c\), and optimizes this
expression (as a polynomial in traces of powers of the symbol \(A\)) for
numerical evaluation by applying Horner's rule in line~\ref{line:horner}.  Up
to this point the numerical coefficients are computed exactly as rational
numbers for numerical stability, but in line~\ref{line:float} they are
converted to floating point numbers for faster numerical evaluation.  The
coefficient of \(\lambda^k\) is then multiplied by the symbolic power \(A^k\)
making use of the Cayley--Hamilton theorem and accumulated as a symbolic
expression \(r\); finally the resulting symbolic polynomial \(r\) in powers of
\(A\) and powers of traces of \(A\) is converted into an anonymous numerical
procedure taking a numerical floating-point matrix \(A\) as its argumentby
constucting \(\lambda\)-expression (this \(\lambda\) has nothing to do with the
variable \(\lambda\) used before).  This anonymous \(\lambda\)-expression may
be compiled at this point, as this computation is only performed once when the
program is initialized, but the resulting procedure is later invoked many
times.  During initialization the matrix dimension \(n=\dim A\) and the degree
\(N-1\) of the Taylor expansion of the exponential are known explicitly, but
the matrix \(A\) is not.

\begin{algorithm}[ht]
  \begin{algorithmic}[1]
    \Procedure{ExpPoly}{$N,n$}
      \State \(cp\assign\algprocname{CharPoly}(n,\lambda,A)\)
      \Comment Algorithm~\ref{alg:charpoly}
      \State \(t\assign1\)
      \For{\(k\in[1,N)\)} 
        \State \(t\assign t+\lambda^k/k!\) \Comment Symbolic Taylor expansion
        \State \(e\assign\mathop{\rm rem}(t,cp,\lambda)\)
        \Comment Remainder of polynomial division \label{line:rem}
      \EndFor
      \State \(r\assign0\)
      \For{\(k\in[0,N)\)}
        \State \(c\assign\mathop{\rm coeff}(e,\lambda^k)\)
        \Comment \(c\) is a polynomial in \(\tr A^j\) \label{line:coeff}
        \State \(c\assign\mathop{\rm horner}(c)\) \label{line:horner}
        \Comment Improve numerical stability using Horner's Rule (see
          eq.~\eqref{eq:horner_app})
        \State \(c\assign\mathop{\rm float}(c)\) \label{line:float}
        \Comment Convert rational numbers to floats
        \State \(r\assign r+c\times A^k\) \Comment \(A\) is symbolic
      \EndFor
      \State \(r\assign\alglambda(A,r)\) \label{line:lambda}
      \Comment Construct and compile anonymous function to evaluate~\(r(A)\)
      \Return \(r\)
    \EndProcedure
  \end{algorithmic}
  \caption{Construct an anonymous function that computes the Taylor expansion
    of \(e^A\) of an \(n\times n\) matrix to order \(N-1\) as a function of an
    explicit floating-point matrix~\(A\).  \(cp\) is the characteristic
    polynomial as a symbolic polynomial in \(\lambda\) and \(\tr A^j\).}
  \label{alg:exp_poly}
\end{algorithm}

There is a further \(\lambda\)-expression that may be useful constructed during
initialization, and that is to compute the gradient of the (extended) potential
function \(V:\GL{n,\R}\to\R\).  We assume that we are provided with a symbolic
expression \(\textit{Vs}\) for this potential as a function of the \(\GL n\)
matrix~\(X\); this potential must be invariant under~\(\K\).  The anonymous
numerical procedure to evaluate the gradient as a function of the
floating-point matrix \(X\) is computed by the procedure
\(\algprocname{AutoDiff}\) in Algorithm~\ref{alg:autodiff}.

\begin{algorithm}[ht]
  \begin{algorithmic}[1]
    \Procedure{AutoDiff}{\textit{Vs}}
      \For{\(i\in[1,n]\)}
        \For{\(j\in[1,n]\)}
          \State \(dV_{ji}\assign\mathop{\rm diff}(\textit{Vs},X_{ij})\)
          \Comment Symbolic derivative of \(V\), note implicit transpose
        \EndFor
      \EndFor
      \State \(f\assign\alglambda(X,dV)\)
      \Comment Construct and compile anonymous function for \(dV(X)\)
      \Return \(f\)
    \EndProcedure
  \end{algorithmic}
  \caption{Pre-compute the gradient of the symbolic potential function
    \(\textit{Vs}:\GL{n,\R}\to\R\).  Note the implicit transpose for
    consistency with \refeq{eq:gradV}.}
  \label{alg:autodiff}
\end{algorithm}

So far all our algorithms are used at initialization, and in they can be
precomputed by hand or machine if so desired.  We now turn to the numerical
algorithms that are used throughout the HMC sampling.  Let us start with the
computation of the matrix exponential of Algorithm~\ref{alg:MatExp}.  The
algorithm use the precomputed Taylor expansion \(r(A)\) if \(\|A\|\) is small
enough, otherwise it computes a suitable \(k\in\NN\) and evaluates
\(r(A/k)^k\).  The formula for \(\delta\) used in line~\ref{line:delta} and
\(k\) used in line~\ref{line:k} are to be found in
\S\ref{sec:scale-and-square}, and depend upon the Taylor expansion degree
\(N-1\), the desired accuracy \(\varepsilon\), and parameter \(\alpha\).
Typical values might be \(N=10\), \(\varepsilon=10^{-6}\), and \(\alpha=0.9\),
but these will depend on the particular application.  In line~\ref{line:delta}
we also need to compute an upper bound on the spectral norm \(\|A\|\); as
discussed in \S\ref{sec:taylor} we may use \(\|A\|\leq\sqrt{\tr AA^T}\) in
general or \(\|A\|\leq\sqrt{\half\tr AA^T}\) for \(A\in\liegroup{SO}(n)\).

\begin{algorithm}[ht]
  \begin{algorithmic}[1]
    \Procedure{Exp}{$A$} \Comment Argument is floating-point matrix \(A\)
    \If{\(\|A\|<\delta\)}
      \Comment \(\delta\) is defined in~\S\ref{sec:scale-and-square}
      \label{line:delta}
      \Return \(r(A)\)
    \Else
      \State Define \(k\assign(\|A\|/\delta)^{\frac N{N-1}}\)
      \Comment See~\S\ref{sec:scale-and-square}
      \label{line:k}
      \Return \(\algprocname{BinaryPowering}(A/k,k)\)
    \EndIf
  \EndProcedure
  \end{algorithmic}
  \caption{Compute matrix exponential, where \(\delta\) can be precomputed
    as it does not depend on \(A\), and \(r\defn\algprocname{ExpPoly}(N,n)\)
    has been precomputed by Algorithm~\ref{alg:exp_poly}.}
  \label{alg:MatExp}
\end{algorithm} 

The procedure \(\algprocname{BinaryPowering}\) evaluates the power \(a^n\) of a
matrix \(a\) efficiently when \(n\) is a non-negative integer.  This is, of
course, trivial to evaluate using matrix multiplication, but it may be
optimized somewhat if we express \(n\) in binary \(n=\sum_{j\geq0} n_j2^j\)
with \(n_j\in\{0,1\}\), as then
\begin{equation*}
  a^n = \prod_{j\geq0} \left\{\begin{array}{ll}
    1 & \mbox{if \(n_j = 0\)} \\
    a^{2^j} & \mbox{if \(n_j = 1\),}
  \end{array}\right.
\end{equation*}
where the powers \(a^{2^j}\) may be computed by repeated squaring.  The
pseudocode for \(\algprocname{BinaryPowering}\) is given in
Algorithm~{\ref{alg:binary}.

\begin{algorithm}[ht]
  \begin{algorithmic}[1]
    \Procedure{BinaryPowering}{$a,n$}
      \State \(k\assign n\) \Comment So as not to modify the argument~\(n\)
      \State \(p\assign1\)
      \Comment Set \(p\) to be the unit element of the algebra
      \State \(ak\assign a\)
      \Comment \(ak\) set to successive squares \(a,a^2,a^4,a^8,\ldots\)
      \If{\(k\bmod2=1\)}{\(p\assign ak\)}
         \Comment Unroll the first iteration of the following loop
      \EndIf
      \While{\(k>0\)}
        \State \(ak\assign ak\cdot ak\) \Comment Square \(ak\)
        \If{\(k\bmod2=1\)}{\(p\assign p\cdot ak\)}\EndIf
        \State \(k\assign\lfloor k/2\rfloor\)
        \Comment Shift \(k\) one bit to the right
      \EndWhile\label{euclidendwhile}
      \Return \(p\)
    \EndProcedure
  \end{algorithmic}
  \caption{Binary powering: compute \(a^n\) for \(a\) in any associative
    algebra and~\(n\in\NN\).}
  \label{alg:binary}
\end{algorithm}

Algorithm~\ref{alg:HMCSamples}
\begin{algorithm}[h]
  \begin{algorithmic}[1]
    \Procedure{HMC}{$N$}
      \State \(Q_0\assign1\)
      \For{\(j\in[1,N]\)}
        \State \(P\assign\algprocname{Gibbs}()\)
        \Comment Algorithm~\ref{alg:Gibbs}
        \State \(Q_j\assign\algprocname{MDMC}(Q_{j-1},P)\)
        \Comment Algorithm~\ref{alg:MDMC}
      \EndFor
      \Return \(Q_1,Q_2,\ldots,Q_N\)
    \EndProcedure
  \end{algorithmic}
  \caption{HMC Algorithm to generate \(N\) samples.}
  \label{alg:HMCSamples}
\end{algorithm}
is the basic HMC algorithm that performs momentum refreshments using a Gibbs
sampler
\begin{algorithm}[h]
  \begin{algorithmic}[1]
    \Procedure{Gibbs}{}
      \State \(M\assign0\)
      \For{\(j\in[1,\dim\p]\)}
        \State \(v\assign\N(0,\lambda^{-1}_j/2)\)
        \Comment Gaussian-distributed random float 
        \State \(M\assign M + vT_j\)
      \EndFor
      \Return \(M\)
    \EndProcedure
  \end{algorithmic}
  \caption{Momentum Refreshment, given the inner product matrix \(T\), and a
    basis of \(n\times n\) matrix generators \(T_i\) of~\(\p\).  Here
    \(\lambda_j\) denotes the eigenvalue of of the kinetic energy matrix \(T\)
    corresponding to the eigenvector \(T_j\).}
  \label{alg:Gibbs}
\end{algorithm}
and MDMC steps
\begin{algorithm}[h]
  \begin{algorithmic}[1]
    \Procedure{MDMC}{$Q,P$}
      \State \(P0\assign P\) \Comment Save initial position in phase space
      \State \(Q0\assign Q\)
      \For{\(\varepsilon,\textit{step}\) in
          \(\algprocname{Trajectory}(\dt,\trjlen)\)}
        \State\(Q,P\assign\textit{step}(\varepsilon,Q,P)\)
        \Comment Perform symplectic integration steps
      \EndFor
      \State \(Q',P'\assign\algprocname{Metropolis}(Q0,P0,Q,P)\)
      \State \(P'\assign-P'\) \Comment This momentum flip is irrelevant
      \label{line:irrelevant}
      \Return \(Q'\)
    \EndProcedure
  \end{algorithmic}
  \caption{MDMC Markov step.  \(\algprocname{Trajectory}(\dt,\trjlen)\) is an
    iterator for any symmetric symplectic integrator with itegrator step size
    \(\dt\) and (mean) trajectory length~\(\trjlen\).}
  \label{alg:MDMC}
\end{algorithm}
to generate \(N\) samples \(Q_1,\ldots,\fairbreak Q_N\).  The Gibbs sampler is
illustrated in Algorithm~\ref{alg:Gibbs} and MDMC in Algorithm~\ref{alg:MDMC}.
In line~\ref{line:irrelevant} of the latter we perform a momentum sign flip,
which does nothing as the momentum is promptly replaced by an independent Gibbs
sample; we include it as it is a necessary step in variants of the algorithm in
which the momenta are only partially refreshed.  The MDMC algorithm is expressed
in terms of an iterator \(\algprocname{Trajectory}\)
\begin{algorithm}[h]
  \begin{algorithmic}[1]
    \Procedure{Trajectory}{$\dt,\trjlen$} 
      \State \Yield \(\dt/2,\algprocname{That}\) \Comment Initial half step
      \State \Yield \(\dt,\algprocname{Vhat}\)
      \For{\(j\in[2,\trjlen/dt]\)} \Comment Assuming that \(\trjlen/\dt\in\NN\)
        \State \Yield \(\dt,\algprocname{That}\)
        \State \Yield \(\dt,\algprocname{Vhat}\)
      \EndFor
      \State \Yield \(\dt/2,\algprocname{That}\) \Comment Final half step
    \EndProcedure
  \end{algorithmic}
  \caption{Trajectory iterator: we use leapfrog as an example with step size
    \(\dt\) and fixed trajectory length~\(\trjlen\).  The \(\Yield\) statement
    provides the next value of the iterator, but the state is saved and
    execution continues with the next statement for the next iteration.  The
    iterator finishes when the end of the procedure is reached.  It is simple
    to modify this for randomly distributed trajectory lengths.}
  \label{alg:trajectory}
\end{algorithm}
that generates pairs \(\varepsilon, \textit{step}\) each consisting of a step
size \(\varepsilon\) and procedure \(\textit{step}\) that performs the
appropriate step.  We provide an example iterator for a simple Leapfrog
integrator in Algorithm~\ref{alg:trajectory}, but we stress that the use of
such an iterator allows the implementation of more sophisticated symmetric
symplectic integrators without modifying the MDMC code itself.  The MDMC
algorithm makes use of the Metropolis algorithm~\ref{alg:Metropolis},
\begin{algorithm}[h]
  \begin{algorithmic}[1]
    \Procedure{Metropolis}{$Q0,P0,Q,P$}
      \Comment Accept \(Q,P\) with probability \(\min(1,e^{-\delta H})\)
      \State \(H0\assign T(P0)+\U(Q0)\) \Comment Initial energy
      \State \(H\assign T(P) + \U(Q)\) \Comment Final energy
      \State \(\delta H\assign H-H0\) \Comment Change in energy
      \State \(U\assign \text{Uniform}[0,1]\)
      \Comment Generate a uniformly distributed random number in~[0,1]
      \If{\(U\leq\exp{(-\delta H)}\)}
        \Return \(Q,P\) \Comment Accept new point in phase space
      \Else
        \Return \(Q0,P0\) \Comment Reject and repeat old point
      \EndIf
    \EndProcedure
  \end{algorithmic}
  \caption{Metropolis algorithm.  We assume the numerical procedures \(T\) and
    \(\U\) compute the kinetic and potential energy respectively.  It would be
    reasonable to define \(\U\assign\alglambda(X,\textit{Vs})\) during
    initialization where \(\textit{Vs}\) is the symbolic expression for the
    potential used in Algorithm~\ref{alg:autodiff}.}
  \label{alg:Metropolis}
\end{algorithm}
and indirectly of the integtator steps \(\hvf V\)
\begin{algorithm}[h]
  \begin{algorithmic}[1]
    \Procedure{Vhat}{$\varepsilon,Q,P$}
      \State \(M\assign\textit{gradV}(Q)\cdot Q\)
      \State \(M\assign\sum_{i=1}^n\tr\bigl(M\cdot T_i\bigr)T^i\)
      \Comment Algorithm~\ref{alg:project} may be used to project \(M\)
        onto \(\liealgebra{SO}(n)\)
      \State \(P'\assign P-\half\varepsilon\,\M\)
      \Return \(Q,P'\)
    \EndProcedure
  \end{algorithmic}
  \caption{Momentum update step of size \(\varepsilon\), which follows the
    integral curve of \(\hvf\U\).  We set \(\textit{gradV}\assign
    \algprocname{Autodiff}(\textit{Vs})\) during initialization.}
  \label{alg:Vhat}
\end{algorithm}
and \(\hvf T\) implemented in Algorithms~\ref{alg:Vhat} and~\ref{alg:That}.
\begin{algorithm}[h]
  \begin{algorithmic}[1]
    \Procedure{That}{$\varepsilon,Q,P$}
      \State \(Q'\assign Q\cdot\algprocname{Exp}(P\varepsilon)\)
      \Comment Algorithm~\ref{alg:MatExp}
      \State \(Q'\assign\algprocname{Project}(Q')\)
      \Comment Reduce floating point errors, Algorithm~\ref{alg:project}
      \label{line:proj}
      \Return \(Q',P\)
    \EndProcedure
  \end{algorithmic}
  \caption{Position update step of size \(\varepsilon\), which follows the
    integral curve of \(\hvf T\).  We call \(\algprocname{Project}\) in
    line~\ref{line:proj} to reduce floating point arithmetic errors: in exact
    arithmetic it does nothing.}
  \label{alg:That}
\end{algorithm}

Finally, Algorithm~\ref{alg:project} is the Gram--Schmidt algorithnm to project
a matrix \(M\) onto \(\liealgebra{SO}(n)\) to reduce floating-point errors.
\begin{algorithm}
  \begin{algorithmic}[1]
    \Procedure{Project}{$M$}
    \Comment Orthonormalise a matrix \(M\in\R^{n\times n}\)
      \State \(M'\assign M\) \Comment So as not to modify argument matrix \(M\)
      \For{\(i\in[1,n]\)}
        \State \(ri\assign\mathop{\rm row}(M',i)\)
        \Comment \(ri\) is row \(i\) of \(M'\)
        \For{\(j\in [1,i-1]\)}
          \State \(rj \assign\mathop{\rm row}(M',j)\)
          \Comment \(rj\) is row \(j\) of \(M'\)
          \State \(c\assign\langle ri,rj\rangle/\langle rj,rj\rangle\)
          \State \(ri\assign ri - c\cdot rj\)
          \Comment Orthogonalise rows \(i\) and \(j\)
        \EndFor
        \State \(c'\assign\sqrt{|\langle ri,ri\rangle|}\)
        \State \(ri\assign ri/c'\)
        \Comment Normalise row \(i\)
        \State \(\mathop{\rm row}(M',i)\assign ri\)
        \Comment Update row \(i\) of \(M'\)
      \EndFor
      \Return\(M'\)
    \EndProcedure
  \end{algorithmic}
  \caption{Gram--Schmidt orthonormalization with respect to the nondegenerate
    inner product~\(\langle\cdot,\cdot\rangle\).  We assume that the matrix
    \(M\) is close to being orthonormal; the procedure can also be applied to
    an arbitrary matrix \(M\), but in that case additional work has to be done
    to suitably adjust the sign of each row.}.
  \label{alg:project}
\end{algorithm}

\section[Parameterization of a 2-sphere]%
        {Parameterization of \(\sphere2\) using~\(\p\)}%
        \label{sec:sphere}

We now describe the parametrisation of the sphere as a homogeneous space.  The
standard action (by matrix product) of \(\liegroup{SO}(3,\R)\) on \(\sphere2
\subset \R^3\) is transitive, and the isotropy group of the ``north pole''
\(p_0 = (1,0,0)\) is \(\diag(1,A)\) for \(A\in \liegroup{SO}(2,\R)\).  From
\refsec{sec:red-hom-space}, any coset \(g\K_{p_0}\) may be identified with the
point \(g\cdot p_0\in\M\).  The sphere is then diffeomorphic to
\(\liegroup{SO}(3,\R)/\liegroup{SO}(2,\R)\), or, in other words, any point on
the sphere can be identified with a rotation about a line in the \(2\)--\(3\)
plane.

As in~\refsec{sec:exp-so3} we choose the generators of \(\liealgebra{so}(3)\)
to be \((T_i)_{jk} = \varepsilon_{ijk}\),
\begin{equation*}
  T_1 = \left(\begin{array}{ccc} 0&0&0 \\ 0&0&1 \\ 0&-1&0 \end{array}\right),
  \:
  T_2 = \left(\begin{array}{ccc} 0&0&-1 \\ 0&0&0 \\ 1&0&0 \end{array}\right),
  \:\mbox{and}\:
  T_3 = \left(\begin{array}{ccc} 0&1&0 \\ -1&0&0 \\ 0&0&0 \end{array}\right).
\end{equation*}

In this case \(\k=\vspan T_1\) generates \(\liegroup{SO}(2,\R)\) and
\(\p=\vspan(T_2, T_3)\) generates~\(\exp(\p)\cdot p_0 = \M\).  In order to
clarify the geometrical meaning of these generators we parameterize the coset
space \(\M\) in terms of real angles \(\theta\) and \(\phi\),
\begin{align}
  S(\theta,\phi) &= \exp\left[\theta(\sin\phi\,T_2 - \cos\phi\,T_3)\right]
  = \exp\left[\theta\left(\begin{array}{ccc}
    0 & -\cos\phi & -\sin\phi \\
    \cos\phi & 0 & 0 \\
    \sin\phi & 0 & 0
  \end{array}\right)\right] \nonumber \\
  &= \left(\begin{array}{ccc}
    \cos\theta & -\sin\theta\cos\phi & -\sin\theta\sin\phi \\
    \sin\theta\cos\phi & (\sin\phi)^2 + \cos\theta(\cos\phi)^2
      & (\cos\theta - 1)\cos\phi\sin\phi \\
    \sin\theta\sin\phi & (\cos\theta - 1)\cos\phi\sin\phi
      & (\cos\phi)^2 + \cos\theta(\sin\phi)^2
  \end{array}\right), \label{eq:S}
\end{align}
and we identify the point
\begin{equation*}
  S(\theta,\phi) \left(\begin{array}{c} 1 \\ 0 \\ 0 \end{array}\right)
  = \left(\begin{array}{c}
    \cos\theta \\ \sin\theta\cos \phi \\ \sin\theta\sin\phi
  \end{array}\right)  \in  \sphere2
\end{equation*}
in spherical polar coordinate.  We may parameterize the isotropy group
\(\K_{p_0}\) as
\begin{equation*}
   \exp(tT_1) = \left(\begin{array}{ccc}
     1 & 0 & 0 \\
     0 & \cos t & \sin t \\
     0 & -\sin t & \cos t
   \end{array}\right);
\end{equation*}
clearly, \(\exp(tT_1)p_0 = p_0\).  \(S(\theta,\phi)\exp(tT_1)\) parameterizes
\(\liegroup{SO}(3,\R)\), and the isotropy group \(K_p\) at \(p\defn
S(\theta,\phi)p_0\) is parameterized by \(S(\theta,\phi) \exp(tT_1)
S(\theta,\phi)^T\).

\begin{figure}
  \begin{center}
    \epsfxsize=0.3\textwidth \epspdffile{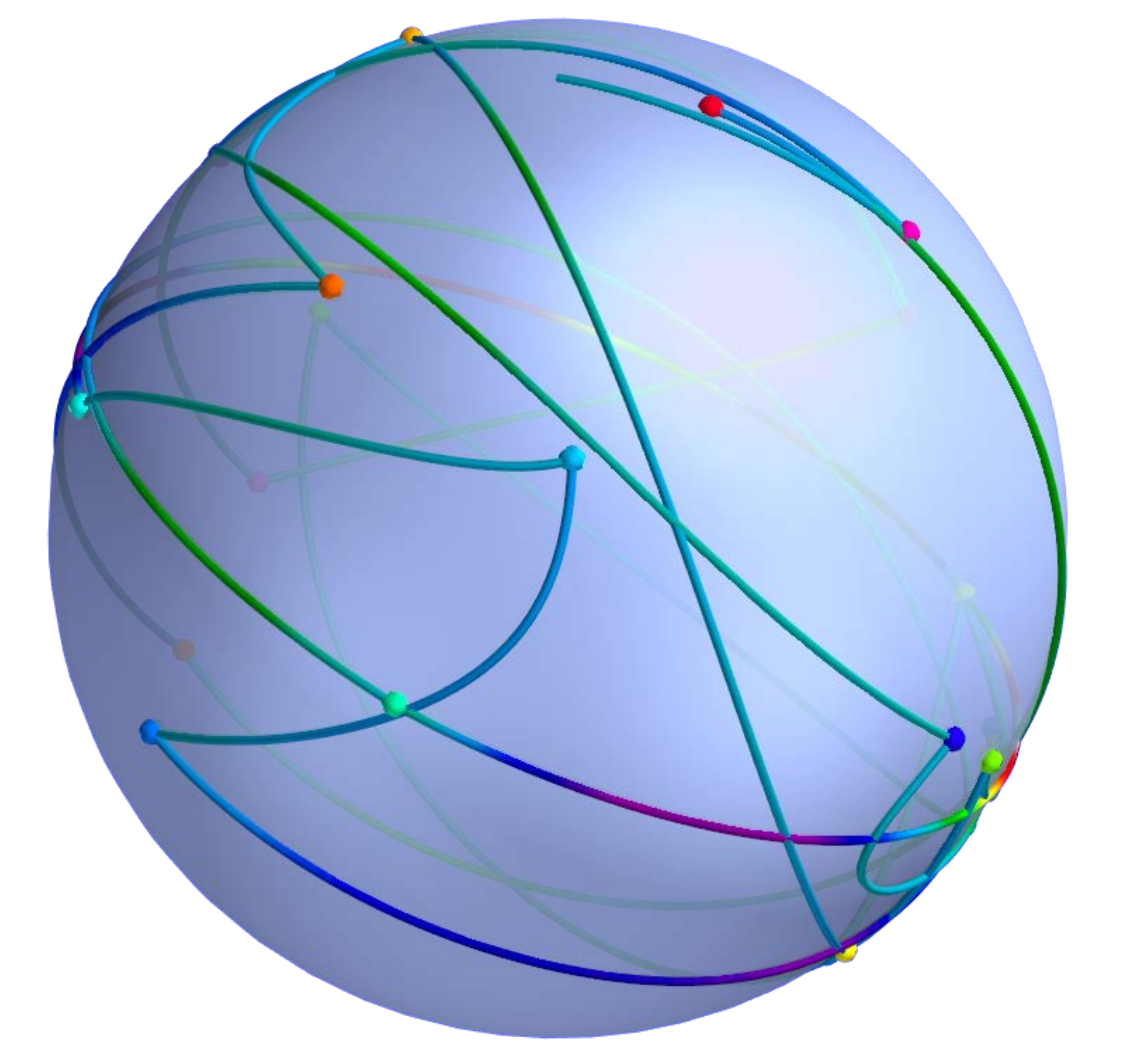}
  \end{center}
  \caption{20 HMC trajectories, chronologically coloured blobs, the
    trajectories are coloured proportionally to the values of \(\delta H\),
    with potential \(\U(x,y,z)=yz^2\exp(x^2)\), \(\dt=0.1\), \(\trjlen=2\).}
\end{figure}

\begin{figure}
  \begin{center}
    \epsfxsize=0.6\textwidth \epspdffile{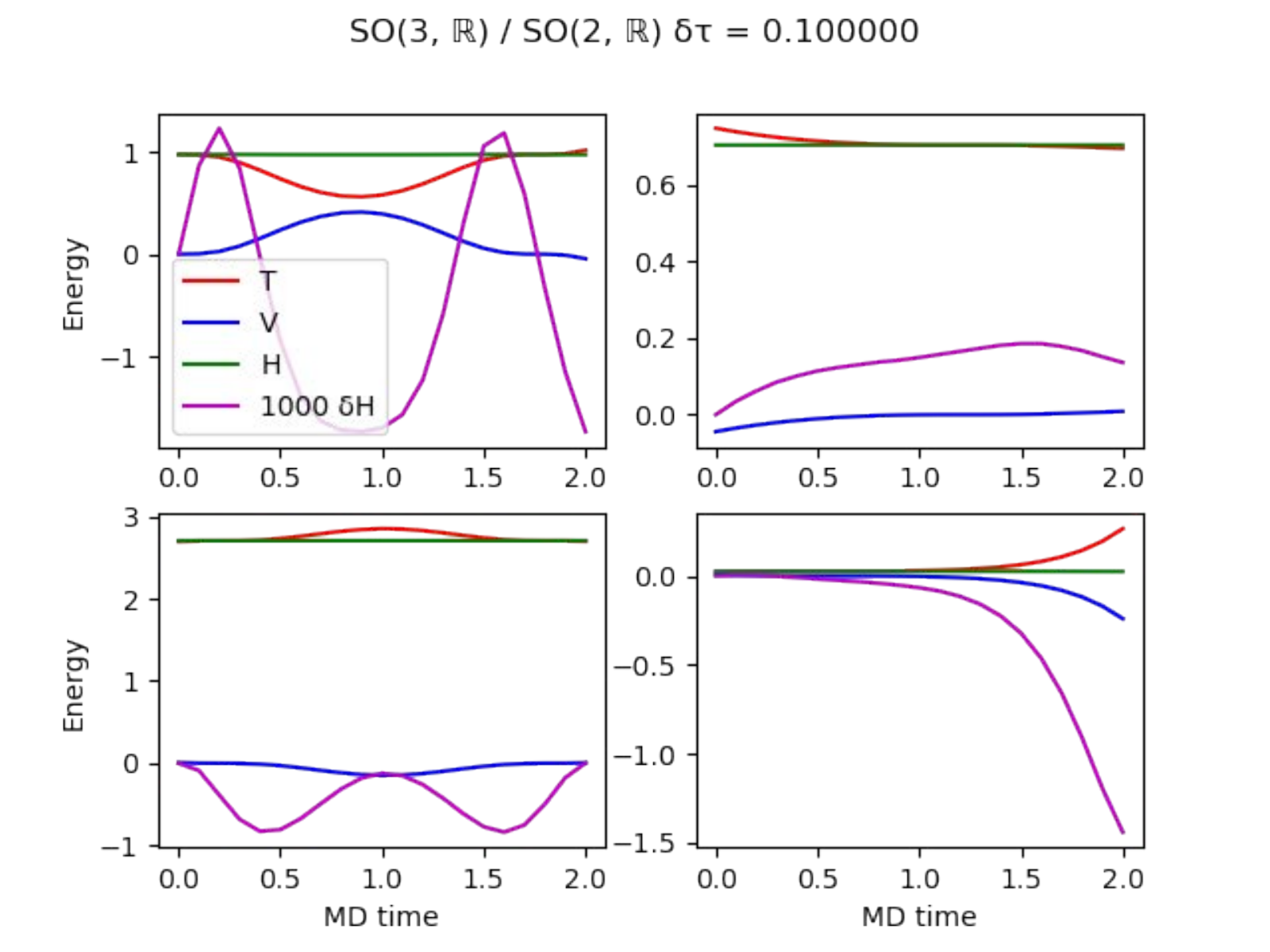}
  \end{center}
  \caption{\(4\) HMC trajectories, with potential is \(\U=y^3\exp(z^2+x^2)\),
    \(\trjlen=2\), \(\dt=0.1\), with a leapfrog integrator.  Notice the
    conservation of energy violation \(\delta H\) is tiny.  The second figure
    has \(\U=y+z^2+\exp(x^2)\), \(\trjlen=0.25\), \(\delta H\) and verifies the
    scaling of integrator with step-size.  The graphs show how the various
    energies change along each trajectory.}
\end{figure}
 
\begin{figure} 
  \epsfxsize=0.49\textwidth
  \epspdffile{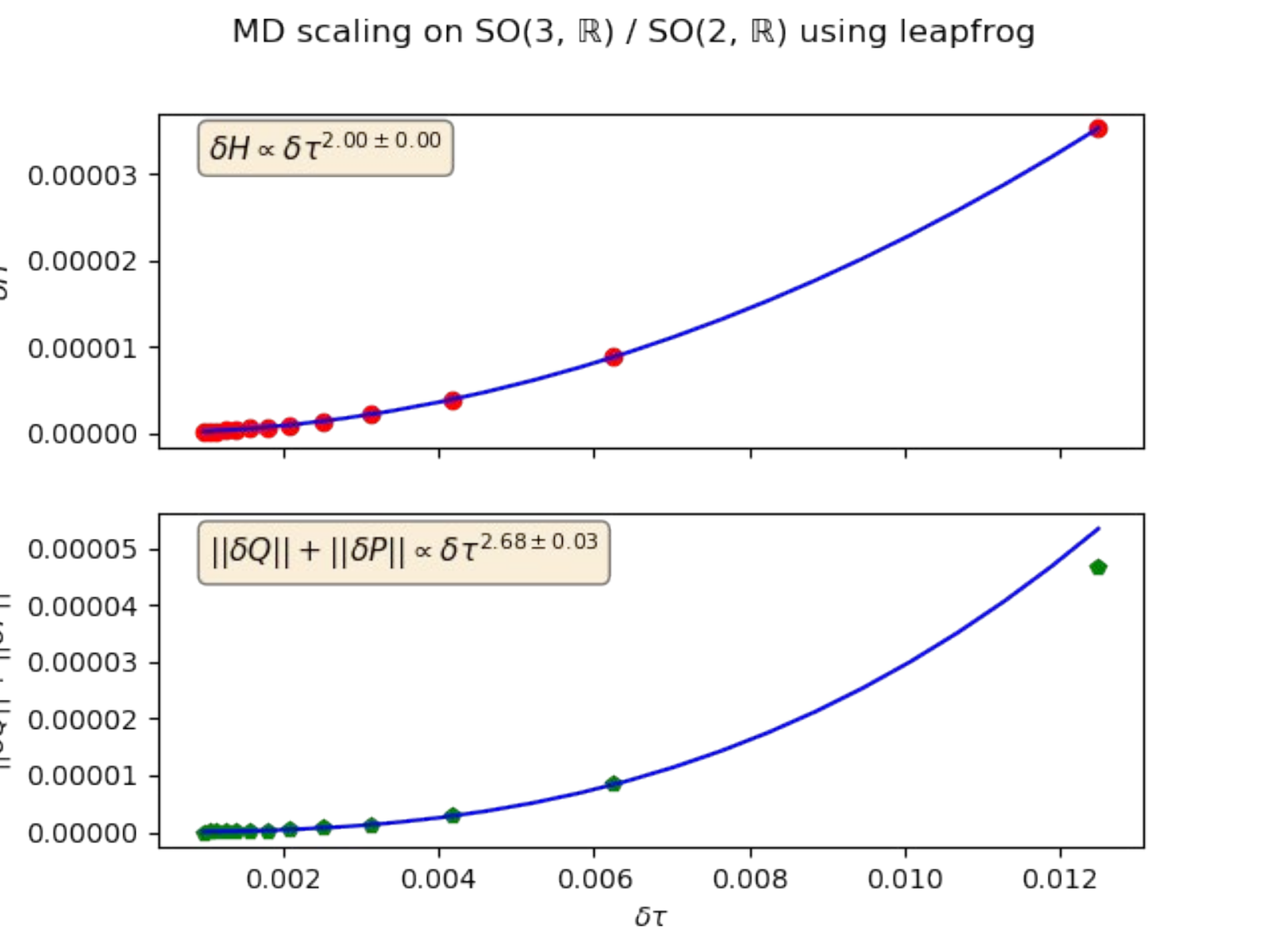}
  \epsfxsize=0.49\textwidth
  \epspdffile{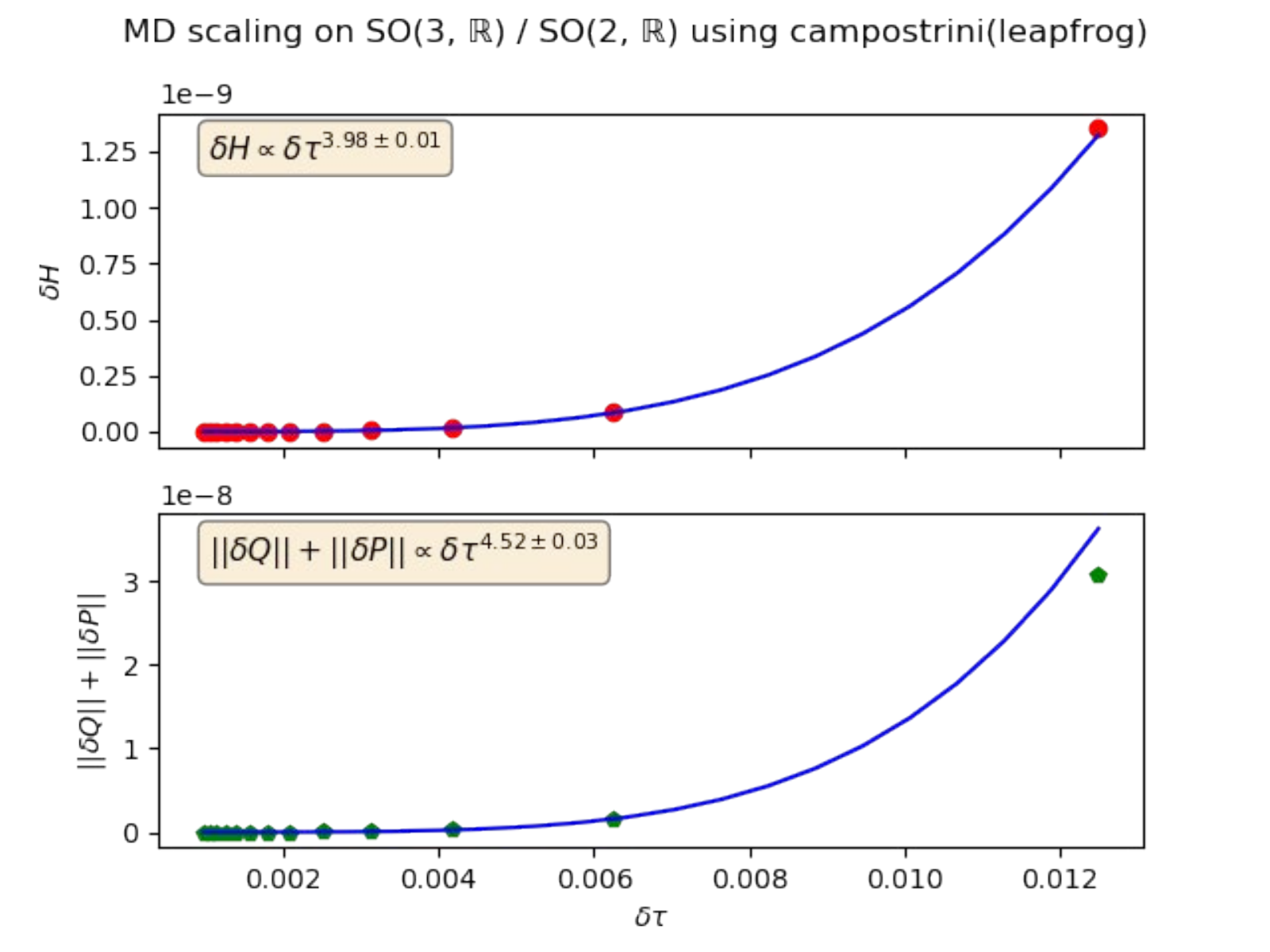}
  \caption{The first figure has \(\U=y+z^2+\exp(x^2)\), \(\trjlen=0.25\),
    \(\delta H\) and verifies the scaling of integrator with step-size.  The
    top graphs verify that the change in energy \(\delta H\) scales as
    \(\dt^2\) for the Leapfrog integrator and \(\dt^4\) for the higher-order
    Campostrini \cite{campostrini89a, creutz89a} integrator.  The bottom graphs
    show the corresponding scaling of the error in the final positions in phase
    space.}
\end{figure} \label{fig:check-scaling}

\begin{figure}
  \begin{center}
    \epsfxsize=0.5\textwidth \epspdffile{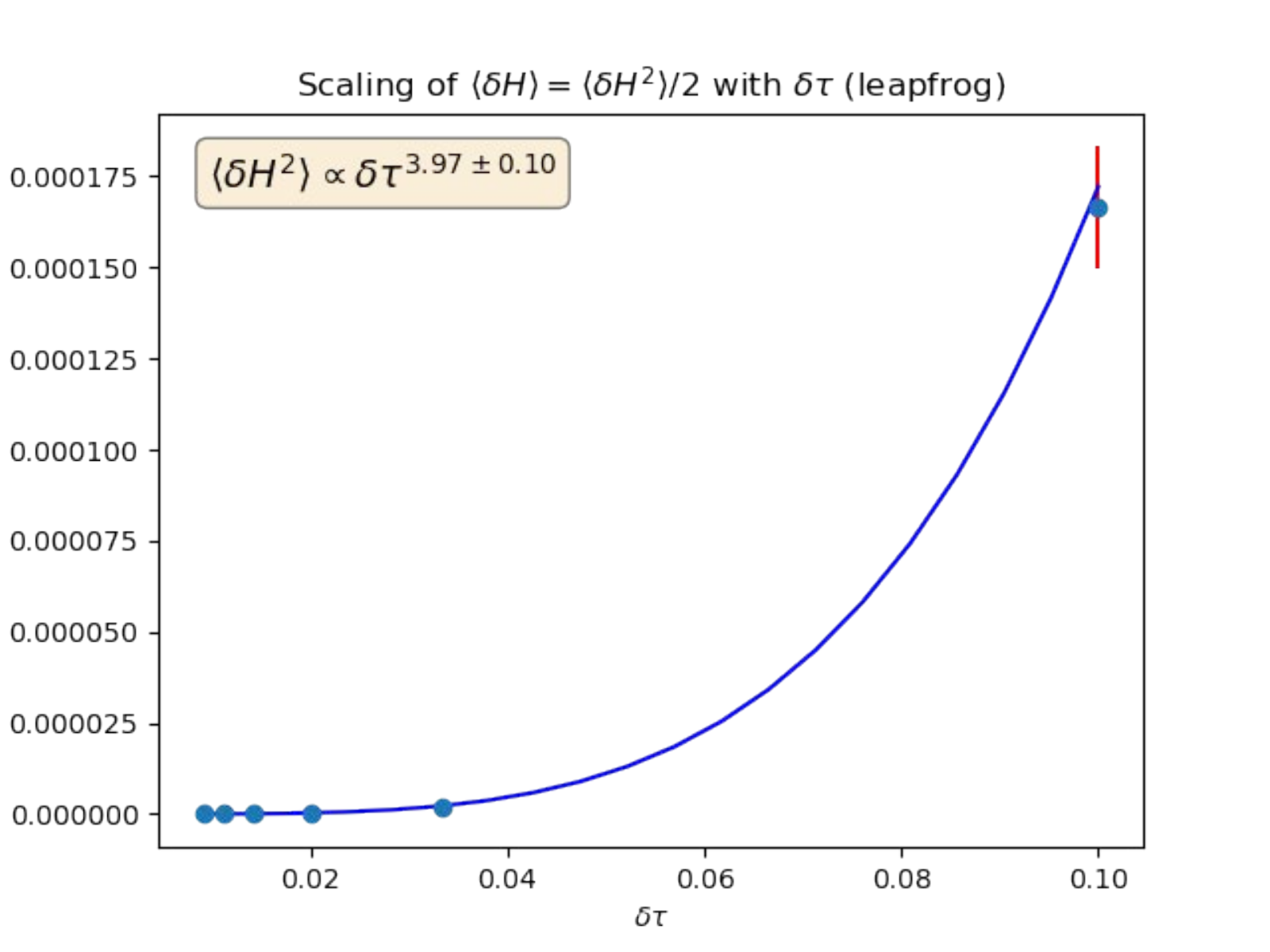}
  \end{center} 
  \caption{Here the potential is \(\U=y\exp(z^2+2x^2)\) with \(\trjlen=1\).  We
    measure the sample average value of \(\delta H^2\) over \(500\) HMC
    samples, and verify that we generate the right distribution.  The value of
    \(\langle\delta H^2\rangle\) scales as \(\dt^4\) for a leapfrog integrator,
    which is consistent with the condition \(\langle e^{-\delta H}\rangle = 1\)
    that must hold if the samples are selected from the correct distribution
    \(\propto e^{-H}\).}
 \end{figure}

\section{Matrix Exponentiation} \label{sec:mat-exp}

The discrete symplectic integration step for \(\hvf T\)~\refeq{eq:leapfrog}
requires numerical evaluation of the exponential map \(\exp:\g\to\G\) from a
Lie algebra to its corresponding Lie group, or more precisely between their
matrix representations.  It is desirable that such numerical integration be
computed exactly, that is close to the precision of the floating point
arithmetic being used.

The numerical analysis of matrix exponentiation is not a new
subject~\cite{moler:2003,higham:2008}, but there are some simplification that
are possible for this application.  We shall consider four approaches: Taylor
series expansion for matrices near the identity~\refsec{sec:taylor}; the
scaling and squaring algorithm~\refsec{sec:scale-and-square}; the use of the
Cayley--Hamilton theorem to transform the matrix to block diagonal
form~\refapp{sec:rodrigues}; and numerical diagonalisation using standard
numerical algorithms such as QR iteration~\refapp{sec:numerical}.  Finally we
discuss the exponentiation of non-normal matrices in~\refapp{sec:abnormal-exp}.

\begin{figure}
  \epsfxsize=0.49\textwidth
  \epspdffile{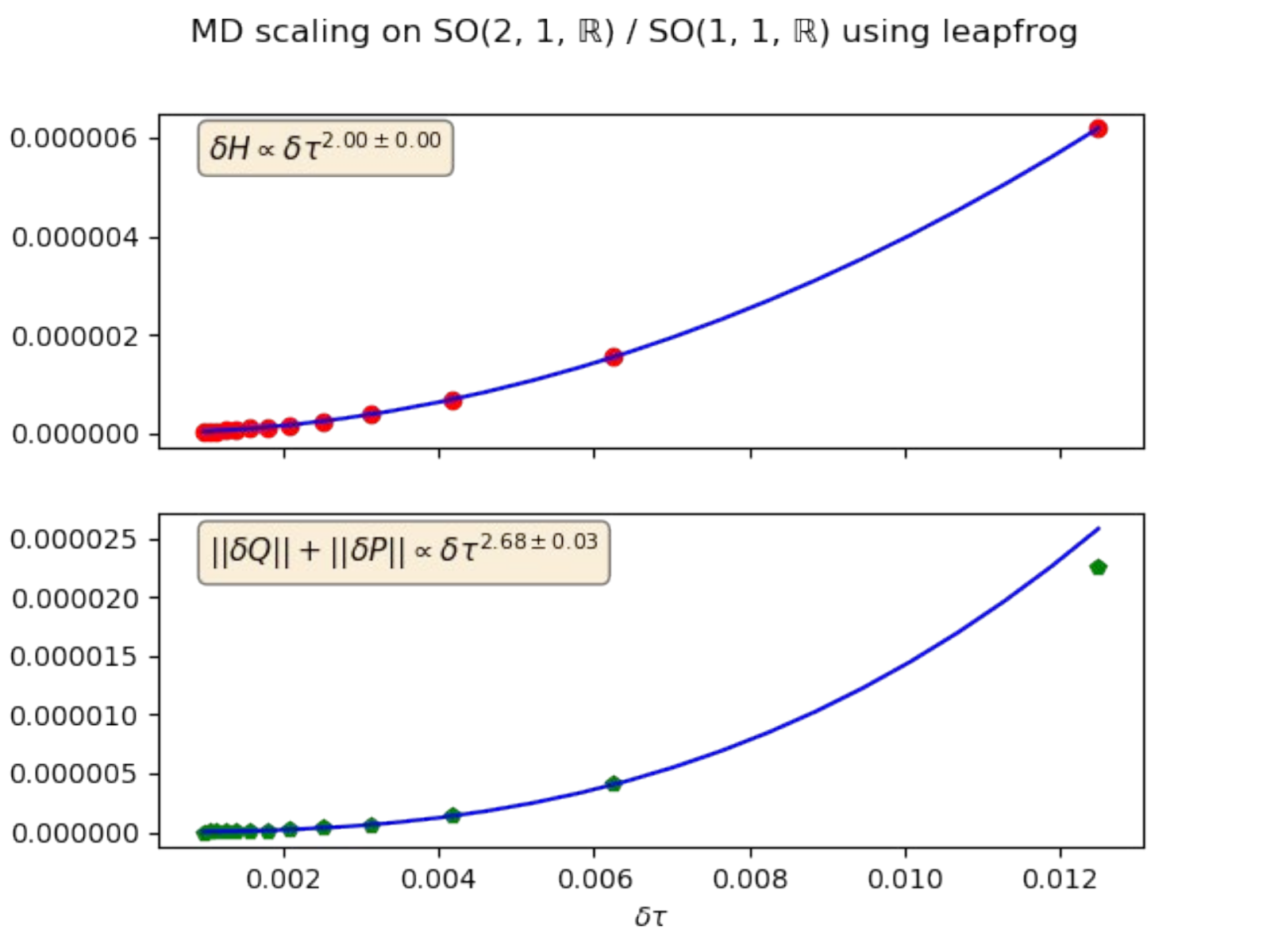}
  \epsfxsize=0.49\textwidth
  \epspdffile{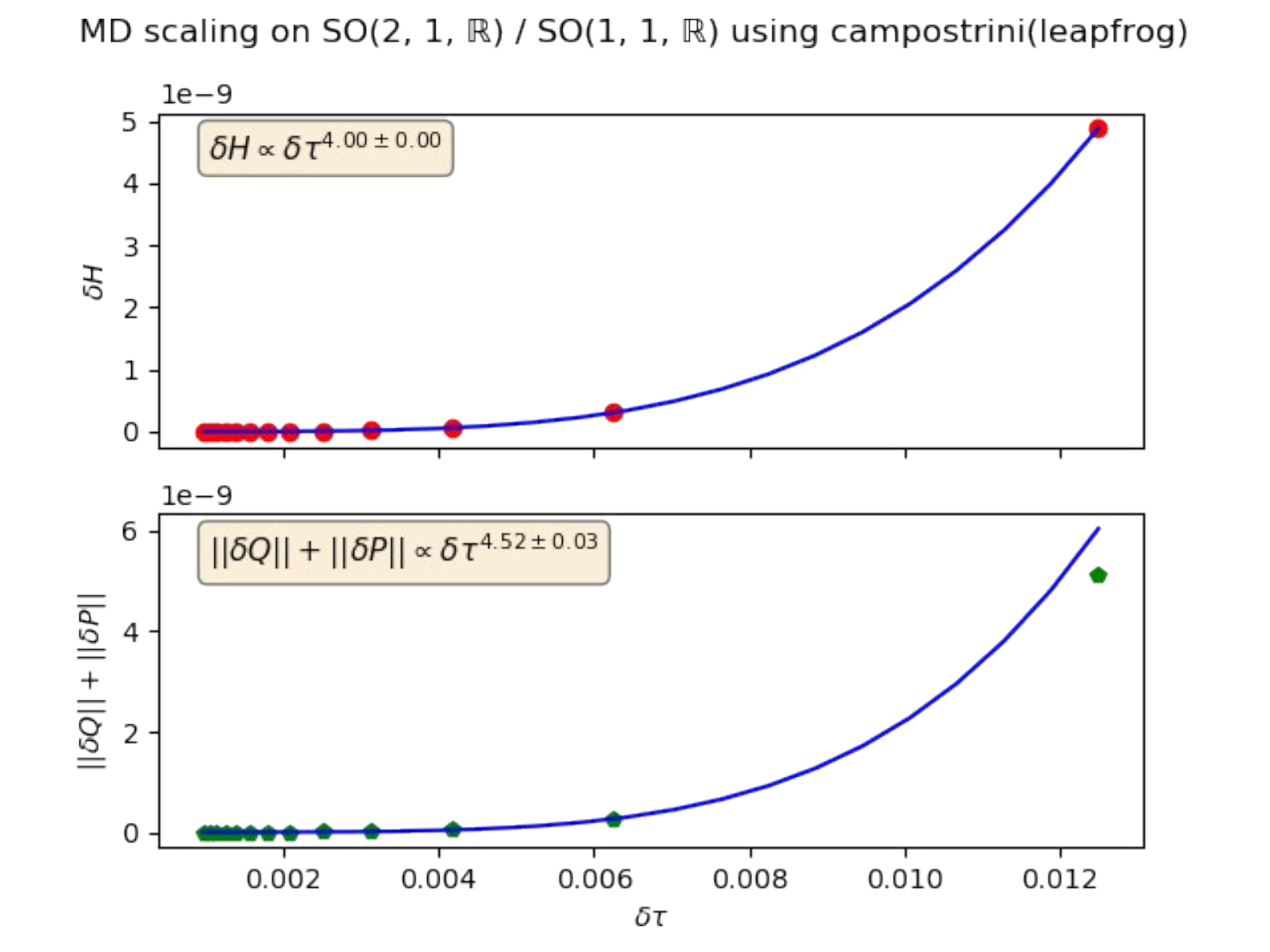} 
  \caption{This is the analogous procedure to figure \ref{fig:check-scaling},
    but on the manifold \(\hyperboloid21\).  This indicates that there is no
    problem with carrying out MDMC on this space, even though there is no Gibbs
    sampler for the momenta.}   
\end{figure}

\subsection{Cayley--Hamilton Theorem} \label{sec:cayley}

The Cayley--Hamilton theorem tells us that every matrix \(A\) satisfies its own
characteristic equation \(c(A) = 0\), where
\begin{equation}
  c(\lambda) \defn \det(A - \lambda\ident).
  \label{eq:cayley-hamilton}
\end{equation}
We may express the determinant of any matrix \(X\) in terms of traces of its
powers using the identity
\begin{equation}
  \det X = \sum_P\prod_{j=1}^n
    \frac{(-1)^{r_j(j+1)}}{r_j!j^{r_j}}\,\left(\tr X^j\right)^{r_j},
  \label{eq:dettr}
\end{equation}
where the sum is over all partitions \(1^{r_1},2^{r_2},\ldots,n^{r_n}\) of the
integer \(n = \dim X\) where~\(\sum_{j=1}^n jr_j=n\).

\subsection{Exponentiation by Taylor expansion} \label{sec:taylor}

If the potential \(\U\) (the log posterior distribution) is steep the stepsize
used by the symplectic integrator must be small so as to avoid the integrator
becoming unstable.  In this case we have to exponentiate matrices \(A\) in the
Lie algebra that are close to the identity.  In this case it suffices to expand
the exponential function
\begin{equation}
  e(A) \defn \sum_{j=1}^{N-1}\frac{A^j}{j!}, \qquad
  \exp A = e(A) + \O\left(\|A\|^N\right)
  \label{eq:exp-series}
\end{equation}
such that the remainder is negligible.  Here \(\|A\|\) is the spectral norm of
the matrix~\(A\), that is \(\|A\|=|\lambda_{\max}|\) where \(\lambda_{\max}\)
is the eigenvalue of \(A\) of largest magnitude.

We next establish a simple bound on \(\|A\|\) to ensure that \(\|e(A)-\exp(A)\|
\leq\varepsilon\).  We have
\begin{align*}
  \sum_{j=N}^\infty \frac{x^j}{j!}
  &\leq\sum_{j=N}^\infty \frac{|x|^j}{j!}
  = \frac{|x|^N}{N!}
    \left(1 + \frac{|x|}{N+1} + \frac{|x|^2}{(N+1)(N+2)} + \cdots\right) \\
  &\leq \frac{|x|^N}{N!}
    \left(1 + \frac{|x|}N + \frac{|x|^2}{N^2} + \cdots\right)
  \leq \frac{|x|^N}{N!} \sum_{j=0}^\infty\left(\frac{|x|}N\right)^j
  = \frac{|x|^N}{N!}\left(1-\frac{|x|}N\right)^{-1}
\end{align*}
where the geometric series converges provided that \(|x|<N\).  The function
\begin{equation*}
  f(x) \defn \frac{x^N}{N!}\left(1-\frac xN\right)^{-1}
\end{equation*}
is positive and monotone increasing for \(0\leq x<N\) so it suffices to find a
\(\delta\) such that \(f(\delta)<\varepsilon\).  Indeed, it is sufficient to
require that both \(\delta/N\leq\alpha\) and \(\delta\leq[\varepsilon
  N!(1-\alpha)]^{1/N}\) for any value \(0<\alpha<1\) that we may care to
specify, for example \(\alpha = 0.9\), as for such a \(\delta\) we have
\(\|r(A)-\exp(A)\|\leq\varepsilon\) whenever~\(\|A\|\leq\delta\).  A simple
bound on the spectral norm is \(\|A\| \leq\sqrt{\tr AA^T}\); for
\(A\in\liegroup{SO}(n,\R)\) we have \(\|A\|\leq\sqrt{\half\tr AA^T}\) since all
the non-zero eigenvalues occur in complex conjugate pairs.

This algorithm may be improved by use of the Cayley--Hamilton
theorem~\refeq{eq:cayley-hamilton}.  Since \(c(A) = 0\) we may divide the
truncated exponential series \(e(x)\) by the characteristic polynomial \(e(x)=
q(x)c(x) + r(x)\) to obtain a polynomial \(r(A)\approx\exp(A)\) with \(\deg
r<n\) where \(n=\dim A\).  The coefficients of \(r\) may be expressed in terms
of traces of powers \(A^k\) where \(k<n\).  The advantage of this approach is
that the only the matrix powers \(A, A^2,\ldots,A^{n-1}\) are required even
for~\(N\gg n\).

As an example let us consider \(A\in\liealgebra{SO}(3)\) with \(N=10\).  The
characteristic polynomial is \(c(x)= -x\bigl(x^2 - \half\tr A^2\bigr)\), as
shown in~\refsec{sec:exp-so3}, so we have
\begin{align} \nonumber
  r(A) & = \left[\half + \left(\rational1{48} + \left(\rational1{2880} +
    \rational1{322560}\,\tr A^2\right) \tr A^2\right) \tr A^2\right] A^2
    \nonumber \\
   &\qquad + \left[1 + \left(\rational1{12} + \left(\rational1{480}
      + \left(\rational1{40320} + \rational1{5806080}\,\tr A^2\right)
      \tr A^2\right)\tr A^2\right) \tr A^2\right] A + \ident \nonumber \\
   & \approx \exp A, \label{eq:horner_app}
\end{align}
where we have used Horner's rule to improve numerical stability.  

\subsection{Scaling and Squaring} \label{sec:scale-and-square}

While the exponential series \refeq{eq:exp-series} converges for all \(A\) the
required value of \(N\) and the necessary floating point precision become
infeasible when \(\|A\|\) is not small.  A simple way of numerically evaluating
\(\exp A\) in this case is to use the identity \(\exp A=\bigl(\exp(A/k)
\bigr)^k\) for \(k\in\NN\) such that \(\exp(A/k)\) may be computed sufficiently
accurately by means of Taylor expansion.

Let us derive a simple lower bound on \(k\) that ensures that the relative
error is small, \(e(A/k)^k = \exp(A)\bigl(1 + \O(\varepsilon)\bigr)\), ignoring
errors due to finite precision floating point arithmetic.  This bound is good
enough for HMC purposes as the update will almost certainly be rejected it the
exponential is large enough for such floating point errors to be significant.

Let \(\delta(\varepsilon)\defn[\varepsilon N!(1-\alpha)]^{1/N}\) as
in~\refsec{sec:taylor}.  Set \(\displaystyle k=\left\lceil\frac{\|A\|}
{\delta(\varepsilon/k)}\right\rceil\) as then \(\|A\|<k\delta(\varepsilon/k)\)
and thus
\begin{align*}
  e\left(\frac{\|A\|}k\right)^k
  &= \left(\exp\frac{\|A\|}k + \frac{\varepsilon'}k\right)^k \\
  &= \exp\|A\|\left(1 + \frac{\varepsilon'}k\,\exp\frac{-\|A\|}k\right)^k
  = \exp\|A\|(1 + \Delta)
\end{align*}
with \(|\varepsilon'|\leq\varepsilon\) and
\begin{equation*}
  |\Delta| \leq \left(1 + \frac{|\varepsilon'|}k\right)^k-1
  = \sum_{j=1}^k \binom kj \left|\frac{\varepsilon'}k\right|^j
  \leq \sum_{j=1}^k \frac{|\varepsilon'|^j}{j!}
  = \O(\varepsilon),
\end{equation*}
where we have used the facts that \(e^s>1\) for \(s>0\) and \(\displaystyle
\binom kj\leq\frac{k^j}{j!}\).  Finally
\begin{equation*}
  \left[\delta\!\left(\frac\epsilon k\right)\right]^N
  = \frac{\delta(\varepsilon)^N}k
  \leq \frac{\delta(\varepsilon/k)\delta(\varepsilon)^N}{\|A\|}
  \leq\left(\frac{\delta(\varepsilon)^N}{\|A\|}\right)^{\frac N{N-1}},
\end{equation*}
from which it follows that
\begin{equation*}
  k\geq\left(\frac{\|A\|}{\delta(\varepsilon)}\right)^{\frac N{N-1}}.
\end{equation*} 

See Algorithm 10.20 in \cite{higham:2008} for more details concerning the
scaling and squaring algorithm, and a discussion of the merits of using Pad\'e
approximants (such as the Cayley transform \cite{bou2009hamilton}) rather than
Taylor series.

The HMC algorithm is valid whatever function is used provided it defines a
symmetric symplectic integrator, but care must be taken in the trade-off
between the cost of the ``exponentiator" and the longer mixing time resulting
from a lower Metropolis acceptance rate.

\section{Conclusions} \label{sec:conclusions}

In this paper we have developed a general method to sample from a large class
of manifolds that often arise in applications of statistics and physics.  The
algorithm samples on any naturally reductive manifold \(\G/\K\) by lifting the
HMC scheme to the associated Lie Group \(\G\), and in particular can be
immediately extended to use higher order integrators and Shadow Hamiltonians \cite{Kennedy:2012}.
As we have seen it does not require introducing local coordinates on the sample
space and automatically handles its geometry and curvature, without having to
impose explicit constraints.

Finally we have provided examples and detailed calculations on the sphere and
its associated hyperbolic spaces, which can be easily visualised.  We note that
the same procedure may be used to sample from many manifolds obtained by
symplectic reduction with a target density given by the reduced potential
energy.
%%
%% The method can be used to perform Geodesic Monte Carlo without computing the
%% inverse metric by exponentiating the lie algebra, and we have described in
%% detail how to do so.
%%

\section*{Acknowledgements}

We would like to thank Jos\'{e} Miguel Figueroa-O'Farrill, Darryl Holm and So
Takao for very useful discussions.  MG was supported by the EPSRC grants
[EP/J016934/3, EP/K034154/1, EP/P020720/1, EP/R018413/1], an EPSRC Established
Career Fellowship, the EU grant [EU/259348] and the Lloyds Register Foundation
Programme on Data-Centric Engineering.  AB was supported by a Roth scholarship
from the Department of Mathematics at Imperial College London.  ADK was
supported by STFC grant ST/P0000630/1 and an Alan Turing Institute Faculty
Fellowship.  This work was also supported by the Enrichment programme of The
Alan Turing Institute and the EPSRC grant [EP/N510129/1].

\appendix

\section{Proofs} \label{proofs}

\subsection{Proof of Equation (\ref{eq:symmetriesfof})}%
           \label{proof-symmetriesfof}

Let \(\beta\in\field{\cotbundle\G}\), \(p_i:\cotbundle\G\to\R\), \(\pi:T^*\G
\to \G\) and \(\fof\in\Lambda^1(\cotbundle\G)\).  By definition of the
pull-back of the tangent map \(\tang g \beta\), \refapp{differential-geometry}
(note we are taking the pull-back of \(\beta\) viewed as a map \(\G \to
\cotbundle{\G}\))
\begin{align*} 
  \pull{(\beta)}\fof(g)
  \defn \pull{\bigl(\tang g\beta\bigr)}\fof_{\beta(g)} 
  = \fof_{\beta(g)}\circ \tang g\beta
  = (p_i\pull\pi\theta^i)_{\beta(g)}\circ \tang g\beta.
\end{align*}
We can write \(\beta = f_i \theta^i\) for some functions \(f_i\in
C^{\infty}(\G)\).  It follows using \refeq{eq:momentum-coord} that
\(p\bigl(\beta(g)\bigr) = L^*_g \beta_g =
L^*_g\big(f_i(g)\theta^i_g\big)=f_i(g) \theta^i(1)\), so
\(p_i\big(\beta(g)\big) = f_i(g)\), and, since \(\pi\circ\beta : g\mapsto g\),
\begin{align*}
  (\pull\pi\theta^i)_{\beta(g)}\circ \tang g\beta
  = \theta^i_g\circ \tang {\beta(g)}\pi\circ \tang g\beta
  = \theta^i_g\circ \tang g(\pi\circ\beta)
  = \theta^i_g.
\end{align*}   
  
Hence \(\pull{(\beta)}\fof(g) = p_i\bigl(\beta(g)\bigr) (\pull\pi
\theta^i)_{\beta(g)}\circ \tang g\beta = f_i(g)\theta^i_g=\beta(g)\).  This
proves that \(\fof\) is the canonical Liouville form (see prop 3.2.11
of~\cite{Abraham:2008}).

\subsection{Proof \(L^*_g\) and \(R_g^*\) are symplectomorphisms}
           \label{proof-symplectomorphism}

Since \(L_g\) is a diffeomorphism of \(\G\), the induced map \(\pull{L_g}:
\cotbundle\G\to\cotbundle\G\) is a diffeomorphism.  Let us denote its pullback
by \(L_g^{**} \defn (L_g^*)^*\).

Let \(\fof=p_i\pull\pi\theta_i\).  Let us check directly that \(L^{**}_g\fof =
\fof\).  Indeed
\begin{equation*}
  L^{**}_g \fof
  = \bigl(p_i \circ L_g^* \bigr)L^{**}_g \pi^*\theta^i
  = p_i \pi^* \theta^i \circ\tang {}L_g^*
  = p_i \theta^i \circ \tang {}\pi \circ \tang {}L_g^*
  = p_i \theta^i \circ \tang {}( \pi \circ L_g^*).
\end{equation*}
Moreover, \(\pi \circ L_g^*: T^*\G \to \G\), and for \((h,\alpha)\in T^*_h\G\),
\(\pi\circ\pull{L_g}(h,\alpha) = \pi(gh,\alpha \circ \tang hL_g) = gh =
L_g\circ \pi(h,\alpha)\).  Thus
\begin{equation*}
  L^{**}_g\fof
  = p_i\theta^i \circ \tang {}( L_g\circ\pi)
  = p_i\theta^i\circ \tang {}L_g\circ \tang {}\pi
  = p_i\pull{L_g}\theta^i\circ \tang {}\pi = p_i\theta^i\circ\tang{}\pi
  = \fof.
\end{equation*}

More generally, using a derivation similar to section~2 in
\cite{alekseevsky:1994}, we can re-write our Liouville form \(\fof\) as the
canonical Liouville form: Let \(\alpha\in\tanbundle{\cotbundle\G}\), with
\(\alpha=(\beta,v_\beta)=\bigl((g,\gamma_g),v_{(g,\gamma_g)}\bigr)\), where
\(\beta=(g,\gamma_g)\in T^*_g\G\).  Let \(\chi:\tanbundle{\cotbundle\G}\to
\cotbundle\G\), \(\pi:\cotbundle\G\to\G\) be the canonical projections.  By
definition \(p(\pi^*\theta)(\alpha)=p(\chi(\alpha))\bigl(\pi^*\theta(\alpha)
\bigr)\) Then
\begin{align*}
  \fof(\alpha)
  = p(\pi^*\theta)(\alpha)=(\theta^{-1}_g)^{*} (\chi(\alpha))
    \bigl(\theta_g\circ \tang {}\pi(\alpha)\bigr)
  &=\chi(\alpha)\circ\theta^{-1}_g\circ\theta_g\circ \tang {}\pi(\alpha) \\
    &=\chi(\alpha) \bigl(\tang {}\pi(\alpha)\bigr),
\end{align*}
which is precisely how the canonical 1-form acts on \(\alpha\) (see for example
\cite{Abraham:2008} theorem 3.2.10) (this gives a frame-independent way of
showing \(\fof\) is the canonical Liouville form).  It then follows from Theorem
3.2.12 \cite{Abraham:2008} that if \(f\) is any diffeomorphism of \(\G\), then
its induced map \(f^{*}\)on \(\cotbundle\G\) is a symplectomorphism.

\subsection[Proof of K-invariance of H]%
           {Proof of \(\K\)-invariance of \(H\)}%
           \label{proof-K-invariance}
  
To show this we extend the \(\Ad_ \G\K\)-invariant inner product on \(\p\) to
an inner product on \(\g\) which is \(\Ad_ \G\K\)-invariant, by defining a
\(\Ad_ \G\K\)-invariant inner product on \(\k\) and setting \(\k \perp \p\).
This corresponds to a \(\G\)-left invariant and \(\K\)-right invariant
Riemannian metric \(\langle \cdot, \cdot \rangle_{\G}\) on \(\G\).  Define the
kinetic energy \(T(\cdot) \defn \half \langle \cdot, \cdot \rangle_{\G}^{-1}\)
using the Riemannian metric on one-forms corresponding to \(\langle \cdot,
\cdot \rangle_{\G}\).  The coset independence of the potential,
\(\U(g)=\U(g\K)\) implies that \(\U\) is constant along a left-invariant vector
field \(X\in\k\) (below \(F^X\) is the flow of \(X\))
\begin{equation*}
  X_g(\U) = \Lie{X_g}\U= \left.\frac{d\U\left(F^X_t(g)\right)}{dt}\right|_{t=0}
  = \left.\frac{d\U\left(ge^{tX_1}\right)}{dt}\right|_{t=0}
  = \left.\frac{d\U(g)}{dt}\right|_{t=0} = 0,
\end{equation*}
thus momentum is conserved in the direction of isotropy (this important fact
will be used throughout the paper).  Now consider the right action of
\(R:\G\times\K\to\G\) of \(\K\) on \(\G\) (by right translations).  Then
\(\U=\U\circ R_k\).  Moreover \(R\) acts by isometries, \(R_k^{*}\langle u,v
\rangle_{\G} = \langle R_{k*}u,R_{k*}v\rangle_{\G} = \langle u,v \rangle_{\G}\)
since \(\langle \cdot, \cdot \rangle_{\G}\) is \(\K\)-right invariant.  If
\(\alpha,\beta\) are the 1-forms corresponding to \(u,v\) by the musical
isomorphism, then \(\langle R^{*}_k\alpha,R^{*}_k\beta \rangle_{\G}^{-1} =
\langle R_{-k*}u, R_{-k*}v \rangle_{\G} = \langle u,v \rangle_{\G} = \langle
\alpha,\beta \rangle^{-1}_{\G}\), so that \(T\circ R^{*}_k = T\).  Defining the
Hamiltonian \(H = T+\U\circ\pi\), where \(\pi:T^{*}\G\to\G\) is the projection,
implies \(H\circ R^{*}_k = H\).  Note that since \(R_k\) is a diffeomorphism of
\(\G\), its lift \(R_k^{*}:T^{*}\G\to T^{*}\G\) acts symplectically
(\refapp{proof-symplectomorphism}).

\subsection[Proof of 1-forms in pullback of zero momentum]% 
           {Proof of 1-forms in \(J^{-1}(0)\)}\label{proof-inverse-at-zero}
 
Note \(J^{-1}(0)\) consists of 1-forms on \(\G\) whose left translation at the
identity vanishes \(\left.\pull{L_g}\alpha_g\right|_\k =0\), in other words
\(\pull{L_g} \alpha_g(v)=0\) for all \(v\in\k\).  Let \(u\in\g\) be the vector
associated to \(\pull{L_g}\alpha_g\) by the inner product \(\langle \cdot,
\cdot \rangle_{\G}\), so \(\langle u,v \rangle_{\g}=0\) for all \(v\in\k\) and
thus \(u\in\p\) (since \(\p \perp \k\)).  It follows that \(J^{-1}( 0)\)
consists of one-forms whose translation at the identity is in the dual
complementary subspace.  Thus \(J^{-1}(0)\diffeo\G\times\p^{*}\diffeo
\G\times\p\).  Thus if we initialise the mechanics using an element of \(\G
\times \p\), the symmetries of the Hamiltonian will ensure that the system
never leave the level set \(J^{-1}(0)\).  This is still true for the
discretised mechanics, since \(\hvf\U = -e_i(\U)\partial_{p_i}\), and
\(-e_i(\U)=0\) for all \(e_i \in \k\) as shown in \refapp{proof-K-invariance}
(in fact this follows from the fact that \(T\) and \(\U\) are both
\(\K\)-invariant, and thus by Noether's theorem their flow is horizontal).

It is worth noting that \(J^{-1}(0)\) is the horizontal space of a connection
\(\A\) on the principal bundle \(\G\to\G/\K\), called the mechanical
connection.
  
\subsection[Proof that pi is an isometry on p]%
           {Proof that \(\pi\) is an isometry on \(\mathfrak p\)}%
           \label{proof-of-isometry}

First recall from \refapp{homogeneous-spaces} that \(\Phi_L\) is the action of
\(\G\) on \(\G/\K\), \(\Phi_L(g,h\K)\defn \Phi_L(g)\big(h\K)\fairbreak\defn
gh\K\).  Consider a \(\G\)-invariant metric \(\langle\cdot,\cdot
\rangle_{\G/\K}\) on \(\G/\K\), i.e., \(\Phi_L(g)^{*}\langle\cdot,\cdot
\rangle_{\G/\K}=\langle \cdot , \cdot \rangle_{\G/\K}\) for any \(g\in \G\).
The \(\G\)-invariance of \(\langle \cdot , \cdot \rangle_{\G/\K}\) on \(\G/\K\)
implies that the value of the inner product of \(X,Y \in \field{T \G/\K}\) can
be calculated at the identity \(\langle X_{g \K}, Y_{g \K} \rangle_{g \K}=
\langle \tang {g\K}\Phi_L(g^{-1})X_{g \K}, \tang {g\K}\Phi_L(g^{-1})Y_{g \K}
\rangle_\K\).  Using the isomorphism \(\tang 1 \pi:\p\to T_\K\G/\K\), we define
an inner product \(\langle \cdot , \cdot \rangle_\p\) on \(\p\) corresponding to
\(\langle\cdot,\cdot\rangle_{\G/\K}\) by \(\langle u,v\rangle_\p\defn\langle
\tang 1\pi(u),\tang 1\pi(v)\rangle_\K.\) There is bijection between
\(\G\)-invariant metric on \(\G/\K\) and \(\Ad_ \G\K\)-invariant inner products
on \(\p\), since
\begin{align*}
  \langle\Ad(k)u,\Ad(k)v\rangle_\p
  &= \langle \tang 1\pi\circ\Ad(k)u,\tang 1\pi\circ \Ad(k)v\rangle_\K \\ 
  &= \langle \tang \K\Phi_L(k)\circ \tang 1\pi u,\tang \K\Phi_L(k)\circ
    \tang 1\pi v\rangle_\K \\ 
  &= \langle \tang 1\pi u,\tang 1\pi v\rangle_{k\K}
  = \langle u,v\rangle_\p, 
\end{align*}
where in the second equality we have used the fact that if \(k \in \K\), then
\(\pi \bigl(I_k(g)\bigr) = k g k^{-1}\K = k g \K = \Phi_L(k)(\pi(g))\), and
thus, taking the differential at the identity implies \(\tang 1 \pi \circ
\Ad(k) =\tang 1(\pi\circ I_k) = \tang \K \Phi_L(k)\circ \tang 1 \pi\).

Let us fix an \(\Ad_ \G\K\)-invariant non-degenerate quadratic form on \(\k\)
and set \(\p\) and \(\k\) to be orthogonal.  Then an \(\Ad_ \G\K\)-invariant
inner product on \(\p\) defines an \(\Ad_ \G\K\)-invariant non-degenerate
quadratic form on \(\g\), which gives rise to a \(\G\)-left invariant and
\(\K\)-right-invariant pseudo-Riemannian metric \(\langle \cdot , \cdot
\rangle_{\G}\) on \(\G\), as explained in \refapp{lie-groups} When \(\G/\K\) is
a symmetric space, the Killing form is negative definite on \(\k\), so we may
negate it to define an inner product on \(\k\).

When the inner product on \(\p\) is the one induced by the \(\G\)-invariant
metric \(\langle \cdot , \cdot \rangle_{\G/\K}\), the resulting
pseudo-Riemannian metric on \(\G\) will be isometric (wrt \(\pi\)) to \(\langle
\cdot , \cdot \rangle_{\G/\K}\) when restricted to \(\p\):
 
By definition \(\langle u_g, v_g \rangle_g \defn \langle \tang {}L_{g^{-1}}u_p,
\tang {}L_{g^{-1}} v_p \rangle_{\g}\).  Then if \(u_g = \tang {}L_g u, v_g =
\tang {}L_g v\), for some \(u,v \in \p\), we have
 \begin{align*}
  \langle \tang g \pi u_g , \tang g \pi v_g \rangle_{g\K}
  &= \langle \tang {g\K}\Phi_L(g^{-1})
    \tang g\pi u_g,\tang {g\K}\Phi_L(g^{-1})\tang g\pi v_g\rangle_{\K} \\
  &= \langle \tang {g}\bigl(\Phi_L(g^{-1})\circ\pi\bigr)u_g,\tang {g}
    \bigl(\Phi_L(g^{-1})\circ \pi \bigr) v_g \rangle_{\K} \\
  &= \langle \tang {g}\bigl(\pi\circ\Phi(g^{-1})\bigr)u_g,\tang
         {g}\bigl(\pi\circ\Phi(g^{-1})\bigr)v_g\rangle_{\K} \\
  &= \langle \tang {1}\pi\circ \tang g\Phi(g^{-1})
    u_g,\tang 1\pi\circ \tang g\Phi(g^{-1})v_g\rangle_{\K} \\
  &= \langle \tang g\Phi(g^{-1})u_g,\tang g\Phi(g^{-1})v_g\rangle_{\p} \\
  &= \langle \tang g L_{g^{-1}}u_g,\tang gL_{g^{-1}}v_g\rangle_{\p} \\
  &= \langle u_g,v_g \rangle_{g}.
\end{align*}

\subsection[Proof of a bundle isomorphism]%
           {Proof \(J^{-1}(0)/\K \diffeo T^*(\G/\K)\)}%
           \label{proof-bundle-isomorphism}

Recall \(\K\) acts on \(\G\) by right action \(R_k:\G\to\G\), which defines a
projection \(\pi:\G\to\G/\K\).  The tangent map is \(\tang {}\pi:T\G\to
T(\G/\K)\), where \(\tang {}\pi(g,v_g) = \bigl(g\K,\tang g\pi(v_g)\bigr)\in
T_{g\K}(\G/\K)\).

The action lifts to an action \(\tang{} R_k:T\G\to T\G\), which defines cosets
\begin{align*}
 (g,v_g) \K = \{\tang {}R_k(g,v_g):k\in\K\}
  &= \left\{\bigl(R_kg,\tang gR_k(v_g)\bigr):k\in\K\right\} \\
  &= \bigl(g\K,\tang gR_{\K}(v_g)\bigr),
\end{align*}
for any \((g,v_g)\in T_g\G\).  The coset space is then
\begin{equation}
  (T\G)/\K \defn \{v\K:v\in T\G\}
  = \left\{\bigl(g\K,\tang gR_{\K}(v_g)\bigr):v\in T\G\right\}, 
\end{equation}
and we denote the projection \(v\mapsto v\K\) by \(\tau:T\G\to(T\G)/\K\).  Now
let \(v=(g,v_g)\in J^{-1}(0)\).  Note \(\tang gR_k(v_g) \in T_{gk}\G\), and
\(\tang {gk}\pi\Bigl(\tang gR_k(v_g)\Bigr) = \tang g\bigl(\pi\circ
R_k\bigr)(v_g)\fairbreak=\tang g\pi(v_g)\) which is independent of \(k\in\K\)
(in particular, observe that when \(k=1\), then \(\tang gR_k(v_g) = v_g\in \tang
1L_g(\p)\)).  Moreover recall that \(\tang g\pi\) is a linear isomorphism
between \(\tang 1L_g(\p)\subseteq T_g\G\) and \(T_{g\K}(\G/\K)\), and \(v_g\neq
u_g\) implies \(\tang gR_{\K}(v_g)\neq \tang g R_{\K}(u_g)\) (note \(R_k\) is a
diffeomorphism, so that \(T_{gk}\G\) will be distinct tangent spaces as \(k\)
varies).  Hence we may identify \(\tang gR_{\K}(v_g)\sim v_g\sim \tang
g\pi(v_g)\), which shows \(J^{-1}(0)/\K\diffeo T^*(\G/\K)\).

This identification is the reason we can project the mechanics onto the
homogeneous spaces using \(\tang {}\pi\) rather than \(\tau\) (and why we have
\(H_0\circ \tang {}\pi=H\) rather than \(H_0\circ\tau=H\) on \(J^{-1}(0)\)).

\subsection{Proof Algorithm Samples from Target} \label{proof-sample-target}

The fact that our algorithm samples from the correct distribution follows from
\(H_0\circ \tang {}\pi=H|_{J^{-1}(0)}\), \(\pull{\tang{}\pi}(\omega_0) =
\omega|_{J^{-1}(0)}\) and the fact the mechanics is constrained to the
submanifold \(J^{-1}(0)\).  Consider local (cotangent-lifted) coordinates
\((q,p)\) in a neighbourhood of a point \((q_0,p_0) \in\cotbundle\M\).  Let
\(G(q)\) be the (local) matrix associated to the \(\G\)-invariant metric
\(\langle\cdot,\cdot\rangle_q\).  From the symplectic reduction and the
identification \(J^{-1}(0)/\K\diffeo\cotbundle{(\G/\K)}\) in
\refapp{proof-bundle-isomorphism}, the projected motion is a Hamiltonian
dynamics induced by the reduced Hamiltonian \(H_0(q,p)=\tilde\U +
p^TG^{-1}(q)p\), where \(\tilde\U\defn-\log(\lambda_\H)\) and \(\lambda_\H\) is
the density of the target measure wrt to the Riemannian volume form (often
called the Hausdorff measure).  In local coordinates the target measure is
\(\lambda_\H\,d\Vol=\lambda_\H\sqrt{\det\bigl(G(q)\bigr)}\,dq\defn
\lambda_\L\,dq\), where \(dq\) is the local Lebesgue measure.  In other words,
the target measure has density \(\lambda_\L\) with respect to the local
Lebesgue measure.  From Liouville's theorem, the reduced mechanics preserves
the differential form \(e^{-H_0}\bigwedge_{j=1}^{\dim\M}\omega_0=
e^{-H_0(q,p)}dq\wedge dp\), and thus the canonical distribution
\begin{equation*}
  \int_Ae^{-H_0}\bigwedge_{j=1}^{\dim\M} \omega_0
  = \int_AF^*_{\hvf{H_0}}\Bigl(e^{-H_0}\bigwedge_{j=1}^{\dim\M}\omega_0\Bigr)
  = \int_{F_{\hvf{H_0}}(A)}e^{-H_0}\bigwedge_{j=1}^{\dim\M}\omega_0,
\end{equation*}

where
\(F_{\hvf{H_0}}\) is the flow of the Hamiltonian vector field associated with
\(H_0\).  Averaging pointwise over momenta (marginalising) 
 
\begin{align*}
\int_{T_q^{*}\M}
e^{-H_0(q,p)}\,dp\,dq &=\lambda_\H(q)\int_{T_q^*\M}e^{-p^TG^{-1}(q)p}\,dp\,dq  \\
&=
\lambda_\H(q)\sqrt{\det\bigl(G(q)\bigr)}\,dq = \lambda_\L(q)\,dq,
\end{align*}
 which is
indeed the target measure (since \(\lambda_\L(q)\,dq=\lambda_\H\,d\Vol\), and
the right hand side is a smooth differential form, this does not depend on a
choice of coordinates).  It is worth mentioning that some authors define their
target measure in a local coordinate chart to be \(e^{-\tilde V(q)}dq\), (that
is \(\lambda_\L\defn e^{-\tilde V}\)), which is not a differential form and
therefore, depends on the choice of coordinates.

Note that at any point \(q\in\M\), the momentum refreshment step clearly
samples from the correct conditional distribution \(p|q\).  Indeed, this step
samples vectors in \(\p\diffeo \tang{} L_{g^{-1}}(T_g\G)\) from \(\mathcal
N(0,T^{-1})\) (in velocity space).  Moreover \(\push L\mathcal N(0,T^{-1})=
\mathcal N(0,LT^{-1}L^T)\) for any linear transformation \(L\) (where \(\push
L\mathbb P\) denotes the push-forward measure), and for \(L=\tang {}\pi\),
\(e^{v^T(LT^{-1}L^T)^{-1}v}=e^{v^T(L^{-1})^T T L^{-1} v}= e^{(L^{-1}v)^T T
  L^{-1} v}=e^{(LL^{-1}v)^TGLL^{-1}v}=e^{v^T G v},\) where we have used
\((Lu)^T G (Lu) = u^T T u\) in the last step (since \(L= \tang {}\pi\)
preserves the inner product), thus \(\push L\mathcal N(0,T^{-1}) = \mathcal
N(0, G^{-1})\), and the projection/transport of the samples will be generated
by the correct conditional distribution.
 
Finally, note that pull-backing the target measure with \(\tang {}\pi\) we
find:\footnote{Recall the restriction commutes with the wedge product since the
  pull-back does.} 
\begin{align*}
 \pull{(\tang {}\pi)} \Bigl(e^{-H_0}\bigwedge_{j=1}^{\dim\M}
\omega_0\Bigr)&=\Bigl(e^{-H_0}\circ \tang {}\pi\Bigr)\pull{(\tang {}\pi)}
\Bigl(\bigwedge_{j=1}^{\dim\M}\omega_0\Bigr)= \\
 &=e^{-H|_{J^{-1}(0)}}
\Bigl(\bigwedge_{j=1}^{\dim\M}\pull{(\tang {}\pi)}\omega_0\Bigr)=
\left.e^{-H}\bigwedge_{j=1}^{\dim\M}\omega\right|_{J^{-1}(0)}.
\end{align*}

The term on the RHS is a measure that is preserved by the mechanics on
\(J^{-1}(0)\), since the flow of \(H\) preserves \(\omega\), leaves
\(J^{-1}(0)\) invariant, and therefore preserves \(\omega|_{J^{-1}(0)}\) (see
also Theorem 4.3.5 in \cite{Abraham:2008}).  

\subsection{Derivation of relation with Killing Fields}%
           \label{proof-of-relation-Killing}

The definition of Killing fields \(Y^{\G/\K}\) is \(Y^{\G/\K}_{g \K} \defn
Y^{\G/\K}(g \K)\defn\left.\frac d{dt}\right|_{t=0} e^{tY}g\K\)
(~\refapp{killing}).  Recall that the vector space isomorphism \(\p\diffeo
T_\K\G/\K\), is given by \(\tang 1\pi\), where \(\pi:\G\to\G/\K:g\mapsto g\K\).
Thus \(\tang 1 \pi(Y_1) = \frac d{dt}\big|_{t=0} \pi \bigl(e^{tY_1}\bigr) =
\frac d{dt}\big|_{t=0} e^{tY_1}\K = Y^{\G/\K}_\K\), thus Killing vectors are
precisely the vectors \(T_\K\G/\K\) identified with vectors in \(\p\) by the
canonical vector space isomorphism.  We can characterised the relation between
the various metrics on \(\G/\K\) and \(\p\) using Killing fields.  First note
we immediately have \(\langle X_1,Y_1 \rangle_\p = \langle X^{\G/\K}_\K,
Y^{\G/\K}_\K \rangle_\K\).  Furthemore
\begin{align*}
  \bigl((\push{\Phi_L(g)}X^{\G/\K})f\bigr)(h\K)
  &= X^{\G/\K}_{g^{-1}h\K}(f\circ\Phi_L(g)) 
  =\frac d{dt}\bigl(f\circ g e^{tX}g^{-1}h\K\bigr)\big|_0 \\
  &=\frac d{dt}\bigl(f\circ e^{t\Ad_ gX}h\K\bigr)\big|_0
  =\bigl(\Ad_ gX\bigr)^{\G/\K}_{h\K}f.
\end{align*}
Thus \(\tang {g\K}\Phi_L(g^{-1})X^{\G/\K}_{g\K}=\bigl(\Ad_{g^{-1}}X
\bigr)^{\G/\K}_\K\) from which it follows that
\begin{equation*}
  \langle X^{\G/\K}_{g \K}, Y^{\G/\K}_{g \K} \rangle_{g \K}
  = \Bigl\langle\bigl(\Ad_ {g^{-1}}X\bigr)^{\G/\K}_\K,
    \bigl(\Ad_ {g^{-1}}Y\bigr)^{\G/\K}_\K\Bigr\rangle_\K
  = \bigl\langle\Ad_ {g^{-1}}X,\Ad_ {g^{-1}}Y\bigr\rangle_\p.
\end{equation*}

By a derivation analogous to the proof that 
\(\tang
{g\K}\Phi_L(g^{-1}) X^{\G/\K}_{g\K}=\bigl(\Ad_ {g^{-1}}X\bigr)^{\G/\K}_\K\), it
can be shown that the isomorphism \(\p \to T_{g\K}\M\) is given by \(u \mapsto
(\Ad_ g u)^{\G/\K}_{g\K}\).

\subsection{Derivation of Equation of Motion} \label{proof-of-motion}

Using equation~\refeq{eq:matrix-group-1}, \(e_i\rho = \tang {}\rho(e_i) =
\rho\cdot T_i\),\footnote{Note \(\tang {}\rho(e_i): g \mapsto \tang g \rho
  (e_i|_g) \in T_{\rho(g)} \rho(\G) \hookrightarrow T_{\rho(g)}\GL{n,\R}
  \diffeo \R^{n^2}\), so that we can view \(\tang {}\rho(e_i)\) as a
  matrix-valued function on \(\G\), and it is in that sense that \(\tang
  {}\rho(e_i) = e_i\rho\) (this is the standard identification of the tangent
  map with the exterior derivative on functions).} where \(\cdot\) is the
matrix product, we can derive matrix-valued Hamiltonian vector fields
corresponding to \(\U \defn \U_{\rho}\circ\rho\) and \(T \defn T_{\rho}\circ
\tang 1\rho\):
\begin{align*}
  \hvf\U\defn\hvf{\U_{\rho}\circ\rho}
  = -e_i(\U_{\rho}\circ\rho)\,\partial_{p_i}
  &= -\tang {}(\U_{\rho}\circ\rho)(e_i)\partial_{p_i}
  = -\tang {}\U_{\rho}(\tang {}\rho(e_i))\partial_{p_i} \\
  &= -\frac{\partial\U_{\rho}}{\partial x_{ab}}(x(\rho))dx_{ab}
    (\rho\cdot T_i)\partial_{p_i}\\
  &= -\tr\bigl(\partial_x\U_{\rho}\cdot\rho\cdot T_i\bigr)\partial_{p_i},
\end{align*}
where \(x_{ab}:\GL{n,\R}\to\R\) are the canonical matrix coordinates, and
\(\bigl(\partial_x\U_{\rho} \bigr)_{ab}\defn\frac{\partial\U_{\rho}}{\partial
  x_{ba}}\).  Observe that when \(\U\) is right \(\K\)-invariant, as will be
the case when we consider homogeneous spaces, then \(e_i(\U)= \L_{e_i}\U=0\)
for any \(e_i \in \k\), as shown in \refapp{proof-K-invariance}.  As a result
\(\hvf \U\) will live in \(\p\).
 
Since the metric is \(\Ad \G\)-invariant, its geodesics are given by the curves
\(t \mapsto g\exp(t p^ie_i)\), and thus \(\hvf T =p^ie_i\) (where the dual
coordinates are defined using the pullback metric \((\tang1\rho)^{*}\langle
\cdot,\cdot\rangle_{\rho}\), see \refapp{riemannian}).
Note that
\begin{equation*}
  T(p) = T (p^ie_i)
  = \half\left\langle \tang  1\rho(p^ie_i),\tang1\rho(p^je_j)\right\rangle_{\rho}
  = \half\langle P,P\rangle_{\rho},
\end{equation*}
where \(P\defn p^iT_i\) are the ``matrix-valued coordinates'' on \(\tang
1\rho(\g)\) associated to the vector \(p^ie_i(1) \in \g\), which is itself
associated to the momentum \(p_i \theta^i(1) \in \g^*\).

%% The basis vectors \(\partial_{p_i}\) may be expressed in terms of the
%% matrix of coordinate basis vectors as
%% \begin{equation*}
%%   \partial_{p_i} = \sum_{a,b=1}^n
%%     \frac{\partial P_{ab}}{\partial p_i}\,\frac\partial{\partial P_{ab}}
%%   = \tr T_i\cdot\partial_P
%%   \qquad\mbox{where}\qquad
%%   (\partial_P)_{ab} \defn \frac\partial{\partial P_{ba}}\,;
%% \end{equation*}
%% observe that these relations are only defined on the image \(\rho(\g)\), and
%% not the whole of~\(\R^{n^2}\).

The Hamiltonian trajectories on \(\cotbundle \G\) are the integral curves of
the vector field \(\hvf H=\hvf T+\hvf\U\), and we shall construct symplectic
integrators built of interleaved segments of the integral curves of \(\hvf T\)
and~\(\hvf\U\).  The rate of change of functions \(P\) and \(\rho\) along a
curve \(c:\R\to\cotbundle\G\) is
\begin{equation*}
  \dot P(t)=\hvf\U P
  = -\tr\bigl(\partial_x \U_{\rho}\cdot\rho(t)\cdot T_i\bigr)T^i
  \qquad\mbox{and}\qquad
  \dot\rho(t) = \hvf T \rho=\rho(t)\cdot P(t)
\end{equation*}
respectively, where we have used \(\partial_{p_i}P = \partial_{p_i}p^kT_k
=g^{ki}T_k \defn T^i\).

\subsection{Derivation of representation-valued Maurer-Cartan form}%
           \label{mc-form}

Using the notation of \refapp{lie-groups} we may consider \(\rho\) to be a
matrix of \(0\)-forms, \(\rho_{ab}:\G\to\R\), that satisfy \(\rho_{ab}(gh) =
\rho_{ac}(g)\rho_{cb}(h)\) (using the summation convention for repeated indices
as usual).  The left action \(L_g\) of \(g\in\G\) on such a \(0\)-form is
\(\pull{L_g}\rho_{ab} = \rho_{ab}\circ L_g\), hence \(\pull{L_g}\rho_{ab}(h) =
\rho_{ab}(L_gh) = \rho_{ab}(gh) = \rho_{ac}(g)\rho_{cb}(h)\) for any
\(h\in\G\); we thus have \(\pull{L_g}\rho_{ab}=\bigl(\rho(g)\rho\bigr)_{ab}\),
or in matrix notation \(\pull{L_g} \rho = \rho(g)\rho\).

If we apply a left invariant basis vector \(e_i\in\X\G\) to this matrix-valued
\(0\)-form (in the sense that \((e_i \rho)_{ab} \defn e_i \rho_{ab}\)) and then
apply \(\pull{L_g}\) to the resulting matrix-valued \(0\)-form we get
\begin{align*}
  \pull{L_g}e_i\rho &= \pull{L_g}(\push{L_g}e_i)\rho
    && \mbox{since \(e_i = \push{L_g}e_i\) is left invariant} \\
  &= \pull{L_g}(\pull{L_g^{-1}}e_i\pull{L_g})\rho
    && \mbox{by the definition of \(\push{L_g}\)} \\
  &= (\pull{L_g}\pull{L_g^{-1}})e_i(\pull{L_g}\rho) \\
  &= e_i\rho(g)\rho
    && \mbox{because \(\rho\) is a representation} \\
  &= \rho(g) e_i\rho.
\end{align*}
Evaluating this at the identity \(\ident\in\G\) gives
\begin{equation*}
  (e_i\rho)(g) = (\pull{L_g} e_i\rho)(\ident)
  = \rho(g)(e_i\rho)(\ident) = \rho(g)\T_i,
\end{equation*}
where \(\T_i\defn(e_i\rho) (\ident)\) is a \definition{generator}
matrix.  As this holds for any \(g\in\G\) we express this result as
\begin{equation}
  e_i\rho = \rho\T_i.
  \label{eq:matrix-group-1}
\end{equation}
From this equation we see that the Lie bracket defined as the commutator of
left invariant vector fields is the same as that given by the commutator of
matrix generators,
\begin{align*}
  [e_i,e_j]\rho &= c^k_{ij}e_k\rho = \rho c^k_{ij}\T_k \\
  &= (e_ie_j - e_je_i)\rho
  = e_i(e_j\rho) - e_j(e_i\rho)
  = e_i\rho\T_j - e_j\rho\T_i
  = \rho(\T_i\T_j - \T_j\T_i) \\
  &= \rho[\T_i,\T_j] 
\end{align*}
whence \([\T_i,\T_j] = c^k_{ij}\T_k\) as a matrix equation.  Using the
representation we identify \(\G \times \g \cong T\rho(\G)\)\footnote{ if
\(\gamma\) is a curve on \(\G\) with \(\rho \big( \gamma(0)\big)=A\), then its
tangent vector at \(A\) is \(\big(\tang 0(\rho \gamma)\big)(\frac{d}{dt}) =
\tang {\gamma(0)} \rho\circ \tang 0\gamma(\frac{d}{dt})\), so
\(T_A(\rho\G)=\tang{}\rho(T_{\gamma(0)}\G)\)}, with \(\tang{}\rho (g,e_i) =
(\rho(g), T_i)\).

Another consequence follows from the observation that \(e_i\rho = d\rho(e_i)\)
where \(d\rho\) is a matrix-valued \(1\)-form, or equivalently a matrix of real
\(1\)-forms.  Equation~\refeq{eq:matrix-group-1} tells us that \(T_i =
\rho^{-1} e_i\rho = \rho^{-1}d\rho(e_i)\) is a constant,\footnote{A left
invariant \(0\)-form \(f\in\form0\G\) satisfies \(f = \pull{L_g}f = f\circ
L_g\) so \(f(h) = f(L_gh) = f(gh)\) for all \(h\in\G\): hence all left
invariant \(0\)-forms are constant functions and \emph{vice versa}, even if
they are matrix-valued.}  so \(\rho^{-1}d\rho\) must be left invariant, and
thus has the expansion
\begin{equation}
  \rho^{-1}d\rho = \theta^j\T_j = \theta
  \label{eq:matrix-maurer-cartan}
\end{equation}
in the basis of left invariant \(1\)-forms, where \(\theta\defn\theta^j\T_j\)
is the left invariant \definition{matrix-valued Maurer--Cartan form}.
 
In terms of the Maurer--Cartan form we may write the Maurer--Cartan structure
relations of equation~\refeq{eq:Maurer-Cartan-relations} in the form
\begin{equation*}
  d\theta + \half[\theta\wedge\theta] = 0,
\end{equation*}
where the quantity \([\theta\wedge\theta] = [T_i,T_j] \theta^i\wedge\theta^j =
c^k_{ij} T_k\theta^i\wedge\theta^j\) is a matrix-valued left invariant
\(2\)-form.

\section{Parameterization of Quotient Spaces} \label{sec:par-quotient}

\subsection[Parameterization of hyperbola]%
           {\(\liegroup{SO}(1,1,\R)\) and parameterization of the hyperbola
             \(\hyperboloid12\)}%
           \label{sec:par-hyper}

We shall briefly consider the real forms of the abelian Lie algebra
\(\liealgebra{SO}(2,\C)\).  Its defining representation acts on \(\C^2\), and
its maximal compact real form \(\liealgebra{SO}(2,\R)\) is obtained by
restricting the ground field to~\(\R\).  The group \(\liegroup{SO}(2,\R)\) is
parameterized by exponentiation of its sole generator~\(L\)
\begin{equation*}
  C(\theta) = e^{\theta L} = \exp\left[\theta\left(\begin{array}{cc}
      0 & 1 \\ -1 & 0
    \end{array}\right)\right]
  = \left(\begin{array}{cc}
    \cos\theta & \sin\theta \\ -\sin\theta & \cos\theta
  \end{array}\right),
\end{equation*}
and the orbit of a point in \(\R^2\) is the unit circle
\begin{equation*}
  C(\theta)\left(\begin{array}{c} x \\ y \end{array}\right)
  = \left(\begin{array}{c}
    x\cos\theta + y\sin\theta \\ -x\sin\theta + y\cos\theta
  \end{array}\right)
\end{equation*}
of points with Euclidean distance \(x^2 + y^2\) from the origin.

We may also construct a non-compact real form by use of the Weyl unitary trick.
In this case this corresponds to multiplying the generator by \(i\), which
preserves the Lie algebra's trivial structure.  The generator \(L'\defn iL\)
maps \((x,iy)\mapsto(-y,-ix)\) for \(x,y\in\R\),
\begin{equation*}
  L'\left(\begin{array}{c} x \\ iy \end{array}\right)
  = \left(\begin{array}{cc} 0 & i \\ -i & 0 \end{array}\right)
    \left(\begin{array}{c} x \\ iy \end{array}\right)
  = \left(\begin{array}{c} -y \\ -ix \end{array}\right),
\end{equation*}
so a simple change of basis by the matrix \(M\defn\diag(1,i)\) gives its
action explicitly on~\(\R^2\) in terms of the real symmetric generator~\(W\defn
M^\dagger L'M\in\liealgebra{SO}(1,1,\R)\)
\begin{equation*}
  W = \left(\begin{array}{cc} 1 & 0 \\ 0 & i \end{array}\right)^\dagger
    \left(\begin{array}{cc} 0 & i \\ -i & 0 \end{array}\right)
    \left(\begin{array}{cc} 1 & 0 \\ 0 & i \end{array}\right)
  = \left(\begin{array}{cc} 0 & -1 \\ -1 & 0 \end{array}\right).
\end{equation*}
The Lie group \(\liegroup{SO}(1,1,\R)\) is parameterized by exponentiation,
\begin{equation*}
  H(\theta) = e^{\theta W} = \exp\left[\theta\left(\begin{array}{cc}
      0 & -1 \\ -1 & 0
    \end{array}\right)\right]
  \left(\begin{array}{cc}
    \cosh\theta & -\sinh\theta \\ -\sinh\theta & \cosh\theta
  \end{array}\right),
\end{equation*}
which preserves the pseudo-Riemannian hyperbolic metric \(M^2=\diag(1,-1)\),
\begin{equation*}
  \left(\begin{array}{cc}
    \cosh\theta & -\sinh\theta \\ -\sinh\theta & \cosh\theta
  \end{array}\right)^T
  \left(\begin{array}{cc} 1 & 0 \\ 0 & -1 \end{array}\right)
  \left(\begin{array}{cc}
    \cosh\theta & -\sinh\theta \\ -\sinh\theta & \cosh\theta
  \end{array}\right)
  = \left(\begin{array}{cc} 1 & 0 \\ 0 & -1 \end{array}\right).
\end{equation*}
The orbits of a point in \(\R^2\) is the hyperbola
\begin{equation*}
  H(\theta)\left(\begin{array}{c} x \\ y \end{array}\right)
  = \left(\begin{array}{c}
    x\cosh\theta - y\sinh\theta \\ -x\sinh\theta + y\cosh\theta
  \end{array}\right)
\end{equation*}
of points with hyperbolic distance \(x^2 - y^2\) from the origin.  Unlike the
compact case the space of points at unit distance from the origin is not
connected but has two components, one reachable by exponentiation from the
point \((1,0)\), the other from the point~\((-1,0)\).

The characterisation of the real Lie groups \(\liegroup{SO}(p,q,\R)\) in terms
of the signature (the number of positive and negative terms in the diagonal
pseudo-Riemannian metric tensor) reflects \definition{Sylvester's law of
inertia}, which states that this signature is preserved by any non-singular
real similarity transformation.

\subsection[Two Sheeted Hyperbolic Space]%
           {Two Sheeted Hyperbolic Space, \(\hyperboloid22 \defn
             \liegroup{SO}(2,1,\R)/\liegroup{SO}(2,\R)\)}%
           \label{sec:h212}

The real form \(\liealgebra{SO}(3,\R)\) obtained by restricting the parameters
of \(\liealgebra{SO}(3,\C)\) to be real is not the only real form of the Lie
algebra.  Other (non-compact) real forms may be found by finding the
\definition{involutive isomorphisms} \(T:\liealgebra{SO}(3,\C)\to
\liealgebra{SO}(3,\C)\) and using Weyl's unitary trick.  \(T\) must be a linear
transformation that satisfies \(T[u,v] = [Tu,Tv]\) for it to be an isomorphism,
and \(T^2=\ident\) for it to be involutive.  As \(T^2\) only possesses the
single eigenvalue~\(1\) \(T\)~can only have eigenvalues \(\pm1\), and thus it
splits \(\liealgebra{SO}(3) = \k\oplus\p\) where \(T\k=\k\) and \(T\p=-\p\);
moreover \([\k,\k]\subseteq\k\), \([\k,\p]\subseteq \p\), and
\([\p,\p]\subseteq\k\) since \(T\) is an isomorphism.

One such involutive isomorphism is provided by the linear mapping
\(T_x:\C^3\to\C^3\) with \(T_x:x\mapsto-x\) on the space \(\C^3\) carrying the
defining representation of \(\liealgebra{SO}(3,\C)\).  This is represented by
the matrix \(M_x = \diag(-1,1,1)\); \(\tr M_x\neq0\) so \(M_x\notin
\liealgebra{SO}(3,\C)\), and \(\det M_x\neq1\) so \(M_x\notin
\liegroup{SO}(3,\C)\), but it provides an \definition{outer automorphism} on
both by \(T_x:g\mapsto T_xg=M_xgM_x^{-1}\) since \([M_xuM_x^{-1},M_xvM_x^{-1}]
= M_x[u,v]M_x^{-1}\) for \(u,v\in\liealgebra{SO}(3,\C)\) and \((M_xgM_x^{-1})
\penalty-500 (M_xhM_x^{-1}) = M_x(gh)M_x^{-1}\) for \(g,h\in\liegroup{SO}(3,
\C)\).  Under this action on the Lie algebra we have \(T_xL_1=L_1\) and
\(T_xL_{2,3} = -L_{2,3}\), so \(\k=\vspan(L_1)\) and \(\p=\vspan(L_2,L_3)\).

The Weyl unitary trick is to restrict the ground field to \(\R\) and to
multiply all elements of \(\p\) by~\(i\).  Under this transformation the Lie
algebra still closes under the Lie bracket operation, as \([\k,\k]
\subseteq\k\), \([\k,i\p]\subseteq i\p\), and \([i\p,i\p] \subseteq\k\).  This
gives the real Lie algebra \(\liealgebra{SO}(2,1,\R)\).  Note that it is the
ground field of the Lie algebra that is real, not the components of the
generators.  Furthermore, all the matrices \(u\in\liealgebra{SO}(3,\C)\) are
real and antisymmetric, \(u^T = -u\), so they become antihermitian, \(u^\dagger
= -u\), when the ground field is restriced to \(\R\) in \(\liealgebra{SO}(3,
\R)\).  The image of the exponential map \(\exp:\liealgebra{SO}(3,\R)\to
\liegroup{SO}(3,\R)\) is thus compact.  The matrices \(v\in\liealgebra{SO}(2,1,
\R)\) are antisymmetric, but while those in \(\k\) remain real and hence
antihermitian those in \(i\p\) become imaginary and hence hermitian.  The image
of \(\exp:\liealgebra{SO}(2,1,\R)\to \liegroup{SO}(2,1,\R)\) is thus
non-compact.

The Weyl trick gives a real form of a Lie algebra for each involutive
automorphism, but what happens to the action of the defining representation
under this transformation?  The generators in the defining representation
become
\begin{equation*}
  T_1' \defn T_1
    = \left(\begin{array}{ccc} 0&0&0 \\ 0&0&1 \\ 0&-1&0 \end{array}\right),
  \quad
  T_2' \defn iT_2
    = \left(\begin{array}{ccc} 0&0&-i \\ 0&0&0 \\ i&0&0 \end{array}\right),
  \end{equation*}
  and
  \begin{equation*}
  T_3' \defn iT_3
    = \left(\begin{array}{ccc} 0&i&0 \\ -i&0&0 \\ 0&0&0 \end{array}\right),
\end{equation*}
and these clearly map map vectors \((ix,y,z) \mapsto(ix',y',z')\) with
\(x,y,z,x',y',z'\in\R\).  This allows the linear space carrying the defining
representation to be restricted to \(\R^3\) by conjugating the each matrix in
the fundamental representation by \(M_x' = \diag(i,1,1)\), so \(M_x'^2 = M_x\)
and \(M_x'^\dagger M_x' = \ident\).  The generators \(W_j \defn M_x'^\dagger
L_j'M_x'\) thus become
\begin{equation*}
  W_1 = \left(\begin{array}{ccc} 0&0&0 \\ 0&0&1 \\ 0&-1&0 \end{array}\right),
  \quad
  W_2 = \left(\begin{array}{ccc} 0&0&-1 \\ 0&0&0 \\ -1&0&0 \end{array}\right),
  \quad
  W_3 = \left(\begin{array}{ccc} 0&1&0 \\ 1&0&0 \\ 0&0&0 \end{array}\right).
\end{equation*}
The matrices in this representation of \(\liealgebra{SO}(2,1,\R)\) are thus
real, but those representing \(\k\) are antisymmetric whereas those
representing \(\p\) are symmetric.

The matrices in the definining representation of \(\liegroup{SO}(3,\C)\) and
\(\liegroup{SO}(3,\R)\) are all orthogonal, \(\Omega^T\Omega = \ident\), but
this is not the case for \(\Omega\in\liegroup{SO}(2,1,\R)\).  In terms of the
Lie algebra element \(\omega\) for which \(\Omega = \exp\omega\) the
orthogonality condition is \(\omega^T = -\omega\), and under the Weyl unitary
transformation \(\omega\mapsto\omega' = M_x\omega M_x^{-1} = M_x\omega M_x^T\)
this becomes \(\omega'^TM_x=M_x\omega^TM_x^T=-M_x\omega M_x^T=-M_x\omega'\).
Under exponentiation this becomes \(\Omega'^TM_x\Omega'=M_x\), so the Lie group
\(\liegroup{SO}(2,1)\) leaves the pseudo-Riemannian metric \(M_x\) invariant.
This metric may also be expressed as \((x,y,z)^T M_x (x,y,z) = -x^2 + y^2 +
z^2\).

We may parameterize the two sheeted hyperbolic space \(\hyperboloid22 =
\liegroup{SO}(2,1,\R)/\fairbreak\liegroup{SO}(2,\R)\) just as we did
for~\(\sphere2\) in~\refsec{sec:sphere} by exponentiating \(\exp\p\) in terms of
two real parameters \(\phi\) and~\(\xi\),
\begin{align*}
  H&(\phi,\xi)
  = \exp\left[\xi\left(\sin\phi\,W_2 - \cos\phi\,W_3\right)\right] \\
  &= \exp\left[\xi\left(\begin{array}{ccc}
      0 & -\cos\phi & -\sin\phi \\
      -\cos\phi & 0 & 0 \\
      -\sin\phi & 0 & 0
    \end{array}\right)\right] \\
  &= \left(
  \begin{array}{ccc}
    \cosh\xi & -\sinh\xi\cos\phi & -\sinh\xi\sin\phi \\
    -\sinh\xi\cos\phi & \cosh\xi(\cos\phi)^2 + (\sin\phi)^2
      & (\cosh\xi-1)\sin\phi\cos\phi \\
    -\sinh\xi\sin\phi & (\cosh\xi-1)\sin\phi\cos\phi
      & \cosh\xi(\sin\phi)^2 + (\cos\phi)^2
  \end{array}\right).
\end{align*}
As in~\refsec{sec:sphere} we only compute this expression in closed form to
illustrate the structure, for HMC computations the exponentiation is computed
numerically using the methods of~\refsec{sec:mat-exp}.  In this case, as
in~\refsec{sec:sphere}, the matrix representing the Lie algebra element is
normal; for the case of \(\sphere2\) this is because it was antisymmetric,
whereas here it is because it is symmetric.

We indentify the point
\begin{equation*}
  H(\phi,\xi)\left(\begin{array}{c} 1 \\ 0 \\ 0 \end{array}\right)
  = \left(\begin{array}{c}
    \cosh\xi \\ -\sinh\xi\cos\phi \\ -\sinh\xi\sin\phi
  \end{array}\right)  \in  \hyperboloid22.
\end{equation*}
The hyperbolic space \(\hyperboloid22\) is not connected, so this only
parameterizes one of the two sheets, the other may be reached from the point
\(-(1,0,0)^T\).  Together these two sheets are the locus of all points at
pseudo-Riemannian distance \(-1\) from the origin.

\begin{figure}
  \begin{center}
    \epsfxsize=0.3\textwidth \epspdffile{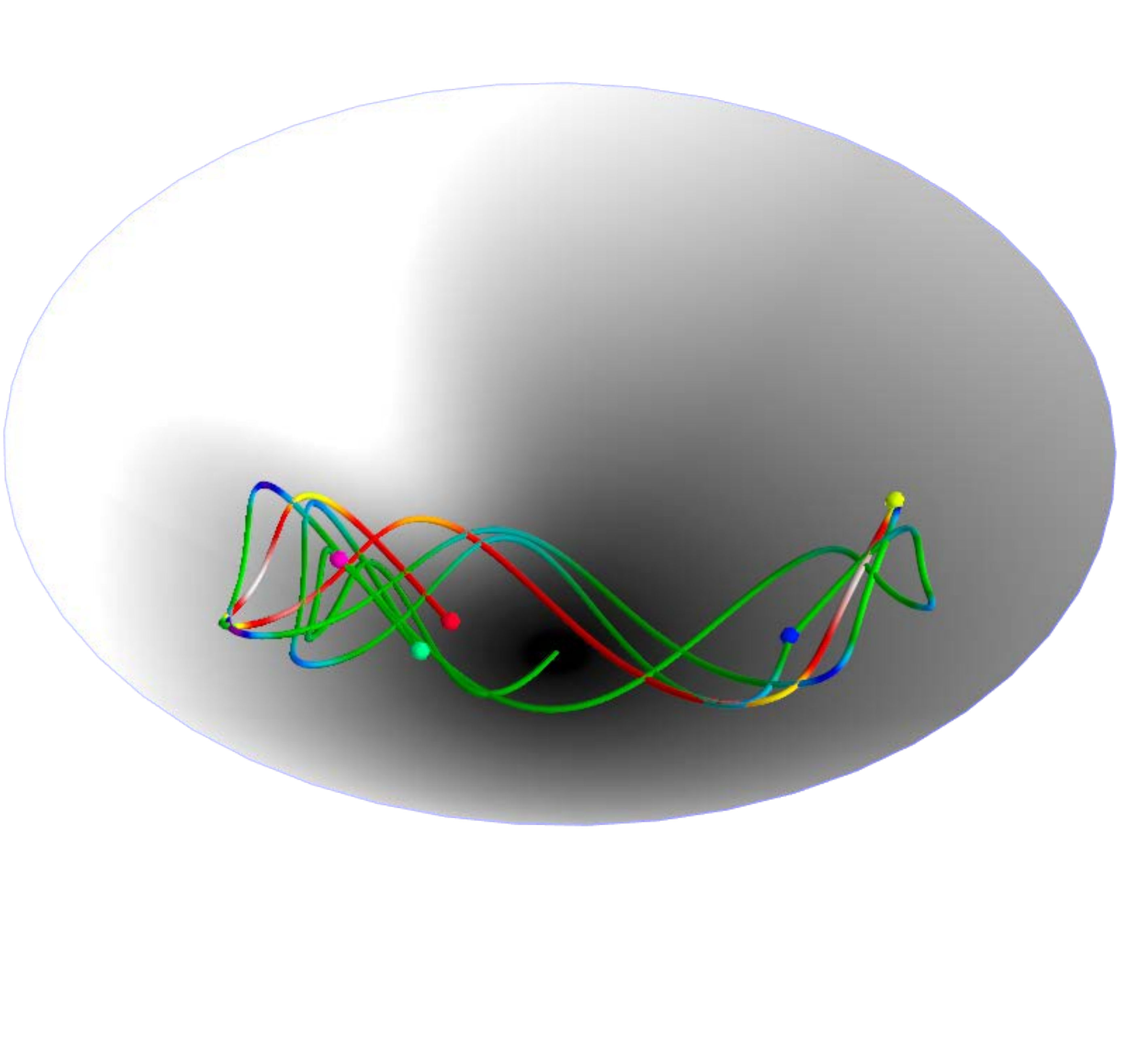}
  \end{center}
  \caption{\(5\) HMC trajectories, chronologically coloured blobs, the
    trajectories are coloured proportionally to the values of \(\delta H\),
    with potential \(\U(x,y,z)=(y^2+4)z^2\exp(x^3)\), \(\dt=0.1\),
    \(\trjlen=2\).}
\end{figure}

\subsection[Single Sheeted Hyperbolic Space]%
           {Single Sheeted Hyperbolic Space, \(\hyperboloid21 \defn
             \liegroup{SO}(2,1)/\liegroup{SO}(1,1)\)}%
           \label{sec:h2111}

There is no reason why we have quotient \(\liegroup{SO}(2,1,\R)\) by the
subgroup \(e^\k\) rather than by the subgroup generated by one of the
generators in~\(\p\).  To this end we consider the involutive isomorphism
induced by the linear map \(T_z:\C^3\to\C^3\) whith \(T_z:z\mapsto-z\).
Following the same line of arguments as in~\refsec{sec:h212} this provides an
outer automorphism of the Lie algebra by \(T_z:g\mapsto T_zg = M_zgM_z^{-1}\)
with \(M_z\defn\diag(1,1,-1)\), which splits \(\liealgebra{so}(3,\C) =
\k\oplus\p\) where now \(\k=\vspan(T_3)\) and \(\p=\vspan(T_1,T_2)\).  The Weyl
unitary trick then maps \(\p\mapsto i\p\), so we have \(T_1'\defn iT_1\),
\(T_2'\defn iT_2\), and \(T_3'\defn T_3\), and the change of basis defined by
\(M_z'\defn\diag(1,1,i)\) then allows us to define the action of
\(\liealgebra{so}(2,1,\R)\) on \(\R^3\) with the real generators \(W_j'\defn
M_z^TT_j'M_z\)
\begin{equation*}
  W_1' = \left(\begin{array}{ccc} 0&0&0 \\ 0&0&-1 \\ 0&-1&0 \end{array}\right),
  \;
  W_2' = \left(\begin{array}{ccc} 0&0&1 \\ 0&0&0 \\ 1&0&0 \end{array}\right),
  \;\mbox{and}\;
  W_3' = \left(\begin{array}{ccc} 0&1&0 \\ -1&0&0 \\ 0&0&0 \end{array}\right).
\end{equation*}
The group \(\liegroup{SO}(2,1,\R)\) preserves the pseudo-Riemannian metric
\((x,y,z)^TM_z\fairbreak(x,y,z) = x^2+y^2-z^2\): it is the same real Lie group
as considered in~\refsec{sec:h212}, we are just considering a different quotient
and hence a different symmetric space.

We may parameterize the single sheeted hyperbolic space \(\hyperboloid21 =
\liegroup{SO}(2,1,\R)/\fairbreak\liegroup{SO}(1,1,\R)\) by exponentiating a
generic element of \(\vspan(W_2',W_3')\) in terms of two real parameters
\(\phi\) and~\(\xi\)
\begin{equation*}
  H'(\phi,\xi) = \exp\left[\xi\left(\sin\phi W_2'-\cos\phi W_3'\right)\right]
  = \exp\left[\xi\left(
    \begin{array}{ccc}
      0 & -\cos\phi & \sin\phi \\ \cos\phi & 0 & 0 \\ \sin\phi & 0 & 0
    \end{array}\right)\right].
\end{equation*}
This matrix is not normal, so it is a little more difficult to evaluate the
exponential, but the methods described in~\refsec{sec:abnormal-exp} may be
used.

As before we will compute the matrix exponential in closed form for expository
purposes, but we remind the reader than in practice the exponential only needs
to be computed numerically for the HMC algorithm.  We first transform \(H'\) to
Schur form by finding its eigenvalues \(\pm\lambda = \pm\sqrt{-\cos2\phi}\) and
eigenvectors, and then applying Gram--Schmidt orthonormalization to find an
unitary matrix \(\Omega\) such that \(H'=\Omega T\Omega^\dagger\),
\begin{align*}
  \Omega& = \dfrac1{\sqrt2} \left(\begin{array}{ccc}
    0 & \sqrt{\dfrac2{1 + \lambda^2}}
      & -\lambda\sqrt{\dfrac2{1 + \lambda^2}} \\[2ex]
    \sqrt{1 + \lambda^2}
      & -\lambda\sqrt{\dfrac{1 - \lambda^2}{1 + \lambda^2}}
      & -\sqrt{\dfrac{1 - \lambda^2}{1 + \lambda^2}} \\[2ex]
    \sqrt{1 - \lambda^2} & \lambda & 1
  \end{array}\right)
  \quad\mbox{and}\\
  T &= \left(
  \begin{array}{ccc}
    0 & \sqrt{1 - \lambda^2} & -\lambda\sqrt{1 - \lambda^2} \\
    0 & \lambda & 1 - \lambda^2 \\
    0 & 0 & -\lambda
  \end{array}\right).
\end{align*}
We may compute \(e^T\) by means of equation~\refeq{eq:abnormal-exp}, and thus
we obtain
\begin{align*}
  H'(\theta,\xi) &= \Omega e^T\Omega^\dagger \\
  &= \left(\begin{array}{ccc}
    \cosh\lambda\xi
    & -\frac{\cos\phi\sinh\lambda\xi}\lambda
    & \frac{\sin\phi\sinh\lambda\xi}\lambda \\[2ex]
    \frac{\cos\phi\sinh\lambda\xi}\lambda
    & \frac{1+\lambda^2-(1-\lambda^2)\cosh\lambda\xi}{2\lambda^2}
    & \frac{\cos\phi\sin\phi(\cosh\lambda\xi-1)}{\lambda^2} \\[2ex]
    \frac{\sin\phi\sinh\lambda\xi}\lambda
    & -\frac{\cos\phi\sin\phi(\cosh\lambda\xi-1)}{\lambda^2}
    & -\frac{1-\lambda^2-(1+\lambda^2)\cosh\lambda\xi}{2\lambda^2}
  \end{array}\right).
\end{align*}

We identify the point
\begin{equation}
  H'(\phi,\xi)\left(\begin{array}{c} 1 \\ 0 \\ 0 \end{array}\right)
  = \left(\begin{array}{c}
    \cosh\lambda\xi \\[2ex]
    \dfrac{\cos\phi\sinh\lambda\xi}\lambda \\[2ex]
    \dfrac{\sin\phi\sinh\lambda\xi}\lambda
  \end{array}\right) \in\hyperboloid21.
  \label{eq:h2111}
\end{equation}
However, while this surface has only a single sheet this parameterization is a
little delicate: firstly, the eigenvalue \(\lambda\in\R\) only for \(\cos2\phi
\leq0\) or \(|\phi+(k+\half)\pi|\leq\quarter\pi\) for \(k\in\Z\).  This maps to
the part of the hyperboloid with \(x>0\).  For \(|\phi+k\pi|<\quarter\pi\) the
eiginvalue \(\lambda\) is imaginary and equation~\refeq{eq:h2111} becomes
\begin{equation*}
  H'(\phi,\xi)\left(\begin{array}{c} 1 \\ 0 \\ 0 \end{array}\right)
  = \left(\begin{array}{c}
    \cos\xi\rho \\[2ex]
    \dfrac{\cos\phi\sin\rho\xi}\rho \\[2ex]
    \dfrac{\sin\phi\sin\rho\xi}\rho
  \end{array}\right) \in\hyperboloid21
\end{equation*}
in manifestly real form where \(\rho = i\lambda\).  This maps to the part of
the hyperboloid with \(-1\leq x<1\).  \(\lambda = 0\) for \(|\phi+k\pi|=
\quarter\pi\), where we have
\begin{equation*}
  H'\left(\quarter(2k+1)\pi,\xi\right)
    \left(\begin{array}{c} 1 \\ 0 \\ 0 \end{array}\right)
    = \left(\begin{array}{c}
      1 \\ (-1)^{k(k+1)/2}\sqrt\half\,\xi \\ (-1)^{k(k-1)/2}\sqrt\half\,\xi
  \end{array}\right) \in\hyperboloid21.
\end{equation*}
We see that it is impossible to represent a point with \(x<-1\) on the single
sheeted hyperboloid as the exponential of an element of the Lie algebra applied
to the point \((1,0,0)\): for non-compact Lie groups not every group element is
the exponential of an element of the Lie algebra.  This is not a problem for
HMC, as every point can be reached by a bounded number (two
\cite{Wuestner:2003}) of steps.

\section{Exponentation of Matrices} \label{sec:sup-exp}

The identity
\begin{equation*}
  \det X = \sum_P\prod_{j=1}^n
    \frac{(-1)^{r_j(j+1)}}{r_j!j^{r_j}}\,\left(\tr X^j\right)^{r_j},
\end{equation*}
follows from the definition of the determinant as a sum over permutations.  The
conjugacy classes of the symmetric group \(\SymmetricGroup n\) consist of
disjoint cycles, and may be labelled by partitions of \(n\) where \(r_j\) is
the number of cycles of length~\(j\).  All the elements of a conjugacy class
give the same product of traces, and the number of element of a conjugacy class
is \(n!/\prod_{j=1}^n (r_j!j^{r_j})\) (there are \(r_j!\) ways of permuting the
\(r_j\) cycles of length \(j\), and \(j\) ways of rotating the elements of a
given cycle).  Each odd permutation has a minus sign, and a permutation is odd
if it has an odd number of cycles of even length: this only occurs if both
\(r_j\) and \(j+1\) are odd.

\subsection{Matrix Polynomials} \label{sec:rodrigues}

Suppose that \(p(x) = \prod_k (x - x_k)\) is a monic polynomial with \(n\)
non-degenerate roots satisfying \(p(x_k) = 0\), and that \(p(X) = 0\) for some
matrix~\(X\).  We define the orthogonal projector onto the subspace
corresponding to the root \(x_j\) to be
\begin{equation*}
  P_j = \prod_{k\neq j} \frac{X - x_k\ident}{x_j - x_k};
\end{equation*}
Clearly \((X - x_j\ident)P_j \propto p(X) = 0\), so \(P_\ell\) and \(P_j\) are
orthogonal \(P_\ell P_j = 0\) if \(\ell\neq j\); moreover
\begin{equation*}
  P_j^2 = \prod_{k\neq j} \frac{X - x_k\ident}{x_j - x_k}\,P_j
  = \prod_{k\neq j} \frac{X - (x_j - x_j + x_k)\ident}{x_j - x_k}\,P_j
  = \prod_{k\neq j} \frac{(x_j - x_k)\ident}{x_j - x_k}\,P_j = P_j,
\end{equation*}
so they are idempotent, and hence \(P_\ell P_j = \delta_{\ell j}P_j\).  A
similar argument shows that \(P_\ell XP_j = P_\ell(X - x_j\ident +
x_j\ident)P_j = x_j\delta_{\ell j}P_j\).

The unique polynomial \(q\) with \(\deg q<n\) taking values \(q(x_j)\) at the
\(n\) distinct points \(x_j\) is given by the Lagrange interpolation formula
\begin{equation*}
  q(x) = \sum_{j=1}^n q(x_j) \prod_{k\neq j} \frac{x - x_k}{x_j - x_k}.
\end{equation*}
If we choose \(q(x_j) = 1\) for \(j=1,\ldots,n\) then we must have \(q(x) = 1\)
as it is unique, and the matrix polynomial satisfies \(q(X) = \sum_{j=1}^n P_j
= \ident\).  This establishes that the set \(\{P_j\}\) of orthogonal projectors
is complete.

\subsection[Exponentiation of so(n)]%
           {\(\exp:\liealgebra{SO}(n)\to\liegroup{SO}(n)\)}%
           \label{sec:expson}

The non-zero eigenvalues of elements of the real Lie algebras
\(\liealgebra{SO}(n)\) (and \(\liealgebra{Sp}(n)\)) appear in pairs, so we
introduce a method for exponentiating them over~\(\R\).
           
All the elements of any irreducible representation of any semisimple Lie
algebra are traceless, \(\tr A = 0\), as a consequence of Schur's lemma.  The
matrix Lie algebra \(\liealgebra{SO}(n)\) consists of antisymmetric matrices
\(A^T = -A\), hence \((A^k)^T = (A^T)^k = (-1)^k A^k\), and thus all odd powers
are traceless
\begin{equation}
  \tr A^{2k+1} = 0.
  \label{eq:trodd}
\end{equation}
Since \(A\) is antisymmetric it is normal, \([A,A^\dagger] = -[A,A] = 0\), and
hence it may be decomposed as \(A = \Omega D\Omega^\dagger\) with
\(\Omega\Omega^\dagger = \ident\) and \(D =\diag(\lambda_1,\ldots,\lambda_n)\).
Since \(A\) is real it follows that \(D^\dagger = (\Omega^\dagger
A\Omega)^\dagger = \Omega^\dagger A^\dagger\Omega = \Omega^\dagger A^\dagger
\Omega = -\Omega^\dagger A\Omega = -D\), so the eigenvalues \(\lambda_j^{*} =
-\lambda_j\) must be purely imaginary.  As \(A\) itself is real this means that
the eigenvalues are either zero or appear in complex conjugate pairs: we shall
order the eigenvalues such that \(\lambda_{n-j} = \lambda_j^{*} = -\lambda_j\).

For \(A\in\liealgebra{SO}(n)\) the characteristic polynomial
\refeq{eq:cayley-hamilton}~is
\begin{equation}
  c(\lambda) = \left\{
  \begin{array}{ll}
    \displaystyle
    \prod_{k=1}^{n/2} (\lambda - \lambda_k)(\lambda + \lambda_k)
      = \prod_{k=1}^{n/2} (\lambda^2 - \lambda_k^2)
    & \mbox{for \(n\) even} \\
    \displaystyle
    \lambda \prod_{k=1}^{(n-1)/2} (\lambda - \lambda_k)(\lambda + \lambda_k)
      = \lambda \prod_{k=1}^{(n-1)/2} (\lambda^2 - \lambda_k^2)
    & \mbox{for \(n\) odd,}
  \end{array}
  \right.
  \label{eq:charpoly}
\end{equation}
so the polynomial
\begin{equation}
  p(\lambda^2) \defn \left\{
  \begin{array}{ll}
    c(\lambda) & \mbox{for \(n\) even} \\
    \lambda c(\lambda) & \mbox{for \(n\) odd}
  \end{array}
  \right.
  \label{eq:pdef}
\end{equation}
may be written as \(p(x) = \prod_{k=1}^{\lceil n/2\rceil} (x - \lambda_k^2)\)
with \(\lambda_k^2\leq0\).  This matrix polynomial\footnote{For our HMC
application the eigenvalues of \(A\) are non-degenerate except on submanifolds
of phase space of codimension greater than zero.} also satisfies~\(p(A^2)=0\).

We define the projector \(P_j\) onto the eigenspace of \(A^2\) belonging to
eigenvalue \(\lambda_j^2\), and we observe that \([A,P_j] = 0\) since \(P_j\)
is a polynomial in~\(A^2\).  If we define \(A_j \defn AP_j\) we have \(A =
A\sum_kP_k = \sum_jAP_j = \sum_jA_j\) and \([A_j,A_k]=[AP_j,AP_k]=A^2[P_j,P_k]
= 0\), so \(\exp A = \prod_{j=1}^{\lceil n/2\rceil} \exp A_j\)~\cite{gallier:2002}.
Moreover \(A_j^3 = \lambda_j^2 A_j = (i|\lambda_j|)^2 A_j\) for each \(j\)
since \((A^2 - \lambda_j^2\ident)P_j = 0\), so we may compute \(\exp A_j\)
using Rodrigues' formula
\begin{align}
  \exp&A_j \defn \sum_{s=0}^\infty \frac{A_j^s}{s!}
  = \ident + \sum_{t=0}^\infty \frac{A_j^{2t+1}}{(2t+1)!}
    + \sum_{t=1}^\infty \frac{A_j^{2t}}{(2t)!} \nonumber \\
  &= \ident + \sum_{t=0}^\infty \frac{(i|\lambda_j|)^{2t+1}}{(2t+1)!}
      \left(\frac{A_j}{i|\lambda_j|}\right)
    + \sum_{t=1}^\infty \frac{(i|\lambda_j|)^{2t}}{(2t)!}
      \left(\frac{A_j}{i|\lambda_j|}\right)^2 \nonumber \\
  &= \ident + \frac{\sin|\lambda_j|}{|\lambda_j|} A_j 
    + \frac{1 - \cos|\lambda_j|}{|\lambda_j|^2} A_j^2
  = \ident + \left(\frac{\sin|\lambda_j|}{|\lambda_j|} A 
    + \frac{1 - \cos|\lambda_j|}{|\lambda_j|^2} A^2\right) P_j.
  \label{eq:rodrigues}  
\end{align}

If the eigenvalues are degenerate we can construct the projectors using only
the distinct eigenvalues.  The above procedure still works since \(A^2\) is
normal\footnote{A matrix \(X\) is \definition{normal} if it commutes with its
adjoint, \([X^\dagger,X] = 0\).  A matrix can be diagonalized by a unitary
transformation iff it is normal.  If \(X = \Omega D\Omega^\dagger\) with
\(\Omega^\dagger\Omega=\ident\) and \(D\) diagonal then \([X^\dagger,X] =
\left[\left(\Omega D\Omega^\dagger\right)^\dagger, \Omega
  D\Omega^\dagger\right] = \left[\Omega D^\dagger\Omega^\dagger, \Omega
  D\Omega^\dagger\right] = \Omega[D^\dagger,D]\Omega^\dagger = 0\).  To
establish the converse we note that any matrix can be transformed to
\definition{Schur form} by a unitary transformation, \(X = \Omega
T\Omega^\dagger\), with \(\Omega^\dagger\Omega=\ident\) and \(T\) (upper)
triangular, \(T_{ij} = 0\) if \(i>j\); hence \([X^\dagger,X] =
\left[\left(\Omega T\Omega^\dagger\right)^\dagger, \Omega
  T\Omega^\dagger\right] = \left[\Omega T^\dagger\Omega^\dagger, \Omega
  T\Omega^\dagger\right] = \Omega[T^\dagger,T]\Omega^\dagger\).  If \(X\) is
normal this implies that \([T^\dagger,T] = 0\) so the diagonal elements are
equal, \((T^\dagger T)_{jj} = (TT^\dagger)_{jj} \implies (e_j,T^\dagger Te_j) =
(e_j,TT^\dagger,e_j) \implies (Te_j,Te_j) = (T^\dagger e_j,T^\dagger e_j)
\implies \|Te_j\|^2 = \|T^\dagger e_j\|^2\).  In other words the columns have
the same norms as the corresponding rows.  Assume that the only non-zero
elements of columns \(1,\ldots,j\) of \(T\) are their diagonal elements: this
is manifestly the case for \(j=1\) since \(T\) is upper triangular.  Since
\(T^\dagger_{jj} = T_{jj}\) row \(j\) must also vanish except for its diagonal
element, and hence the same must hold for column \(j+1\), thus establishing the
induction hypothesis and proving that \(T\) is diagonal.}  and thus has a basis
of eigenvectors.

This algorithm may exhibit numerical instability when some eigenvalues are
almost degenerate.  This will occur when the matrix is small \(\|A\|\ll1\),
where \(\|\cdot\|\) is some matrix norm such as \(\|A\|^2 = |\tr A^2|\); when
this occurs it is recommended to switch to the Taylor expansion method
of~\refsec{sec:taylor}.  It may also occur ``accidentally'' even if \(A\) is
not small, and this may lead to a violation of the reversibility of the
symmetric symplectic integrator: it is to be hoped that this is an infrequent
occurence if sufficient numerical precision is employed.

\subsubsection[Exponentiation of so(3,R)]%
              {\(\exp:\liealgebra{SO}(3,\R)\to\liegroup{SO}(3,\R)\)}%
              \label{sec:exp-so3}

For \(n=3\) equation~\refeq{eq:dettr} is \(\det X = \rational16\left(2\tr X^3 -
3\tr X^2\tr X + (\tr X)^3\right)\); using equations~\refeq{eq:trodd},
\refeq{eq:charpoly}, and~\refeq{eq:pdef} we obtain \(p(A^2) = -\bigl(A^2 -
\half(\tr A^2) \ident\bigr) A^2 = 0\).  The projectors onto the eigenspaces of
\(A^2\) belonging to \(\lambda_1^2 = 0\) and \(\lambda_2^2 = \half\tr A^2\) are
\begin{equation*}
  P_1 = \frac{A^2 - \half\tr A^2}{-\half\tr A^2} \qquad\mbox{and}\qquad P_2 =
  \frac{A^2}{\half\tr A^2},
\end{equation*}
and the projections of \(A^2\) onto these eigenspaces are \(A_1^2 = 0\) and
\(A_2^2 = A^2\); hence \(\exp A_1 = \ident\) and Rodrigues' formula
\refeq{eq:rodrigues} gives
\begin{equation}
  \exp A_2 = \exp A
  = \ident + \frac{\sin\xi}{\xi}\,A + \frac{1-\cos\xi}{\xi^2}\,A^2
  \quad\mbox{with}\quad \xi \defn \sqrt{-\half\tr A^2}.
  \label{eq:expso3}
\end{equation}

Although the intention is to evaluate the matrix exponential numerically, as an
illustration consider the element \(A = \xi(\cos\theta\,L_1 + \sin\theta
\cos\phi\,L_2 + \sin\theta\sin\phi\,L_3)\fairbreak\in\liealgebra{SO}(3)\) with
the matrix representation of the generators \((L_i)_{jk} = \varepsilon_{ijk}\)
in terms of the Levi-Civita tensor,
\begin{equation*}
  A = \xi \left(\begin{array}{ccc}
    0 & \sin\theta\sin\phi & -\sin\theta\cos\phi \\
    -\sin\theta\sin\phi & 0 &\cos\theta \\
    \sin\theta\cos\phi & -\cos\theta & 0
  \end{array}\right).
\end{equation*}
We have \(\xi = \sqrt{-\half\tr A^2}\), so we obtain
\begin{align*}
  \exp&A = \ident + \frac{\sin\xi}{\xi}\,A + \frac{1 - \cos\xi}{\xi^2}\,A^2 \\
  & = \left(\begin{array}{ccc}
    \scriptstyle -c_\theta^2c_\xi + c_\theta^2 + c_\xi
    & \scriptstyle s_\theta
      \left(-c_\phi c_\theta c_\xi + c_\phi c_\theta + s_\xi s_\phi\right) 
    & \scriptstyle -s_\theta
      \left(s_\phi c_\theta c_\xi + s_\xi c_\phi - s_\phi c_\theta\right) \\[2ex]
    \scriptstyle -s_\theta \left(c_\phi c_\theta c_\xi
      - c_\phi c_\theta + s_\xi s_\phi\right) 
    & \scriptstyle c_\theta^2c_\phi^2c_\xi - c_\theta^2c_\phi^2
      - c_\phi^2c_\xi + c_\phi^2 + c_\xi
    & \begin{array}{c}
        \scriptstyle s_\phi c_\theta^2c_\phi c_\xi - s_\phi c_\theta^2c_\phi \\
        \scriptstyle - s_\phi c_\phi c_\xi + s_\phi c_\phi + s_\xi c_\theta
      \end{array} \\[3ex]
    \scriptstyle s_\theta
      \left(-s_\phi c_\theta c_\xi + s_\xi c_\phi + s_\phi c_\theta\right) 
    & \begin{array}{c}
        \scriptstyle s_\phi c_\theta^2c_\phi c_\xi - s_\phi c_\theta^2c_\phi \\
        \scriptstyle - s_\phi c_\phi c_\xi + s_\phi c_\phi - s_\xi c_\theta
      \end{array}
    & \begin{array}{c}
        \scriptstyle - c_\theta^2c_\phi^2c_\xi + c_\theta^2c_\phi^2 + c_\phi^2c_\xi \\
        \scriptstyle + c_\theta^2c_\xi - c_\phi^2 - c_\theta^2 + 1
      \end{array}
  \end{array}\right)
\end{align*}
from equation~\refeq{eq:expso3}, where \(c_\theta = \cos\theta\), \(s_\phi =
\sin\phi\), and so forth.

\subsubsection[Exponentiation of so(5,R)]%
              {\(\exp:\liealgebra{SO}(5,\R)\to\liegroup{SO}(5,\R)\)}%
              \label{sec:expso5}

For a less trivial example we shall consider the case where \(n=5\), here
equation~\refeq{eq:dettr} gives
\begin{align*}
  \det X = \frac1{120}\Bigl(24\tr X^5 & - 30\tr X^4\tr X - 20\tr X^3\tr X^2
      + 20\tr X^3(\tr X)^2 \\
    &+ 15(\tr X^2)^2\tr X - 10\tr X^2(\tr X)^3 + (\tr X)^5\Bigr);
\end{align*}
using equations~\refeq{eq:trodd} and~\refeq{eq:pdef} we obtain \(p(\xi) =
\rational18 \Bigl(2\tr A^4 - (\tr A^2)^2 + 4\xi\tr A^2 - 8\xi^2\Bigr)\xi\).
The discriminant of the quadratic factor is proportional to \(\Delta = 4\tr A^4
- (\tr A^2)^2\), so we may factor \(p\) into linear factors over the splitting
field generated by \(\sqrt\Delta\), \(p(\xi) = -\quarter(\xi - \lambda_{+}^2)
(\xi - \lambda_{-}^2)\xi\) where \(\lambda_\pm \defn \half\sqrt{\tr A^2 \pm
  \sqrt\Delta}\).  Using~\refeq{eq:charpoly} we obtain we obtain
\begin{equation*}
  p(A^2) = -\quarter \left(A^2 - \lambda_{+}^2\ident\right)
    \left(A^2 - \lambda_{-}^2\ident\right) A^2 = 0.
\end{equation*}
The projectors onto the eigenspaces of \(A^2\) belonging to \(\lambda_1^2 = 0\)
and \(\lambda_\pm^2\) are
\begin{equation*}
  P_1 = \frac{16\left(A^2 - \lambda_{-}^2\ident\right)
                \left(A^2 - \lambda_{+}^2\ident\right)}
             {\Delta - \left(\tr A^2\right)^2}
  \qquad\mbox{and}\qquad
  P_\pm =
    \pm\frac{8A^2\left(A^2 - \lambda_{\mp}^2\ident\right)}
            {\sqrt\Delta\bigl(\tr A^2 \pm \sqrt\Delta\bigr)},
\end{equation*}
and the projections of \(A^2\) onto these eigenspaces are \(A_1^2 = 0\) and
\(A_\pm^2 = \lambda_\pm^2 P_\pm\); hence \(\exp A_1 = \ident\) and Rodrigues'
formula~\refeq{eq:rodrigues} gives
\begin{equation*}
  \exp A_\pm = \ident + \left(\frac{\sin|\lambda_\pm|}{|\lambda_\pm|}\,A
    + \frac{1 - \cos|\lambda_\pm|}{|\lambda_2|^2}\,A^2\right) P_\pm
\end{equation*}
Combining these results we obtain \(\exp A = \exp A_{+} \exp A_{-}\).

\subsection{Exponentiation by explicit diagonalization} \label{sec:numerical}

The method described in~\refsec{sec:rodrigues} requires finding the roots of
the characteristic polynomial, and becomes computationally expensive and
numerically unstable for large matrices.

In this case it may be preferable to decompose the Lie algebra matrix \(A\)
into diagonal form \(A = \Omega D\Omega^\dagger\) where \(D = \diag(\lambda_1,
\ldots, \lambda_n)\) and \(\Omega\Omega^\dagger = \ident\).  Once this has be
done the matrix exponential is easy to evaluate, \(\exp A = \Omega \exp D
\Omega^\dagger\).  Numerical algorithms for decomposing a matrix into Schur
normal form, which is a diagonal matrix if \(A\) is normal, are well
known~\cite{golub:1996}.  A good choice is the \(QR\) iteration: the matrix is
first transformed to Hessenberg form (\(H_{ij} = 0\) if \(i-j\geq2\)) by means
of a sequence Householder reflections or Givens rotations; these are simple
orthogonal transformations.  A matrix that is Hessenberg and symmetric or
antisymmetric is tridiagonal.  The tridiagonal matrix is then decomposed by
means of Givens rotations \(Q = Q_nQ_{n-1}\cdots Q_1\) into a product \(H =
QR\) where \(Q\) is an orthogonal matrix and \(R\) is upper triangular.  The
matrix \(H' \defn RQ = Q'R'\) is likewise decomposed, whence \(H = QR =
QRQQ^\dagger = QQ'R'Q^\dagger = QH'Q^\dagger\).  This procedure is iterated
until the sequence of tridiagonal matrices converges to a diagonal matrix: it
converges because is may be thought of as an orthonormalized version of the
power method.  The convergence may be significantly improved by shifting \(H\)
at each step by a multiple of the unit matrix so as make some eigenvalue small,
and deflating when appropriate.

\subsection{Exponentiation of non-normal matrices} \label{sec:abnormal-exp}

\subsubsection{Numerical exponentiation} \label{sec:numexp}

In some cases, such as for the single sheeted hyperbolic space
of~\refsec{sec:h2111}, we are required to exponentiate a non-normal matrix.

In this case we may transform the matrix to upper triangular Schur form \(T\)
by a unitary transformation, \(A = \Omega T\Omega^\dagger\) with
\(\Omega^\dagger\Omega = \ident\).  This decomposition is not unique, and one
possible method is to orthonormalize the eigenspaces using the Gram--Schmidt
procedure.

As a simple example consider the non-normal matrix
\begin{equation*}
  A = \left(\begin{array}{cc} 1 & 1 \\ 4 & 1 \end{array}\right)
\end{equation*}
whose eigenvalues are \(\lambda_{\pm} = 1\pm2\) with corresponding eigenvectors
\begin{equation*}
  u_{\pm} = \left(\begin{array}{c} \mp\half \\ 1\end{array}\right).
\end{equation*}
Gram--Schmidt orthonormalization gives the columns of the orthogonal matrix
\begin{equation*}
  \Omega = \frac1{\sqrt5}
    \left(\begin{array}{cc} 1 & -2 \\ 2 & 1 \end{array}\right)
\end{equation*}
which gives the Schur decomposition
\begin{equation*}
  A = \Omega A\Omega^\dagger
  = \left(\begin{array}{cc} -1 & -3 \\ 0 & 3 \end{array}\right).
\end{equation*}

For numerical matrices the QR iteration~\cite{golub:1996} provides a more
efficient and stable algorithm to construct such a Schur form.

Clearly, the diagonal elements of the exponential of a triangular matrix \(T\)
are just the exponentials of its diagonal entries, \((\exp T)_{jj} =
\exp(T_{jj})\), and the off-diagonal entries are then easily found from the
fact that \(\exp T\) commutes with \(T\).

For our simple example we have
\begin{equation*}
  \exp(\xi T) =
    \left(\begin{array}{cc} e^{-\xi} & T_{12} \\ 0 & e^{3\xi} \end{array}\right),
\end{equation*}
so \([e^{\xi T},T] = 0\) requires that \(T_{12} = -\rational32 e^\xi \sinh
2\xi\).  We thus obtain
\begin{equation*}
  \exp(\xi A) = \Omega^\dagger \left(\begin{array}{cc}
    e^{-\xi} & -\rational32 e^\xi\sinh2\xi \\ 0 & e^{3\xi}
  \end{array}\right) \Omega = e^\xi \left(\begin{array}{cc}
    \cosh2\xi & \half\sinh2\xi \\ 2\sinh2\xi & \cosh2\xi
  \end{array}\right).
\end{equation*}

More generally the off-diagonal elements of the exponential of a triangular
matrix with distinct eigenvalues may be expressed in terms of
\definition{divided differences},
\begin{equation*}
  f[a_1,\ldots,a_k] = \left\{
  \begin{array}{cl}
    f(a_1) & \mbox{if \(k = 1\), and} \\[0.5ex]
    \displaystyle \frac{f[a_1,\ldots,a_k] - f[a_2,\ldots,a_{k-1}]}{a_k-a_1}
      & \mbox{for \(k  > 1\).}
  \end{array}\right.
\end{equation*}
by the equations~\cite{vanloan:1975}
\begin{align}
  (\exp T)_{jj} &= \exp(T_{jj}) \defn \lambda_j \label{eq:abnormal-exp} \\
  \mbox{and}\qquad (\exp T)_{ij} & = \sum_{{s_0,\ldots,s_k}\in S_{ij}}
    T_{s_0s_1}T_{s_1s_2}\cdots T_{s_{k-1}s_k}
    \exp\left[\lambda_{s_0},\ldots,\lambda_{s_k}\right]
  \quad\mbox{for \(i < j\),} \nonumber
\end{align}
where \(S_{ij}\) is the set of all strictly increasing sequences of integers
\(i=s_0 < s_1 < \cdots < s_{k-1} < s_k = j\).

A more efficient algorithm for exponentiating triangular matrices is to use
Parlett's recurrence~\cite{Parlett:1976}.  This is based on the observation
that \(F\defn\exp T\) commutes with \(T\) (since the matrix exponential is
defined by its series expansion) and the matrix elements of \(TF-FT\) are
\(\sum_{k=1}^j(T_{ik}F_{kj} - F_{ik}T_{kj}) = 0\) for \(i<j\).  Note that
\(T\) and \(F\) are both upper triangular (all powers of an upper
triangular matrix are themselves upper triangular).  We obtain
\begin{equation*}
  F_{ij} = \frac{\displaystyle T_{ij}(F_{ii}-F_{jj})
    + \sum_{k=i+1}^{j-1} (F_{ik}T_{kj}-T_{ik}F_{kj})}{T_{ii}-T_{jj}},
\end{equation*}
and this may be computed either a superdiagonal at a time, starting with the
diagonal, or a column at a time, starting from the second to last column and
moving up the columns.  When the eigenvalues of \(T\) are degenerate or nearly
degenerate it is better to apply this method to block submatrices
\cite{davies:2003,higham:2008}.

Although \(\exp A\) is real if \(A\) is real, the eigenvalues of \(A\) may be
complex.  It is possible to modify the algorithms to work entirely with real
matrices if desired~\cite{golub:1996,higham:2008}.  

\subsubsection{Avoiding exponentiating Non-normal matrices} \label{sec:avoid}

The HMC algorithm requires a reversible symplectic approximation of the
classical trajectory, when we are only interested in sampling from a target
density \(e^{-V}\), and do not need the free motion to follow accurately the
geodesics, it can be convenient to replace the \(\hvf T\) integrator step, by
another symmetric update step that only involves exponentiating normal
matrices.  This can be done by expanding the matrix \(M\) we wish to
exponentiate in the terms of basis of normal matrices.  For example if
\(M=A+B\), where \(A,B\) are normal, we have that from
Baker--Campbell--Hausdorf \cite{Kennedy:2012} formula \(\exp({(A+B)\delta t)}
\approx \exp({A \frac{\delta t}{2}})\exp({ B \delta t})\exp({A \frac{\delta
    t}{2}}) =\exp\bigl((A+B)\dt + \O(\delta t^3)\bigr)\).  This replacement
simplifies the exponentiation.  However when we are using higher order
integrator than leapfrog, this approximation can make the numerical trajectory
less accurate, which could lead to a smaller acceptance rate.

\section{Differential Geometry} \label{differential-geometry}

A \definition{manifold} \(\M\) is a topological space, which we shall also call
\(\M\), with an atlas of \definition{charts} \(\{(U,\phi_U)\}\) where
\(\{U\subseteq\M\}\) is an open cover of \(\M\) and \(\phi:U\to\R^n\) is a
injective homeomorphism map giving the local coordinates of points in~\(U\).
The charts in an atlas are related by smooth maps \(\phi_V^{-1}\circ\phi_U
\subset\smooth{\R^n \to\R^n}\), and endow the manifold with its differentiable
structure.  All the maps on and between manifolds will implicitly be taken to
be smooth.

A \definition{curve} \(c:\R\to\M\) maps a parameter, which will we generically
call ``time'', into the manifold.

A \definition{fibre bundle} \((\B,\M,\pi)\) is a manifold \(\B\), the
\definition{total space}, with a projection \(\pi:\B\to\M\) onto its
\definition{base space}~\(\M\).  Above each point in \(x\in\M\) is a
\definition{fibre} \(\F\diffeo\pi^{-1}x\).  A bundle is locally trivial,
meaning that there is a neighbourhood \(U\ni x\) for which \(\pi^{-1}U =
U\times\F \subseteq\B\).  We will usually abuse notation by referring to the
bundle and its total space by the same name.

A \definition{section} \(\psi\in\field\B\) of a bundle is a map
\(\psi:\M\to\B\) such that \(\pi\circ\psi = \id\M\).

A \(0\)-form \(f\in\form0\M\) is an \(\R\)-valued function on~\(\M\),
\(\form0\M = \smooth{\M\to\R}\).  In the language of bundles we may say that it
is a section of the line bundle over \(\M\) with fibre~\(\R\).

%% The tangent bundle \(\tanbundle\M\) is a bundle over \(\M\) whose fibre
%% above \(p\) is the vector space of linear derivations \(X_p:\smooth\M|_p
%% \to\R\) on the algebra of germs of smooth functions \(\smooth\M\).  A
%% derivation means \(X_p(fg)=f(p)X_pg + g(p)X_pf\), and elements of
%% \(\smooth\M|_p\) are equivalence classes \([f]\equiv f\) of functions that
%% agree on some open neighbourhood of~\(p\).  Note \(f\in\smooth\M|_p\) has a
%% well-defined value at~\(p\), and it can be checked that \(X_p\) is a local
%% differential operator (see Cambridge lecture notes for a proof).  A vector
%% field is a section of~\(\tanbundle\M\).  It is convenient to also view
%% vector fields as linear derivation\(X:\smooth\M\to\smooth\M\), with
%% \(Xf(p)\defn X_pf\).

A \definition{vector field} is a linear differential operator acting on
\(0\)-forms; it may also be considered as a section of the \definition{tangent
bundle}~\(\tanbundle\M\), whose fibre \(T_p\M \defn \pi^{-1}p\) is the space
of linear derivations\footnote{These linear derivation act on
\definition{germs} of smooth functions in a neighbourhood of~\(p\)
\cite{Choque1982analysis}.} \(v(fg) = (vf)g + f(vg)\) at~\(p\), which is
isomorphic as a linear space the \(n\)-dimensional linear space~\(\R^n\).  We
shall use the notation\footnote{We could write \(vf\) and \(vp\) in both cases
relying on the context to resolve any ambiguity, but this implicit Currying
causes confusion, especially for the authors.  We note in passing that defining
\(v|_pf\defn(vf)(p)\) provides the isomorphism between vector fields viewed as
differential operators and as sections of the tangent bundle.}  \(vf\) to
indicate the application of the differential operator \(v\) to the
\(0\)-form~\(f\), and \(v|_p\) for the value of the vector field at the point
\(p\in\M\).

Observe that the \definition{commutator} of two vector fields is itself a
vector field, \([u,v] = uv-vu\in\field{\tanbundle\M}\), because the second
derivative terms cancel.

A \(k\)-form (field) \(\alpha\in\form k\M\) is an antisymmetric multilinear map
\(\alpha:\field{\tanbundle\M}^k\fairbreak\to\R\).  In particular a \(1\)-form
field is a section of the \definition{cotangent bundle}, whose fibre \(T_p^{*}\M
\defn \pi^{-1}p\) is the dual space to that of the tangent bundle, i.e., a
smooth map \(\M \to T^{*} \M\) .

We may construct an \definition{exterior algebra} on the space of forms by
introducing the associative and antisymmetric wedge product, for
\(\alpha\in\form j\M\) and \(\beta\in\form k\M\) we have \(\alpha\wedge\beta
\in\form{j+k}\M\) with \(\alpha\wedge\beta = (-1)^{jk}\beta\wedge\alpha\).  We
may also introduce the \definition{exterior derivative} which is an nilpotent
antiderivation \(d:\form k\M\to\form{k+1}\M\).  For \(k=0\) we have \(df(v) =
vf\) for any \(0\)-form \(f\) and any vector field~\(v\).  For \(k=1\) we have
\(d\alpha(u,v) = u\alpha(v) - v\alpha(u) - \alpha([u,v])\): it is easy to
verify that \(d^2f(u,v) = u\,df(v) - v\,df(u) - df([u,v]) = uvf - vuf - [u,v]f
= 0\).  In general \(d\) is the coboundary operator for the
Chevalley--Eilenberg cohomology of Lie algebras, in which a form
\(\alpha\in\form{k+1}\M\) is \definition{closed} if \(d\alpha=0\) and is
\definition{exact} if \(\alpha = d\beta\) for some \(\beta\in\form k\M\).

A map between manifolds \(\phi:\M\to\N\) induces the \definition{pull-back} map
\(\pull\phi:\form0\N\to\form0\M\) between \(0\)-forms defined by \(\pull\phi:f
\mapsto f\circ\phi\); that is, for any function \(f\in\form0\N\) the function
\(\pull\phi f\in\form0\M\) satisfies \((\pull\phi f)(x)=f\bigl(\phi(x)\bigr)\)
for every~\(x\in\M\).  

The \definition{tangent map} of \(\phi\) at a point \(p\) is the linear map
\(\tang p\phi:T_p\M \to T_{\phi(p)}\N\), defined by \(\tang p\phi(v) =
v\circ\phi^*\), i.e., \(\tang p\phi(v)(f) = v(f\circ\phi)\); the
\definition{pull-back} of this tangent map at \(p\) is the a map
\(\phi^{*}:T^{*}_{\phi(p)} \N \to T^{*}_p \M\), with \(\phi^{*} \alpha \defn
(\tang p\phi)^{*} \alpha =\alpha \circ \tang p \phi\).  Note that when we take
the pullback of a linear map, its domain of definition is restricted to linear
functions (instead of being the space of 0-forms); this means the pullback of a
linear function is defined as its pullback in the category of vector spaces,
rather than manifolds (in this context, the pullback is sometimes referred as
the adjoint or transpose).

Above, the tangent map and pull-back were maps between vectors and covectors.
If \(\phi\) is a diffeomorphism it also has an inverse, which satisfies
\(\pull{\phi^{-1}} = (\pull\phi)^{-1}\), and we can turn these into maps
between vector fields and covector fields.  If \(\psi:\MA\to\M\) then the
pull-back \(\pull{(\phi\circ\psi)}:\form0\N\to\form0\MA\) of the composite map
\(\phi\circ\psi:\MA\to\N\) satisfies \(\pull{(\phi\circ\psi)} = \pull\psi\circ
\pull\phi\) because \(\pull{(\phi\circ\psi)}f = f\circ(\phi\circ\psi) =
(f\circ\phi)\circ\psi = (\pull\phi f)\circ\psi = \pull\psi(\pull\phi f)\) for
all~\(f\in\form0\N\).

In this case it also induces a \definition{push-forward} map of vector fields
\(\push\phi:\field{\tanbundle\M}\fairbreak\to\field{\tanbundle\N}\) defined by
\(\push\phi v = \pull{\phi^{-1}}\circ v\circ\pull\phi\).  The final (leftmost)
map \(\pull{\phi^{-1}}\) is necessary because \(\push\phi\) has to map
functions in \(\form0\N\) to functions in \(\form0\N\); even if \(\M\diffeo\N\)
it is required so that \(\push\phi v\) is a local differential operator.  This
map commutes with the commutation of vector fields, that is \(\push\phi[u,v] =
[\push\phi u, \push\phi v]\): the proof is simple
\begin{align*}
  \push\phi [u,v] &= \pull{\phi^{-1}}[u,v]\pull\phi
  = \pull{\phi^{-1}}(uv - vu)\pull\phi
  = (\pull{\phi^{-1}}uv\pull\phi - \pull{\phi^{-1}}vu\pull\phi) \\
  &= \pull{\phi^{-1}}u(\pull\phi\pull{\phi^{-1}})v\pull\phi
    - \pull{\phi^{-1}}v(\pull\phi\pull{\phi^{-1}})u\pull\phi \\
  &= (\pull{\phi^{-1}}u\pull\phi) (\pull{\phi^{-1}}v\pull\phi)
    - (\pull{\phi^{-1}}v\pull\phi) (\pull{\phi^{-1}}u\pull\phi) \\
  &= (\push\phi u) (\push\phi v) - (\push\phi v) (\push\phi u)
  = [\push\phi u,\push\phi v].
\end{align*}
The push-forward \(\push{(\phi\circ\psi)}:\field{\tanbundle\MA}\to
\field{\tanbundle\N}\) of the composite map \(\phi\circ\psi:\MA\to\N\) of two
diffeomorphisms \(\psi:\MA\to\M\) and \(\phi:\M\to\N\) satisfies
\(\push{(\phi\circ\psi)} = \push\phi\circ\push\psi\) because
\begin{align*}
  \push{(\phi\circ\psi)}v
  &= \pull{(\phi\circ\psi)^{-1}}\circ v\circ\pull{(\phi\circ\psi)}
  = \pull{(\psi^{-1}\circ\phi^{-1})}\circ v\circ\pull{(\phi\circ\psi)} \\
  &= (\pull{\phi^{-1}}\circ\pull{\psi^{-1}})\circ v\circ(\pull\psi\circ\pull\phi)
  = \pull{\phi^{-1}}\circ(\pull{\psi^{-1}}\circ v\circ\pull\psi)\circ\pull\phi\\ 
  &= \pull{\phi^{-1}}\circ\push\psi v\circ\pull\phi
  = \push\phi(\push\psi v) = (\push\phi\circ\push\psi)v
\end{align*}
for all~\(v\in\field{\tanbundle\MA}\).

Since \(\phi\) is a diffeomorphism, the \definition{pull-back map} of 1-forms
\(\pull\phi:\form1\N\to\form1\N\) can be simply written as \(\pull\phi\alpha =
\alpha\circ\push\phi\).  The ambiguity between this map and that on \(0\)-forms
of the same name may be resolved by context.  The pull-back
\(\pull{(\phi\circ\psi)}:\form1\N\to\form1\MA\) of the composite map
\(\phi\circ\psi:\MA\to\N\) of two diffeomorphisms \(\psi:\MA\to\M\) and
\(\phi:\M\to\N\) satisfies \(\pull{(\phi\circ\psi)} = \pull\psi\circ\pull\phi\)
since \(\pull{(\phi\circ\psi)}\alpha = \alpha\circ\push{(\phi\circ\psi)} =
\alpha\circ(\push\phi\circ\push\psi) = (\alpha\circ\push\phi)\circ\push\psi =
(\pull\phi\alpha)\circ\push\psi = \pull\psi(\pull\phi\alpha) =
(\pull\psi\circ\pull\phi)\alpha\) for all~\(\alpha\in\form1\N\).

An \definition{integral curve} of a vector field \(v\in\field{\tanbundle\M}\)
is a map \(c:\R\to\M\) such that \(\dot c=v|_c\); in other words it is a curve
whose tangent \(\dot c\) is always equal to the value of \(v\) at the point
\(c(t)\in\M\) for all time~\(t\).  To specify such an integral curve we must
specify an ``initial condition''such as the value of~\(c(0)\).

The \definition{interior product} \(i_v\alpha\) of a vector \(v\) and a
\((k+1)\)-form \(\alpha\) is the \(k\)-form obtained by setting the first
argument of \(\alpha\) to \(v\): more precisely \(i_v\alpha(u_1,\ldots,u_k) =
\alpha(v,u_1,\ldots,u_k)\).  Moreover we set \(i_vf = 0\) for any
\(0\)-form~\(f\).  It is obvious that \(i_v^2=0\) because \(i_v^2\alpha(u_1,
\ldots,u_{k-1}) = \alpha(v,v,u_1,\ldots,u_{k-1}) = 0\) by antisymmetry.

We may consider the collection of curves passing through different points at
time zero.  This is the \definition{flow} of the vector field \(v\) which is
the map \(\flow:\R\times\M\to\M\) such that\footnote{We use the notation
\(\partial_1f\) here to denote the partial derivative of a function \(f\) with
respect to its first argument.  Alternatively we could write this as \(\partial
f/\partial t\), but this assumes that the first argument has the name~\(t\).
We shall also write \(\partial_1f|_0\) to mean \(\partial_1f\) with its first
argument zero.  For functions \(f\) of a single argument we use \(\dot f\) for
the same purpose: i.e., \(\dot f = \partial_1f\).}  \(\partial_1\flow|_0 = v\).
The integral curve be considered before is then just \(c(t) = \flow(t,c(0))\).
In general it is not guaranteed that the either the integral curves or flow of
a vector field exists for all times.  We introduce the notation
\(\exp(tv):\M\to\M\) for the flow of \(v\) at time~\(t\).

The \definition{Lie derivative} of a \(0\)-form \(f\) with respect to a vector
field \(v\) if the derivative of \(f\) along the flow \(\flow\) of~\(v\),
\(\Lie vf = \partial_1(f\circ F)|_0 = df(\partial_1F|_0) = df(v) = vf\).  The
Lie derivative of a vector field \(u\) with respect to \(v\) is \(\Lie vu =
\partial_1(\push\flow u)|_0 = [v,u]\), and the Lie derivative of a \(1\)-form
\(\alpha\) with respect to \(v\) is \(\Lie v\alpha =
\partial_1(\pull\flow\alpha)|_0 = d\bigl(\alpha(v)\bigr) - (d\alpha)(v).\) The
definition may be generalized easily to arbitrary forms, for which the
identities \(di_v + i_vd = \Lie v\) and \([\Lie v,i_w] = i_{[v,w]}\)
hold.  See~\cite{Szekeres:2004,Holm2009geometric} for more details.

\section{Hamiltonian Dynamics and HMC} \label{hamiltonian-dynamics}

A \definition{symplectic manifold} \((\MA,\ftf)\) is an (even-dimensional)
manifold \(\MA\) together with a closed non-degenerate \(2\)-form~\(\ftf\).
Given a \(0\)-form \(f\in\form0\MA\) we define the corresponding
\definition{Hamiltonian vector field} \(\hvf f\in\field{\tanbundle\MA}\) by
\(df\defn i_{\hvf f}\ftf\), where we recall that the interior product is
\(i_v\ftf(u) = \ftf(v,u)\) for all vector fields~\(u\in\field{\tanbundle\MA}\).

If \(H\in\smooth\MA\) is the \definition{Hamiltonian} function then the
integral curves of the corresponding Hamiltonian vector field \(\hvf H\)
represent the trajectory of the system through phase space.  Such an integral
curve \(c:\R\to\MA\) is always tangential to the vector field, \(\dot c = \hvf
H|_c\).  The triple \((\MA,\ftf,H)\) is called a \definition{Hamiltonian
system}.

The cotangent bundle \(\cotbundle\M\) over a manifold \(\M\) may be given a
canonical symplectic structure, which is built from the \definition{Liouville
  form} (q.v., \cite{Abraham:2008} Theorem 3.2.10).  For example, if
\(\M=\R^n\) then \(\MA=\cotbundle\R^n\diffeo\R^{2n}\) with the canonical
fundamental \(2\)-form \(\ftf = dq_i\wedge dp_i\) (with implicit summation over
the index~\(i\)) and Hamiltonian \(H:\cotbundle\R^n\to\R\) then \(dH =
(\partial_{q_i}H) dq_i + (\partial_{p_i}H) dp_i = i_{\hvf H}\ftf\) and thus
\(\hvf H = \hvf H_{q_i} \partial_{q_i} + \hvf H_{p_i} \partial_{p_i} =
(\partial_{p_i}H) \partial_{q_i} - (\partial_{q_i}H) \partial_{p_i}\).  If the
Hamiltonian is the sum of kinetic and potential energy functions, \(H =
T(p)+\U(q)\), then an integral curve of the Hamiltonian vector field satisfies
\(\dot c = \hvf Hc\), which are Hamilton's equations
\begin{align*}
  \dot Q_i(t) &= \hvf H_{q_i}\bigl(Q(t), P(t)\bigr)
    = \partial_{p_i}T\bigl(P(t)\bigr),\\
  \dot P_i(t) &= \hvf H_{p_i}\bigl(Q(t),P(t)\bigr)
    = -\partial_{q_i}\U\bigl(Q(t)\bigr),
\end{align*}
for the coordinates of the curve~\(c(t) = \bigl(Q(t), P(t)\bigr)\).  We build
symplectic integrators by performing separate updates for the positions and
momenta, that is solving these equations for the cases \(T=0\) and \(\U=0\)
separately, with solutions
\begin{equation*}
  Q_i(t) = Q_i(t_0) + (t - t_0)\partial_{p_i}T\bigl(P(t_0)\bigr),\qquad
  p_i(t) = p_i(t_0) - (t - t_0)\partial_{q_i}\U\bigl(Q(t_0)\bigr),
\end{equation*}
respectively.  The Hamiltonian system is reversible, which means there exists
an antisymplectic involution, (\(\mu : \cotbundle \M \to \cotbundle \M\) with
\(\mu^{*} \omega = -\omega\) and \(\mu^2 = id\)), \cite{Abraham:2008} which
is given by~\(\mu(p)=-p\).

When \(\MA=\cotbundle\M\) there is a canonical measure induced by the
symplectic structure \(\mu \defn \bigwedge^n \omega\).  For HMC we are
interested in sampling from a measure \(\propto e^{-\U(q)}\mu_\M\) on the
space~\(\M\).  This differential form can be pull-backed to \(\cotbundle\M\)
and there exists a measure \(\propto e^{-H(q,p)}\measure\) on phase space which
push-forwards to \(\propto e^{-\U(q)}\mu_\M\)~\cite{Betancourt:2017a}.

The HMC sampler is constructed out of two Markov steps, both of which preserve
the density \(\propto e^{H(q,p)}\).  The first step samples momenta using a
Gibbs sampler (heath bath) with density \(\propto e^{-K(p)}\) while keeping the
position \(q\) fixed; and the second step constructs a candidate point \(x' =
(q',p')\) in phase space using a symmetric symplectic integrator, and then
accepts it with Metropolis probability \(\min(1, e^{-\delta H})\), where
\(\delta H = H(x') - H(x)\).  In many interesting cases the composition of
these steps can be shown to be ergodic.

\section{Lie Groups} \label{lie-groups}

An element \(g\in\G\) of a Lie group can act on \(\G\) using the left action
\(L_g:\G\to\G\) by \(L_g:h\mapsto gh\).  It also has the right action
\(R_g:\G\to\G\) defined by \(R_g:h\mapsto hg\), and this commutes with the left
action \([L_g,R_{g'}] = 0\) since \(L_g(R_{g'}(h)) = L_g(hg') = g(hg') = (gh)g'
= R_{g'}(gh) = R_{g'}(L_g(h))\).

A vector field \(v\in\field{\tanbundle\G}\) is \definition{left invariant} if
\(\push{L_g}v=v\) for all \(g\in\G\).  We shall denote the space of such left
invariant fields by~\(\X\G\).  The ``product'' \(uv\) of two vector fields
\(u,v\in\X\G\) defined by composition of differential operators is not a vector
field in general, so \(\X\G\) is not an associative algebra.  However, their
commutator \([u,v]\in\X\G\), moreover it is trivial to establish the
\definition{Jacobi identity} \([u,[v,w]] + [v,[w,u]] + [w,[u,v]] = 0\) for all
\(w\in\X\G\), hence \(\X\G\) is a \definition{Lie algebra} with the commutator
as its \definition{Lie product} or \definition{Lie bracket}.

Any vector \(\xi_1\in\tanbundle\G_1\) in the tangent space at the identity
\(1\in\G\) has a unique extension to a left invariant vector field defined by
\(\xi\defn\push{L_g}\xi_1\).  As there are \(n\) linearly independent vectors
in \(\tanbundle\G_1\) there must be exactly \(n\) linearly independent left
invariant vector fields.  The space \(\X\G\) of left invariant vectors is thus
an \(n\)-dimensional linear space.  Furthermore, this construction shows that
the tangent space at the identity may also be extended to be a Lie algebra
\(\g\) by defining the Lie product on vectors \(x_1,y_1\in\g\) to be that of
the corresponding left invariant vector fields, and thus \(\X\G\diffeo\g\) are
isomorphic Lie algebras.  Let \(e_1,\ldots,e_n\) be a linear space basis for
the Lie algebra \(\g\), then we have \([e_i,e_j] = c^i_{jk} e_i\) where
\(c^i_{jk}\) are the \definition{structure constants} of~\(\g\).

The elements of the dual basis \(\theta^1, \ldots,\theta^n\) that satisfy
\(\theta^i(e_j) = \delta^i_j\) form the \definition{Maurer-Cartan frame}.  The
frame \(\theta^i\) are left invariant, \(\pull{L_g}\theta^i = \theta^i\) for
all \(g\in\G\), because \(\pull{L_g}\theta^i(e_j) = \theta^i(\push{L_g} e_j) =
\theta^i(e_j) = \delta^i_j = \theta^i(e_j)\) for \(j=1,\ldots,n\).
  
The \definition{Maurer-Cartan form} is defined as \(\theta = e_i \otimes
\theta^i\), and can also be defined in a frame-independent manner as \(\theta :
g\mapsto\tang{g}L_{g^{-1}}\).
  
The dual basis elements satisfy the \definition{Maurer--Cartan} relations
\begin{equation*}
  d\theta^i(e_j,e_k) = e_j\theta^i(e_k) - e_k\theta^i(e_j) - \theta^i([e_j,e_k])
  = e_j\delta^i_k - e_l\delta^i_j - \theta^i(c^\ell_{jk} e_\ell) = -c^i_{jk},    
\end{equation*}
and expanding the \(2\)-form \(d\theta^i=\Theta^i_{j'k'}\theta^{j'}\wedge
\theta^{k'}\) in basis \(2\)-forms we have
\begin{align*}
  d\theta^i(e_j,e_k) &= \Theta^i_{j'k'} \theta^{j'}\wedge\theta^{k'}(e_j,e_k)
  = \Theta^i_{j'k'} \left(\theta^{j'}(e_j)\theta^{k'}(e_k)
    - \theta^{j'}(e_k)\theta^{k'}(e_j)\right) \\
  &= \Theta^i_{j'k'} (\delta^{j'}_j\delta^{k'}_k - \delta^{j'}_k\delta^{k'}_j)
  = \Theta^i_{jk} - \Theta^i_{kj} = -c^i_{jk}
\end{align*}
giving
\begin{equation} \label{eq:Maurer-Cartan-relations}
  d\theta^i = -\half c^i_{jk}\theta^j\wedge\theta^k.
\end{equation}

Riemannian metric tensors on Lie groups are often required to be left or
bi-invariant.  Recall a metric on \(\G\) is called \definition{left invariant}
if\footnote{Note that if \(M\) is a left-invariant metric on \(\G\) then by
definition \(\pull{L_h}M=M\) and thus left translations are isometries of
\(\G\).  It follows that right-invariant vector fields are killing fields,
since they are infinitesimal generators of isometries (i.e., their flow is an
isometry).}  \(M(u,v)=M(\tang {}L(u),\tang {}L(v))\).  Similarly for right
invariance and bi-invariance.  It is easy to check (i)~there is a bijection
between left-invariant metrics on \(\G\) and inner products on \(\g\)
(ii)~there is a bijection between bi-invariant metrics on \(\G\) and
\definition{Ad-invariant} (which means \(M\bigl(\Ad_ g(u), \Ad_ g(v)\bigr) =
M(u,v)\) for all \(g\in\G\) and \(u,v\in\g\)) inner products on~\(\g\) (for
definition of \(\Ad\) see \refapp{ad-rep}).\footnote{To prove (i): if
\(\metric\) is a left invariant metric on \(\G\), then \(\metric(X(p),Y(p)) =
\metric(X(1),Y(1))\) when \(X,Y\in\g\).  Thus \(\metric\) gives a well-defined
inner product on the vector space of left invariant fields \(\g\).  Conversely
if \(\langle \cdot,\cdot\rangle\) is an inner product on \(\g\), then
\(\metric(v(p),u(p))\defn\langle(\tang {}L)_{p^{-1}}v(p), (\tang {}L)_{p^{-1}}
u(p)\rangle\) is a left invariant metric on \(\G\), since \(\metric(\tang
{}L_gu(h), \tang {}L_gv(h)) = \langle \tang {}L_{(gh)^{-1}}u(h), \tang
{}L_{(gh)^{-1}}v(h) \rangle = \langle \tang {}L_{h^{-1}}u(h), \tang
{}L_{h^{-1}}v(h) \rangle = g(u(h),v(h)\).  For (ii): if \(\langle \cdot,\cdot
\rangle\) is \(\Ad\)-invariant, then \(\metric(u(p),v(p)) \defn \langle \tang
{}L_{p^{-1}}u(p), \tang {}L_{p^{-1}}v(p)\rangle\) is bi-invariant.  Indeed
from~(i) we see \(\metric\) is left-invariant.  Showing right-invariance is
easy once we notice that \(L_a\) and \(R_b\) commute, and then use
\(\Ad\)-invariance.  Conversely if \(g\) is bi-invariant, by~(i) it gives a
well-defined inner product on \(\g\), which is clearly \(\Ad\)-invariant since
\(\Ad_ g = \tang {1}(R_{g^{-1}} L_g) = \tang gR_{g^{-1}} \tang 1 L_g\)}

Similarly we will say an inner product on \(\g\) is
\definition{\(\Ad_{\G}\K\)-invariant} if the same relation holds for any
\(g\in\K \subset \G\).  As a result it is often convenient to define metrics on
\(\G\) through inner products on~\(\g\).\footnote{There is also a bijection
between \(\Ad_\G \K\)-invariant quadratic forms on \(\g\) and \(\G\)-left
invariant and \(\K\)-right-invariant pseudo-metrics on \(\G\).}

A \definition{matrix Lie group} has elements that are a collection of \(n\times
n\) matrices with group multiplication being matrix multiplication.  We may
consider these matrices as providing the faithful (defining) representation of
the corresponding abstract Lie group.  A \definition{representation} is a map
\(\rho:\G\to\GL n\) into the space of non-singular linear operators on \(\R^n\)
(or \(\C^n\)) that is a group morphism, \(\rho(gh) = \rho(g)\rho(h)\) where the
composition of linear operators on the right is just matrix multiplication.

\subsection{Adjoint Representation and Lie Group Exponential} \label{ad-rep}

Let \(F^X\) denotes the flow (see \ref{differential-geometry}) of \(X\) and
\(\gamma_X\) the integral curve of \(X\) that passes through the identity
\(1\), \(\gamma_X(t) = F^X(t,1)\).  We define the \definition{Lie group
exponential} map by \(\exp: \g\to\G: \exp(X_1) = \gamma_X(1)\), from which it
follows\footnote{Indeed, if \(\gamma_{tX}\) is the integral curve of \(tX\),
then \(\tang c \gamma_{tX}(\frac d{ds})=tX_{\gamma(c)}\), where \(s\) is a
global coordinate on \(\R\).  Then setting \(\gamma_X \defn \gamma_{tX}\circ
\frac 1 t\), we have \( X_{\gamma(c)} =\tang c \gamma_{tX}(\frac1t \frac d
{ds})= \tang c \gamma_{X}( \frac d {ds})\), and so \(\gamma_X\) is the integral
curve of \(X\) and \( \gamma_X(t) = \gamma_{tX}(1)\).}  that \(\exp(tX_1) =
\gamma_X(t)\).  The flow of left-invariant vector fields generates right
translations.  To see this, first note that by definition the curve \(t \mapsto
\exp(tX_1)\) is tangent to \(X_1\) at \(t=0\), or in other words \(\tang 1
\exp(X_1) = X_1\).  Similarly the curve \(t \mapsto g \exp(tX_1)\) is tangent
to \(X_g\) at \(t=0\), since \(\tang 1 \bigl(L_g\circ \exp)(X_1) = \tang g
L_g\circ \tang 1 \exp(X_1) = \tang g L_g X_1 = X_g\), since \(X\) is
left-invariant, and thus is the integral curve of \(X\) through \(g\).  But
this is precisely the meaning of the curve \(t \mapsto F^X(t,g)\), hence
\(F^X(t,g) = g \exp(tX)= R_{\exp(tX)} (g)\).

The left and right action of a group on itself are not the only actions:
another useful automorphism is conjugation by some fixed group element.  Such
an automorphism is called an \definition{inner automorphism}, \(\In h:\G\to\G\)
where \(\In h:g\mapsto hgh^{-1}\).  This is an automorphism because \(\In hg\In
hg' = (hgh^{-1})(hg'h^{-1}) = hgg'h^{-1} = \In h(gg')\); it also satisfies the
identity \(\In h\circ\In{h'}g = \In h(h'g{h'}^{-1}) = h(h'g{h'}^{-1})h^{-1} =
(hh')g(hh')^{-1} = \In{hh'}g\).

It follows that its tangent map at the identity \(\Ad_ g\defn \tang 1\In g\) is
a Lie algebra automorphism.  The map \(\Ad:\G\to\Aut\g\) with \(g\mapsto\Ad_
g\) is called the \definition{adjoint representation} of \(\G\).  It is a
representation of \(\G\) since \(\Ad_ h\circ\Ad_ {h'} = \tang 1\In h\circ \tang
1\In{h'} = \tang 1 \bigl(\In h\circ\In{h'}\bigr) = \tang 1\In{hh'} = \Ad_
{hh'}\).  Its tangent map at the identity \(\ad:X\to (\tang 1\Ad) X\) is a
representation of the Lie algebra.  Note \(\ad:\g\to\End\g\) where \(\End\g\) is
the space of endomorphisms of~\(\g\).  From example 4.1.25 in
\cite{Abraham:2008} we have \(\ad_ XY =[X,Y]\) from which it is easy to prove
\(\ad\) is a representation using the Jacobi identity.

\section{Homogeneous Spaces} \label{homogeneous-spaces}

A \definition{\(\G\)-space} is a manifold \(\M\) upon which a group \(\G\) acts
as a group morphism: this means that there is an \definition{action} \(\Phi:
\G\times\M\to\M\) such that \(\Phi(g,\Phi(h,x)) = \Phi(gh,x)\).  In the more
compact notation \(\Phi:(g,x)\mapsto gx\) this may be written as \(g(hx) =
(gh)x\).  In such a case \(\G\) is called a \definition{transformation group}
acting on~\(\M\).  The \definition{orbit} of \(\G\) containing \(x\in\M\) is
the set of points \(\G x\defn\{gx\,|\,g\in\G\}\).  If there is only one orbit,
\(\G x=\M\), then \(\G\) is said to act \definition{transitively} on \(\M\):
there is always some group element that takes any point in \(\M\) to any other
point.  In this case \(\M\) is called a \definition{homogeneous space}.

The \definition{stabilizer} is the subgroup \(\K_x\subset\G\) that fixes
\(x\in\M\), \(k\in\K_x\iff kx=x\), with respect to the action \(\Phi\); it is
also called the \definition{isotropy group} or \definition{little
group}.  \footnote{It is easy to see that stabilizers are related by
conjugation: if \(y=gx\) and \(kx=x\) then \((gkg^{-1})y = (gkg^{-1})gx = gkx =
gx = y\), so \(\K_{gx} = g\K_xg^{-1}\).  This means that all the stabilizers
are isomorphic: the isomorphism \(\phi:\K_x\to K_{gx}\) being \(\phi:k\mapsto
gkg^{-1}\), so for all \(k,k'\in\K_x\) we have \(\phi(k)\phi(k') =
(gkg^{-1})(gk'g^{-1}) = g(kk') g^{-1} = \phi(kk')\).}

Let us assume the action is transitive and fix some point \(p\in\M\) whose
stabilizer we shall call \(\K\defn\K_p\); we may construct the quotient space
\(\G/\K = \{g\K\}\) of left cosets.  In general this is not a group because the
stabilizer is not a normal subgroup,~\(\K\notnormalin\G\).  Nevertheless, this
space is interesting because there is a diffeomorphism \(\psi:\G/\K\to\M\)
given by \(\psi:g\K\mapsto \Phi(g\K,p)=g\K p\), since \(\psi(g\K)=(g\K)p = g(\K
p) = gp\) and \(\G p=\M\).  We induce a left action of \(\G\) on \(\G/\K\) by
\(\Phi_L:\G\times\G/\K \to\G/\K\) where \(\Phi_L:(g,g'\K)\mapsto gg'\K\), and
the stabilizer of the coset \(g\K\in\G/\K\) under this left action is the
conjugate\footnote{Pedants may wish to write \(\K_{g\K}\) and \(\K_\K\) rather
than \(\K_g\) and \(\K\).}  \(\K_g = g\K g^{-1}\) since \(\Phi_L:(g\K
g^{-1},g\K)\mapsto g\K g^{-1}g\K=g\K\).

The map \(\psi\) is \definition{equivariant}, \(\psi\bigl(\Phi_L(g,g'\K)\bigr)
= \Phi\bigl(g,\psi(g'\K)\bigr)\in\M\) for any \(g,g'\in\G\); this means that
the following diagram commutes
\begin{equation*}
  \begin{CD}
    \G\times\G/\K            @>\Phi_L>>   \G/\K \\
    @VV\id\G\otimes\,\psi V               @VV\psi V \\
    \G\times\M               @>\Phi>>     \M
  \end{CD}
\end{equation*}
which means \(\M\) and \(\G/\K\) are isomorphic in the category of
\(\G\)-manifolds.

There is also a right action of \(\G\) on \(\G/\K\) given by \(\Phi_R:(g,g'\K)
\mapsto g'\K g\) whose stabilizer is \(\K\), since \((\K,g'\K)\mapsto
g'\K\K=g'\K\); this right action is a morphism of the opposite group, namely
\(\Phi_R\bigl(g,\Phi_R(h,g'\K)\bigr) = \Phi_R(g, g'\K h) = g'\K hg =
\Phi_R(hg,g'\K)\).

The canonical projection \(\pi:\G\to\G/\K\) with \(\pi:g\mapsto g \K\) together
with the above free (that is without any non-trivial fixed points) right action
\(\G\times\K\to\G\) defines a principal \(\K\)-bundle on~\(\G\).

So far our discussion of homogeneous spaces is applicable for any group \(\G\),
but let us now restrict our attention to the case where \(\G\) is a Lie group
and the action \(\Phi\) is smooth.  In this case the identification \(\G/\K\)
with \(\M\) is also a diffeomorphism, and this induces a corresponding
identification of the tangent bundles \(\tanbundle{(\G/\K)}\) and
\(\tanbundle\M\).  In particular the tangent spaces at \(p\in\M\) and at
\(\K\in\G/\K\) can be identified with the quotient vector space \(\g/\k\) where
\(\k\) is the Lie algebra of~\(\K\).

The homogeneous manifold \(\M\) is \definition{reductive},
if there exists a decomposition of the Lie algebra \(\g =
\k\oplus\p\) with \(\Ad_ \G(\K)(\p)\subseteq\p\) (see section \ref{ad-rep}), which is true if
\([\k,\p]\subseteq\p\).\footnote{Since \(\K\) is a Lie subgroup of \(\G\) its
  Lie algebra \(\k\) is a subalgebra of \(\g\), \([\k,\k]\subseteq\k\).}  With
this assumption the tangent space at \(p\in\M\) is isomorphic to~\(\p\).

\section{\(\G\)-Invariant Metrics on \(\G/\K\)} \label{metric}

Any representation \(\rho:\G\to\GL V\) defines a symmetric bilinear form on
\(\g\) by \((x,y)\mapsto\tr\push\rho(x)\push\rho(y)\).  The Cartan--Killing
form is the one corresponding to the adjoint representation \(\Ad\).
Since\footnote{This follows from \(\ad_ {\psi(X)}Y = [\psi(X),Y] =
[\psi(X),\psi\circ \psi^{-1}Y] = \psi [X,\psi^{-1}Y] = \psi\circ\ad_ X\circ
\psi^{-1}(Y)\)} \(\ad_ {\Psi(X)}= \psi\circ\ad_ X\circ\psi^{-1}\) for any
\(X\in\g\) and any automorphism \(\psi\) of \(\g\), the Killing form is
invariant under any automorphism of \(\g\), and thus it is in particular
\(\Ad\)-invariant\footnote{Indeed \(\tr\bigl(\ad_ {\Ad_gu}\circ\ad_{\Ad_gv}\bigr)
= \tr\bigl(\Ad_ g\circ\ad_ u\circ\Ad_ g^{-1}\Ad_ g\circ\ad_ v\circ\Ad_
g^{-1}\bigr) =\tr\bigl(\Ad_ g\circ\ad_ u\circ\ad_ v\circ\Ad_ g^{-1}\bigr) =
\tr\bigl(\ad_ u\circ \ad_ v \bigr)\)} \[(\Ad_gu,\Ad_ gv) = (u,v) \quad\text{and
thus} \quad (\ad_ wu,v)+(u,\ad_ wv)=0.\] Thus whenever the Cartan-Killing form
is an inner product on \(\g\), its restriction to \(\p\) will define an
\(\Ad_\G\K\)-invariant inner product.  The components of the Killing form are
given by \(M_{ij}=c^k_{i\ell}c^\ell_{jk}\).  On \(\liealgebra O(n)\) (or
\(\liealgebra{SO}(n)\)) it is given by \(M(X,Y)=(n-2)\tr(XY)\).

When \(\K\) is a closed subgroup of \(\G\) and \(\Ad\K\) is a compact subset of
\(\GL\g\), (these conditions are satisfied when \(\K\) is the isotropy group of
the action of the isometry group on a symmetric space
(\cite{Holmelin2005symmetric} Theorem 3.2)), it is always possible to construct
\(\G\)-invariant metrics on the reductive homogeneous space \(\G/\K\): we start
with any inner product \(\langle \cdot, \cdot \rangle_\p\) on \(\p\), and
average it over \(\Ad\K\) using Weyl's unitary trick\footnote{Sometimes called
Weyl's unitarian trick, but Unitarian is a religious denomination.  The trick
is apparently due to Adolf Hurwitz.}.  This can be done since \(\Ad\K\) is a
compact Lie group\footnote{Since \(\GL V\) is a manifold, it must be Hausdorff,
from which it follows that \(\Ad(\K)\) is a Lie group since compact subset of
Hausdorff spaces are closed, and closed subgroup of \(\GL V\) are Lie groups},
and \(\Ad\K\p\subset\p\).  This defines an \(\Ad\K\)-invariant inner
product\footnote{The \(\Ad\K\)-invariance of \(( \cdot,\cdot)_\p\) follows from
the fact \(k\mapsto\Ad k\) is a homomorphism: \((\Ad_h u,\Ad_h v)_\p =
\int_{\Ad\K}dk\,\langle\Ad_k\circ\Ad_h u,\Ad_k\circ\Ad_hv\rangle_\p =
\int_{\Ad\K}dk\,\langle\Ad_{kh}u,\Ad_{kh} v\rangle_\p =\int_{\Ad\K}dk'\,
\langle\Ad_{k'}u,\Ad_{k'}v\rangle_\p = (u,v)_\p\), since the right action
\(R_h\) is a bijection.}  \((u,v)_\p\defn\int_{\Ad\K} dk\,\langle\Ad k(u),\Ad
k(v)\rangle_\p\) on \(\p\), (here \(dk\) is the Haar-measure on \(\Ad\K\)),
which in turn induces a \(\G\)-invariant metric on \(\G/\K\).

\section{Killing Fields on Homogeneous Space} \label{killing}

As we did for left-invariant vector fields, it is easy to check the flow of a
right-invariant vector field \(Y\), is given by \(t,g \mapsto e^{tY}g=
L_{e^{tY}}g\), that is right invariant vector fields generate left
translations.  It follows that these are isometries of left-invariant metric on
\(\G\).  A Killing field is a vector field on a Riemannian manifold whose flow
generates isometries of the metric~\cite{Paulette:1987}, hence right-invariant
vector fields are Killing fields.  The vector field \(Y\) on \(\G\) descends to
a vector field \(Y^{\G/\K} \in \field{T\G/\K}\), with
\begin{equation*}
  Y^{\G/\K}_{g \K}\defn\left.\frac d{dt}\right|_{t=0} e^{tY}g\K 
\end{equation*}
Note the flow of \(Y^{\G/\K}\) is \((t,g\K) \mapsto \Phi_L(e^{tY}, g \K)\).  In
particular if we have an action \(\Phi\) of \(\G\) on \(\M\) by isometries,
then \(\G\) acts by isometries on \(\G/\K\) (the action being \(\Phi_L\)) and
thus \(Y^{\G/\K}\) is a Killing field.

\section{Riemannian Manifolds and the Isometry Group} \label{riemannian}

If \(\Phi:\G\to\Diff\M\) is a left action of a Lie group \(\G\) on a manifold
\(\M\), where \(\Diff\M\) is the space of diffeomorphisms~\(\M\to\M\), and
\(X\in\g\), then we may define \(\Phi_X:\R\times\M\to\M\) where \(\Phi_X:(t,x)
\mapsto\Phi(e^{tX})(x)\) is a \(\R\)-action on \(\M\) (a flow).  The associated
vector field \(X_\M\in\field{\tanbundle\M}\), called the infinitesimal
generator of the action associated to \(X\), is given by
\begin{equation*} X_\M (q) =
  \frac d{dt}|_0 \Phi_q(\exp(tX)) = \tang 0 \Phi_q(X) 
\end{equation*}
In general if \(f:H\to\G\) is a group homomorphism, then \(\tang 1 f: \mathfrak
h \to\g\) is a Lie algebra homomorphism.  The map \(\g\to\field{T\M}\) with
\(X\mapsto X_\M\) above,\footnote{The Lie algebra of the group of
diffeomorphism is the algebra of vector fields} is precisely the
(anti)-homomorphism induced by the left action.  \footnote{The map
\((X,p)\mapsto(-X)_\M(p)\) is often called the action of \(\g\) on~\(\M\).}
 
Recall a \definition{pseudo-Riemannian manifold} is a manifold \(\M\) endowed
with a \definition{pseudo-Riemannian metric}, which is a map
\(\metric:\field{\tanbundle\M} \times\field{\tanbundle\M}\to\form0\M\) that is
bilinear \(\metric(eu,fv) = ef\metric(u,v)\), symmetric \(\metric(u,v) =
\metric(v,u)\), and non-degenerate \(\metric(u,v)=0 \quad \forall v\iff u=0\),
where \(e,f\in\form0\M\) are functions and \(u,v\in\field{\tanbundle\M}\) are
vector fields.

An \definition{isometry} is a diffeomorphism \(\sigma:\M\to\M\) that preserves
this metric, \((\pull\sigma\metric)(u,v)\defn\pull\sigma \bigl(\metric(\push
\sigma u,\push\sigma v)\bigr) = \metric(u,v)\).  Such isometries form a group
called the \definition{isometry group} \(\isog\M\) of \(\M\); a theorem of
Myers and Steenrod shows the isometry group is a Lie group~\cite{Myers:1939}.

In this case the induced vector fields \(X_\M\) are Killing vector fields,
i.e., the Lie derivative of the metric in their direction vanishes:
\(\Lie{X_\M} \langle \cdot,\cdot\rangle =0\).  The space of Killing fields
forms a Lie subalgebra of \(\field{T\M}\), and the map \(X\mapsto X_\M\) is an
anti-isomorphism between \(\g\) and the Lie algebra of complete Killing vector
fields~\cite{Paulette:1987}.  In particular when \(\M\) is complete, then every
Killing vector field is complete and \(X\mapsto X_\M\) is an anti-isomorphism.

\section{Maurer-Cartan form on \(\sphere2\)} \label{mc-form-on-s2}

In the parameterization \refeq{eq:S} \(S(\theta,\phi)\) defines a map \(\sigma:
\sphere2\to\liegroup{SO}(3)\) that is constant over cosets, and thus reduces to
a map \(\sphere2\to\liegroup{SO}(3)/\liegroup{SO}(2)\).  Moreover
\(H(t)\defn\exp(tT_1)\) parameterizes the isotropy group, and together \(\sigma
\defn S(\theta,\phi)H(t)\) can be used to parameterize \(\liegroup{SO}(3)\).%
\footnote{ The projection \(\pi:\liegroup{SO}(3,\R)\to\sphere2\) is a principal
\(\liegroup{SO}(2,\R)\) bundle.  If we let \(\rho:\sphere3\to
\liegroup{SO}(3,\R)\) be the standard double cover (i.e., two-to-one surjective
homomorphism) of \(\liegroup{SO}(3,\R)\) (here \(\sphere3\) is the three
sphere), then \(\pi \circ\rho:\sphere3\to\sphere2\) is the \definition{Hopf
  fibration}.}  We can pull-back Maurer--Cartan on \(\liegroup{SO}(3)\) using
\(\sigma\) which gives
\begin{align*} 
  \sigma ^{-1}d\sigma
  & = (S(\theta,\phi) H(t))^{-1}d(S(\theta,\phi) H(t))
  = H^{-1}S^{-1} dS H + H^{-1}S^{-1}S dH \\
  &= H^{-1}S^{-1}dS H+H^{-1}dH.
\end{align*}
We have 
\begin{align*}
  &H^{-1}dH = T_1dt \\
  &H^{-1}T_1 H = T_1 \\
  &H^{-1}T_2 H = \sin tT_3 + \cos tT_2 \\
  &H^{-1}T_3 H = \cos tT_3 -\sin tT_2
\end{align*}

Moreover
\begin{align*}
  S^1dS = T_1 (1-\cos(\theta)) d\phi & + T_2 \bigl(\sin(\phi)\,d\theta +
  \sin(\theta)\cos(\phi)\, d\phi\bigr) \\
  &+ T_3\bigl(\sin(\phi)\sin(\theta)\,
  d\phi - \cos(\phi)\,d\theta\bigr).
\end{align*}

Writing \(S^1\,dS\defn\omega T_1 + e^2 T_2 + e^3T_3\), we finally find 
\begin{align*}
  \sigma^1\,d\sigma = T_1\bigl(\omega + dt \bigr) + T_2\bigl(e_2\cos(t) -
  e_3\sin(t)\bigr) + T_3\bigl(e_3\cos(t) + e_2\sin(t)\bigr).
\end{align*}

\bibliographystyle{plain}
\bibliography{CMSS}

\begin{thebibliography}{10}

\bibitem{Abraham:2008}
Ralph Abraham and Jerrold~E. Marsden.
\newblock {\em Foundations of Mechanics}.
\newblock American Mathematical Society, second edition, May 2008.
\newblock with the assistance of {Tudor} {Ratiu} and {Richard} {Cushman}.

\bibitem{alekseevsky:1994}
Dmitri~V. Alekseevsky, Janusz Grabowski, Giuseppe Marmo, and Peter~W. Michor.
\newblock Poisson structures on the cotangent bundle of a {Lie} group or a
  principle bundle and their reductions.
\newblock {\em Journal of Mathematical Physics}, 35(9):4909--4927, 1994.

\bibitem{Barp:2018}
Alessandro Barp, {Fran\c{c}ois}-Xavier Briol, Anthony~D. Kennedy, and Mark
  Girolami.
\newblock {Geometry and Dynamics for Markov Chain Monte Carlo}.
\newblock {\em Annual Review of Statistics and its Application}, 2018.

\bibitem{BarpOates:2018}
Alessandro Barp, Chris Oates, Emilio Porcu, Mark Girolami, et~al.
\newblock A {Riemannian--Stein} kernel method.
\newblock {\em arXiv preprint arXiv:1810.04946}, 2018.

\bibitem{betancourt2013general}
Michael Betancourt.
\newblock A general metric for {Riemannian} manifold {Hamiltonian Monte Carlo}.
\newblock In {\em Geometric science of information}, pages 327--334. Springer,
  2013.

\bibitem{Betancourt:2017b}
Michael Betancourt.
\newblock A conceptual introduction to {Hamiltonian} {Monte} {Carlo}.
\newblock preprint arXiv:1701.02434, 2017.

\bibitem{Betancourt:2018}
Michael Betancourt.
\newblock The convergence of {Markov} {Chain} {Monte} {Carlo} methods: From the
  {Metropolis} method to {Hamiltonian} {Monte} {Carlo}.
\newblock {\em Ann. Phys.}, page 1700214, 2018.

\bibitem{Betancourt:2017a}
Michael Betancourt, Simon Byrne, Sam Livingstone, Mark Girolami, et~al.
\newblock The geometric foundations of {Hamiltonian} {Monte} {Carlo}.
\newblock {\em Bernoulli}, 23(4A):2257--2298, 2017.

\bibitem{betancourt2013generalizing}
Michael~J Betancourt.
\newblock Generalizing the no-{U}-turn sampler to {Riemannian} manifolds.
\newblock {\em arXiv preprint arXiv:1304.1920}, 2013.

\bibitem{bou2009hamilton}
Nawaf Bou-Rabee and Jerrold~E Marsden.
\newblock {Hamilton}--{Pontryagin} integrators on {Lie} groups part {I}:
  Introduction and structure-preserving properties.
\newblock {\em Foundations of Computational Mathematics}, 9(2):197--219, 2009.

\bibitem{bou2017randomized}
Nawaf Bou-Rabee, Jes{\'u}s~Mar{\'\i}a Sanz-Serna, et~al.
\newblock Randomized hamiltonian monte carlo.
\newblock {\em The Annals of Applied Probability}, 27(4):2159--2194, 2017.

\bibitem{Brubaker:2012}
Marcus~A. Brubaker, Mathieu Salzmann, and Raquel Urtasun.
\newblock A family of {MCMC} methods on implicitly defined manifolds.
\newblock In Neil Lawrence and Mark Reid, editors, {\em JMLR Workshop and
  Conference Proceedings}, volume~22, pages 161--172, 2012.

\bibitem{Simon:2013}
Simon Byrne and Mark Girolami.
\newblock {Geodesic Monte Carlo on Embedded Manifolds}.
\newblock {\em Scandinavian Journal of Statistics, Theory and Applications},
  2013.

\bibitem{callaway82a}
David J.~E. Callaway and Aneesur Rahman.
\newblock Microcanonical ensemble formulation of {Lattice} {Gauge} {Theory}.
\newblock {\em Phys. Rev. Lett.}, 49(9):613--616, 1982.

\bibitem{campostrini89a}
Massimo Campostrini and Paolo Rossi.
\newblock A comparison of numerical algorithms for dynamical fermions.
\newblock {\em Nucl. Phys.}, B329:753, 1990.

\bibitem{Choque1982analysis}
Yvonne Choquet-Bruhat, {C\'ecile} DeWitt-Morette, and Margaret Dillard-Bleick.
\newblock Analysis, manifolds and physics, revised edition, 1982.

\bibitem{creutz89a}
Michael Creutz and Andreas Gocksch.
\newblock Higher order {Hybrid} {Monte} {Carlo} algorithms.
\newblock {\em Phys. Rev. Lett.}, 63:9, 1989.

\bibitem{davies:2003}
Philip~I. Davies and Nicholas~J. Higham.
\newblock {A Schur--Parlett algorithm for computing matrix functions}.
\newblock {\em SIAM J. Matrix Anal. Appl.}, 25(2):464--485, 2003.

\bibitem{Deng:2012}
Shaoqiang Deng.
\newblock {\em {Homogeneous Finsler Spaces}}.
\newblock Springer-Verlag New York, 2012.

\bibitem{Diaconis:2013}
Persi Diaconis, Susan~P. Holmes, and Mehrdad Shahshahani.
\newblock Sampling from a manifold.
\newblock {\em Advances in Modern Statistical Theory and Applications: A
  Festschrift in honor of {Morris L. Eaton}}, 2013.

\bibitem{Duane:1987}
Simon Duane, A.~D. Kennedy, B.~J. Pendleton, and D.~Roweth.
\newblock {Hybrid Monte Carlo}.
\newblock {\em Phys. Lett. B}, 195(2):216--222, 1987.

\bibitem{Zdenek:2008}
Zdenek Dusek.
\newblock Survey on homogeneous geodesics.
\newblock {\em Note di Matematica}, 2008.

\bibitem{figueroa2005homogeneity}
Jos{\'e} Figueroa-O'Farrill, Simon Philip, and Patrick Meessen.
\newblock Homogeneity and plane-wave limits.
\newblock {\em Journal of High Energy Physics}, 2005(05):050, 2005.

\bibitem{gallier:2002}
Jean Gallier and Dianna Xu.
\newblock {Computing Exponentials of Skew-Symmetric Matrices and Logarithms of
  Orthogonal Matrices}.
\newblock {\em International Journal of Robotics and Automation}, 18(1):10--20,
  2002.

\bibitem{Girolami:2011}
M.~Girolami and B.~Calderhead.
\newblock {Riemann Manifold Langevin and Hamiltonian Monte Carlo methods}.
\newblock {\em J. Royal Stat. Soc. Series B: Stat. Method.}, 73(2):123--214,
  2011.

\bibitem{golub:1996}
Gene~H. Golub and Charles F.~Van Loan.
\newblock {\em {Matrix Computations}}.
\newblock The Johns Hopkins University Press, 1996.

\bibitem{Hairer:2006}
E.~Hairer, C.~Lubich, and G.~Wanner.
\newblock {\em {Geometric numerical integration algorithms for ordinary
  differential equations}}.
\newblock Springer, 2006.

\bibitem{hamelryck2006sampling}
Thomas Hamelryck, John~T Kent, and Anders Krogh.
\newblock Sampling realistic protein conformations using local structural bias.
\newblock {\em PLoS Computational Biology}, 2(9):e131, 2006.

\bibitem{hartmann2008ergodic}
Carsten Hartmann.
\newblock An ergodic sampling scheme for constrained hamiltonian systems with
  applications to molecular dynamics.
\newblock {\em Journal of Statistical Physics}, 130(4):687--711, 2008.

\bibitem{Helgason:2001}
Sigur$\eth$ur Helgason.
\newblock {\em {Differential Geometry, Lie Groups and Symmetric Spaces}}.
\newblock American Mathematical Society, 2001.

\bibitem{higham:2008}
Nicholas~J. Higham.
\newblock {\em {Functions of Matrices}}.
\newblock SIAM, 2008.

\bibitem{Holbrook:2017}
Andrew Holbrook, Shiwei Lan, Alexander Vandenberg-Rodes, and Babak Shahbaba.
\newblock {Geodesic Lagrangian Monte Carlo} over the space of positive definite
  matrices: with application to {Bayesian} spectral density estimation.
\newblock {\em Journal of Statistical Computation and Simulation}, 2017.

\bibitem{Holm2009geometric}
Darryl~D Holm, Tanya Schmah, and Cristina Stoica.
\newblock {\em Geometric mechanics and symmetry: from finite to infinite
  dimensions}, volume~12 of {\em Oxford texts in applied and engineering
  mathematics}.
\newblock Oxford University Press, 2009.

\bibitem{Holmelin2005symmetric}
Peter Holmelin.
\newblock {Symmetric Spaces}.
\newblock {Master' thesis, Lund Institute of Technology, Lund University},
  2005.

\bibitem{izaguirre2004shadow}
Jes{\'u}s~A Izaguirre and Scott~S Hampton.
\newblock Shadow hybrid {Monte Carlo}: an efficient propagator in phase space
  of macromolecules.
\newblock {\em Journal of Computational Physics}, 200(2):581--604, 2004.

\bibitem{kennedy88b}
A.~D. Kennedy and Pietro Rossi.
\newblock {Classical Mechanics on Group Manifolds}.
\newblock {\em Nucl. Phys.}, B327:782--790, 1989.

\bibitem{Kennedy:2012}
A.~D. Kennedy, P.~J. Silva, and M.~A. Clark.
\newblock {Shadow Hamiltonians, Poisson Brackets, and Gauge Theories}.
\newblock {\em Physical Review}, D87(3):034511, 2013.

\bibitem{kunze2004bingham}
Karsten Kunze and Helmut Schaeben.
\newblock The {Bingham} distribution of quaternions and its spherical radon
  transform in texture analysis.
\newblock {\em Mathematical Geology}, 36(8):917--943, 2004.

\bibitem{Kunze:2004}
Karsten Kunze and Helmut Schaeben.
\newblock {The Bingham Distribution of Quaternions and its Spherical Radon
  Transform in Texture Analysis}.
\newblock {\em Mathematical Geology}, 36(8):917--943, Nov 2004.

\bibitem{Lan:2014}
Shiwei Lan, Bo~Zhou, and Babak Shahbaba.
\newblock {Spherical Hamiltonian Monte Carlo for Constrained Target
  Distributions}.
\newblock In Eric~P. Xing and Tony Jebara, editors, {\em 31st International
  Conference on Machine Learning, Beijing, China}, volume~32 of {\em JMLR
  workshop and conference proceedings}, pages 629--637, 2014.

\bibitem{Leimkuhler:1996}
B.~Leimkuhler and G.~W. Patrick.
\newblock {A symplectic integrator for Riemannian Manifolds}.
\newblock {\em Journal of Nonlinear Science}, 1996.

\bibitem{lelievre2018hybrid}
Tony Leli{\`e}vre, Mathias Rousset, and Gabriel Stoltz.
\newblock {Hybrid Monte Carlo} methods for sampling probability measures on
  submanifolds.
\newblock {\em arXiv preprint arXiv:1807.02356}, 2018.

\bibitem{Paulette:1987}
Paulette Libermann and Charles-Michel Marle.
\newblock {\em {Symplectic Geometry and Analytical Mechanics}}.
\newblock Springer, 1987.

\bibitem{vanloan:1975}
Charles F.~Van Loan.
\newblock A study of the matrix exponential.
\newblock Technical report, University of Manchester, 1975.
\newblock {Numerical Analysis Report No.~10, University of Manchester,
  Manchester, UK, August 1975. Reissued as MIMS EPrint 2006.397, Manchester
  Institute for Mathematical Sciences, The University of Manchester, UK,
  November 2006}.

\bibitem{Mardia:1999}
Kanti~V. Mardia and Peter~E. Jupp.
\newblock {\em {Directional Statistics}}.
\newblock Wiley Series in Probability and Statistics, 1999.

\bibitem{Marsden:1974}
Jerrold Marsden and Alan Weinstein.
\newblock Reduction of symplectic manifolds with symmetry.
\newblock {\em Rep. Math. Phys.}, 5(1):121--130, 1974.

\bibitem{mezzadri2006generate}
Francesco Mezzadri.
\newblock How to generate random matrices from the classical compact groups.
\newblock {\em arXiv preprint math-ph/0609050}, 2006.

\bibitem{moler:2003}
Cleve Moler and Charles~Van Loan.
\newblock {Nineteen Dubious Ways to Compute the Exponential of a Matrix,
  Twenty-Five Years Later}.
\newblock {\em SIAM Review}, 45(1):3--49, 2003.

\bibitem{Myers:1939}
S.~B. Myers and N.~E. Steenrod.
\newblock {The Group of Isometries of a Riemannian Manifold}.
\newblock {\em Annals of Mathematics}, 1939.

\bibitem{Neal:2012}
Radford~M. Neal.
\newblock {MCMC} using {Hamiltonian} dynamics.
\newblock In Steve Brooks, Andrew Gelman, Galin Jones, and Xiao-Li Meng,
  editors, {\em {Handbook of Markov Chain Monte Carlo}}, chapter~5. {Chapman \&
  Hall/CRC Press}, 2011.

\bibitem{Parlett:1976}
Beresford~N. Parlett.
\newblock A recurrence among the elements of functions of triangular matrices.
\newblock {\em Linear Algebra Appl.}, 14(2):117--121, 1976.

\bibitem{Mayavi:2011}
P.~Ramachandran and G.~Varoquaux.
\newblock {Mayavi}: 3d visualization of scientific data.
\newblock {\em {IEEE} Computing in Science \& Engineering}, 13(2):40--51, 2011.

\bibitem{skeel2001practical}
Robert~D Skeel and David~J Hardy.
\newblock Practical construction of modified {Hamiltonians}.
\newblock {\em SIAM Journal on Scientific Computing}, 23(4):1172--1188, 2001.

\bibitem{Szekeres:2004}
Peter Szekeres.
\newblock {\em {A Course in Modern Mathematical Physics}}.
\newblock Cambridge University Press, 2004.

\bibitem{Wuestner:2003}
Michael W\"ustner.
\newblock {A Connected Lie Group Equals the Square of the Exponential Image}.
\newblock {\em Journal of Lie Theory}, 13:307--309, 2003.

\end{thebibliography}

\end{document}